\documentclass{acmsmall} 

\usepackage[latin1]{inputenc}
\usepackage[T1]{fontenc}

\usepackage{latexsym}
\usepackage{amsmath}
\usepackage{amssymb}
\usepackage{amsfonts}
\usepackage{mathrsfs}
\usepackage{stmaryrd}
\usepackage{textcomp}

\usepackage{times}

\usepackage{enumerate}

\usepackage{color}

\usepackage[all]{xy}
\usepackage{tikz}

\usepackage{graphicx}

\usepackage[small]{caption}
\usepackage{wrapfig}
\usepackage[small]{subfigure}
\newcommand{\subfiggap}{\hspace{\subfigtopskip}}

\usepackage{varioref}

\usepackage[pagewise]{lineno}

\usepackage{changebar}

\usepackage{substr}

\usepackage{microtype}

\newcommand{\getsub}[1]{\BeforeSubString{|}{#1}}

\newcommand{\gettxt}[1]{\BehindSubString{|}{#1}}

\newcommand{\showtxtsub}[1]{\ensuremath{\gettxt{#1}_{\getsub{#1}}}}

\newcommand{\txtstyname}[1]{\textsc{#1}}

\newcommand{\txtname}[2][]
	{{\ifx&#1&\txtstyname{#2}\else\txtstyname{#2}\txtsub{#1}\fi}}

\newcommand{\txtsub}[1]{$_{\text{\textnormal{#1}}}$}

\newcommand{\mthstyname}[1]{\mathcal{#1}}

\newcommand{\mthname}[2][]
	{{\ensuremath{\ifx&#1&\mthstyname{#2}\else\mthstyname{#2}\showtxtsub{#1}\fi}}}

\newcommand{\mthstystr}[1]{\mathfrak{#1}}

\newcommand{\mthstr}[2][]
	{{\ensuremath{\ifx&#1&\mthstystr{#2}\else\mthstystr{#2}\showtxtsub{#1}\fi}}}

\newcommand{\mthstycls}[1]{\mathbb{#1}}

\newcommand{\mthcls}[2][]
	{{\ensuremath{\ifx&#1&\mthstycls{#2}\else\mthstycls{#2}\showtxtsub{#1}\fi}}}

\newcommand{\mthstyfam}[1]{\mathscr{#1}}

\newcommand{\mthfam}[2][]
	{{\ensuremath{\ifx&#1&\mthstyfam{#2}\else\mthstyfam{#2}\showtxtsub{#1}\fi}}}

\newcommand{\mthstyset}[1]{\mathrm{#1}}

\newcommand{\mthset}[2][]
	{{\ensuremath{\ifx&#1&\mthstyset{#2}\else\mthstyset{#2}\showtxtsub{#1}\fi}}}

\newcommand{\mthstyfun}[1]{\mathsf{#1}}

\newcommand{\mthfun}[2][]
	{{\ensuremath{\ifx&#1&\mthstyfun{#2}\else\mthstyfun{#2}\showtxtsub{#1}\fi}}}

\newcommand{\mthstyrel}[1]{\mathit{#1}}

\newcommand{\mthrel}[2][]
	{{\ensuremath{\ifx&#1&\mthstyrel{#2}\else\mthstyrel{#2}\showtxtsub{#1}\fi}}}

\newcommand{\mthstysym}[1]{\mathsf{#1}}

\newcommand{\mthsym}[2][]
	{{\ensuremath{\ifx&#1&\mthstysym{#2}\else\mthstysym{#2}\showtxtsub{#1}\fi}}}

\newcommand{\mthstynot}[1]{\mathtt{#1}}

\newcommand{\mthnot}[2][]
	{{\ensuremath{\ifx&#1&\mthstynot{#2}\else\mthstynot{#2}\showtxtsub{#1}\fi}}}

\newcommand{\mthstyelm}[1]{#1}

\newcommand{\mthelm}[2][]
	{{\ensuremath{\ifx&#1&\mthstyelm{#2}\else\mthstyelm{#2}\showtxtsub{#1}\fi}}}

\newcommand{\AName}[1][]{\mthname[#1]{A}}

\newcommand{\GName}[1][]{\mthname[#1]{G}}
\newcommand{\HName}[1][]{\mthname[#1]{H}}

\newcommand{\NName}[1][]{\mthname[#1]{N}}

\newcommand{\TName}[1][]{\mthname[#1]{T}}
\newcommand{\UName}[1][]{\mthname[#1]{U}}

\newcommand{\ASet}[1][]{\mthset[#1]{A}}

\newcommand{\DSet}[1][]{\mthset[#1]{D}}

\newcommand{\FSet}[1][]{\mthset[#1]{F}}

\newcommand{\PSet}[1][]{\mthset[#1]{P}}
\newcommand{\QSet}[1][]{\mthset[#1]{Q}}
\newcommand{\RSet}[1][]{\mthset[#1]{R}}
\newcommand{\SSet}[1][]{\mthset[#1]{S}}
\newcommand{\TSet}[1][]{\mthset[#1]{T}}
\newcommand{\USet}[1][]{\mthset[#1]{U}}
\newcommand{\VSet}[1][]{\mthset[#1]{V}}
\newcommand{\WSet}[1][]{\mthset[#1]{W}}
\newcommand{\XSet}[1][]{\mthset[#1]{X}}
\newcommand{\YSet}[1][]{\mthset[#1]{Y}}
\newcommand{\ZSet}[1][]{\mthset[#1]{Z}}

\newcommand{\LFun}[1][]{\mthfun[#1]{L}}

\newcommand{\fFun}[1][]{\mthfun[#1]{f}}
\newcommand{\gFun}[1][]{\mthfun[#1]{g}}
\newcommand{\hFun}[1][]{\mthfun[#1]{h}}

\newcommand{\mFun}[1][]{\mthfun[#1]{m}}

\newcommand{\uFun}[1][]{\mthfun[#1]{u}}
\newcommand{\vFun}[1][]{\mthfun[#1]{v}}
\newcommand{\wFun}[1][]{\mthfun[#1]{w}}

\newcommand{\zFun}[1][]{\mthfun[#1]{z}}

\newcommand{\CRel}[1][]{\mthrel[#1]{C}}

\newcommand{\RRel}[1][]{\mthrel[#1]{R}}
\newcommand{\SRel}[1][]{\mthrel[#1]{S}}

\newcommand{\ASym}[1][]{\mthsym[#1]{A}}
\newcommand{\BSym}[1][]{\mthsym[#1]{B}}
\newcommand{\CSym}[1][]{\mthsym[#1]{C}}
\newcommand{\DSym}[1][]{\mthsym[#1]{D}}

\newcommand{\PSym}[1][]{\mthsym[#1]{P}}

\newcommand{\RSym}[1][]{\mthsym[#1]{R}}
\newcommand{\SSym}[1][]{\mthsym[#1]{S}}

\newcommand{\fSym}[1][]{\mthsym[#1]{f}}

\newcommand{\pSym}[1][]{\mthsym[#1]{p}}
\newcommand{\qSym}[1][]{\mthsym[#1]{q}}

\newcommand{\sSym}[1][]{\mthsym[#1]{s}}

\newcommand{\vSym}[1][]{\mthsym[#1]{v}}
\newcommand{\wSym}[1][]{\mthsym[#1]{w}}
\newcommand{\xSym}[1][]{\mthsym[#1]{x}}
\newcommand{\ySym}[1][]{\mthsym[#1]{y}}
\newcommand{\zSym}[1][]{\mthsym[#1]{z}}

\newcommand{\BNot}[1][]{\mthnot[#1]{B}}

\newcommand{\aElm}[1][]{\mthelm[#1]{a}}

\newcommand{\cElm}[1][]{\mthelm[#1]{c}}
\newcommand{\dElm}[1][]{\mthelm[#1]{d}}
\newcommand{\eElm}[1][]{\mthelm[#1]{e}}
\newcommand{\fElm}[1][]{\mthelm[#1]{f}}

\newcommand{\lElm}[1][]{\mthelm[#1]{l}}
\newcommand{\mElm}[1][]{\mthelm[#1]{m}}
\newcommand{\nElm}[1][]{\mthelm[#1]{n}}

\newcommand{\pElm}[1][]{\mthelm[#1]{p}}
\newcommand{\qElm}[1][]{\mthelm[#1]{q}}

\newcommand{\sElm}[1][]{\mthelm[#1]{s}}
\newcommand{\tElm}[1][]{\mthelm[#1]{t}}

\newcommand{\wElm}[1][]{\mthelm[#1]{w}}
\newcommand{\xElm}[1][]{\mthelm[#1]{x}}
\newcommand{\yElm}[1][]{\mthelm[#1]{y}}
\newcommand{\zElm}[1][]{\mthelm[#1]{z}}

\newcommand{\set}[2]
	{\{ \arga{#1} \allowbreak : \allowbreak \arga{#2} \}}

\newcommand{\card}[1]
	{\vert \arga{#1} \vert}

\newcommand{\pow}[1]
	{2^{\arga{#1}}}

\newcommand{\class}[2]
	{(\arga{#1} \allowbreak /\!\! \allowbreak \arga{#2})}

\newcommand{\defeq}
	{\triangleq}

\newcommand{\der}[2][]
	{{\widehat{\arga{#2}}_{#1}}}
\newcommand{\adj}[2][]
	{{\widetilde{\arga{#2}}_{#1}}}
\newcommand{\trn}[2][]
	{{\overline{\arga{#2}}_{#1}}}
\newcommand{\flip}[2][]
	{{\widehat{\arga{#2}}_{#1}}}
\newcommand{\dual}[2][]
	{{\overline{\arga{#2}}_{#1}}}

\newcommand{\arga}[1]
	{#1}
\newcommand{\argb}[2]
	{#1, \allowbreak #2}
\newcommand{\argc}[3]
	{#1, \allowbreak #2, \allowbreak #3}
\newcommand{\argd}[4]
	{#1, \allowbreak #2, \allowbreak #3, \allowbreak #4}
\newcommand{\arge}[5]
	{#1, \allowbreak #2, \allowbreak #3, \allowbreak #4, \allowbreak #5}
\newcommand{\argf}[6]
	{#1, \allowbreak #2, \allowbreak #3, \allowbreak #4, \allowbreak #5,
	\allowbreak #6}
\newcommand{\argg}[7]
	{#1, \allowbreak #2, \allowbreak #3, \allowbreak #4, \allowbreak #5,
	\allowbreak #6, \allowbreak #7}
\newcommand{\argh}[8]
	{#1, \allowbreak #2, \allowbreak #3, \allowbreak #4, \allowbreak #5,
	\allowbreak #6, \allowbreak #7, \allowbreak #8}

\newcommand{\tupleb}[2]
	{\langle \argb{#1}{#2} \rangle}
\newcommand{\tuplec}[3]
	{\langle \argc{#1}{#2}{#3} \rangle}
\newcommand{\tupled}[4]
	{\langle \argd{#1}{#2}{#3}{#4} \rangle}
\newcommand{\tuplee}[5]
	{\langle \arge{#1}{#2}{#3}{#4}{#5} \rangle}
\newcommand{\tuplef}[6]
	{\langle \argf{#1}{#2}{#3}{#4}{#5}{#6} \rangle}
\newcommand{\tupleg}[7]
	{\langle \argg{#1}{#2}{#3}{#4}{#5}{#6}{#7} \rangle}
\newcommand{\tupleh}[8]
	{\langle \argh{#1}{#2}{#3}{#4}{#5}{#6}{#7}{#8} \rangle}

\newcommand{\SetN}
	{\mthcls{N}}

\newcommand{\SetNI}
	{\der{\SetN}}

\newcommand{\numcc}[2]
	{[\argb{#1}{#2}]}
\newcommand{\numco}[2]
	{[\argb{#1}{#2}[\:}
\newcommand{\numoc}[2]
	{\:]\argb{#1}{#2}]}
\newcommand{\numoo}[2]
	{\:]\argb{#1}{#2}[\:}

\newcommand{\emptyfun}{\varnothing}

\newcommand{\pto}{\rightharpoonup}

\newcommand{\dom}[1]{\mthfun{dom}(\arga{#1})}

\newcommand{\cod}[1]{\mthfun{cod}(\arga{#1})}

\newcommand{\rng}[1]{\mthfun{rng}(\arga{#1})}

\newcommand{\cmp}{\circ}

\newcommand{\umrg}{\Cup}

\newcommand{\rst}{\upharpoonright}

\newcommand{\AOmicron}[1]
	{\mthnot{O}(\arga{#1})}

\newcommand{\pfx}[1]
	{\mthfun{pfx}(\arga{#1})}

\newcommand{\fst}[1]
	{\mthfun{fst}(\arga{#1})}
\newcommand{\lst}[1]
	{\mthfun{lst}(\arga{#1})}

\newcommand{\FOL}[1][]{\txtname{Fol\small\ifx&#1&\else\textnormal{[#1]}\fi}}

\newcommand{\MSOL}[1][]{\txtname{Msol\small\ifx&#1&\else\textnormal{[#1]}\fi}}

\newcommand{\MTL}[1][]{\txtname{Mtl\small\ifx&#1&\else\textnormal{[#1]}\fi}}

\newcommand{\MPL}[1][]{\txtname{Mpl\small\ifx&#1&\else\textnormal{[#1]}\fi}}

\newcommand{\PML}[1][]{\txtname{Pml\small\ifx&#1&\else\textnormal{[#1]}\fi}}

\newcommand{\MuCalculus}[1][]
	{\txtname{$\mu$Calculus\small\ifx&#1&\else\textnormal{[#1]}\fi}}

\newcommand{\AMuCalculus}{\txtname{A}\MuCalculus}

\newcommand{\LTL}[1][]{\txtname{Ltl\small\ifx&#1&\else\textnormal{[#1]}\fi}}

\newcommand{\QPTL}[1][]{\txtname{QPtl\small\ifx&#1&\else\textnormal{[#1]}\fi}}

\newcommand{\X}{{\mthsym{X}\:}}

\newcommand{\U}{{\mthsym{U}\:}}

\newcommand{\R}{{\mthsym{R}\:}}

\newcommand{\F}{{\mthsym{F}\:}}

\newcommand{\G}{{\mthsym{G}\:}}

\newcommand{\CTL}[1][]{\txtname{Ctl\small\ifx&#1&\else\textnormal{[#1]}\fi}}

\newcommand{\CTLP}[1][]
	{\txtname{Ctl\!$^{+}$\small\ifx&#1&\else\textnormal{[#1]}\fi}}

\newcommand{\CTLS}[1][]
	{\txtname{Ctl\!$^{\ast}$\small\ifx&#1&\else\textnormal{[#1]}\fi}}

\providecommand{\E}{\mthsym{E}}

\providecommand{\A}{\mthsym{A}}

\newcommand{\ATL}[1][]{\txtname{Atl\small\ifx&#1&\else\textnormal{[#1]}\fi}}

\newcommand{\ATLP}[1][]
	{\txtname{Atl\!$^{+}$\small\ifx&#1&\else\textnormal{[#1]}\fi}}

\newcommand{\ATLS}[1][]
	{\txtname{Atl\!$^{\ast}$\small\ifx&#1&\else\textnormal{[#1]}\fi}}

\newcommand{\SL}[1][]{\txtname{Sl\small\ifx&#1&\else\textnormal{[#1]}\fi}}

\newcommand{\OGSL}[1][]{\SL[\txtname{1g}\ifx&#1&\else\textnormal,~#1\fi]}

\newcommand{\BGSL}[1][]{\SL[\txtname{bg}\ifx&#1&\else\textnormal,~#1\fi]}

\newcommand{\NGSL}[1][]{\SL[\txtname{ng}\ifx&#1&\else\textnormal,~#1\fi]}

\newcommand{\TBSL}{\txtname{Tb-}\SL}

\newcommand{\TBOGSL}{\txtname{Tb-}\OGSL}

\newcommand{\TBBGSL}{\txtname{Tb-}\BGSL}

\newcommand{\TBNGSL}{\txtname{Tb-}\NGSL}

\newcommand{\CHPSL}{\txtname{CHP}-\SL}

\providecommand{\emodels}{\models_{\txtname{E}}}

\providecommand{\eimplies}{\Rightarrow_{\txtname{E}}}

\providecommand{\eequiv}{\equiv_{\txtname{E}}}

\newcommand{\CaRet}[1][]{\txtname{CaRet\small\ifx&#1&\else\textnormal{[#1]}\fi}}

\providecommand{\Tt}{\mthstr{t}}

\providecommand{\Ff}{\mthstr{f}}

\providecommand{\Opr}[1][]{{\mthsym[#1]{Op}\:}}

\providecommand{\Qnt}[1][]{{\mthsym[#1]{Qn}\:}}

\renewcommand{\implies}
	{\Rightarrow}

\newcommand{\pnf}{\textit{pnf}}
\newcommand{\enf}{\textit{enf}}

\newcommand{\GL}{\txtname{Gl}}

\newcommand{\CL}{\txtname{Cl}}

\newcommand{\APSet}
	{\mthset{A\!P}}

\newcommand{\VarSet}
	{\mthset{Var}}

\newcommand{\AsgSet}[1][]{\mthset[#1]{Asg}}

\newcommand{\asgFun}[1][]{\mthfun[#1]{\chi}}

\newcommand{\BndSet}[1][]{\mthset[#1]{Bnd}}

\newcommand{\bndFun}[1][]{\mthfun[#1]{\zeta}}

\newcommand{\fun}[3][]{\mthfun[#1]{#2}(\arga{#3})}

\newcommand{\lng}[2][]{\fun[#1]{lng}{#2}}

\newcommand{\sub}[2][]{\fun[#1]{sub}{#2}}
\newcommand{\snt}[2][]{\fun[#1]{snt}{#2}}

\newcommand{\psnt}[2][]{\fun[#1]{psnt}{#2}}

\newcommand{\alt}[2][]{\fun[#1]{alt}{#2}}

\newcommand{\free}[2][]{\fun[#1]{free}{#2}}

\newcommand{\Exs}[1]{\langle \arga{#1} \rangle}
\newcommand{\All}[1]{\lbrack \arga{#1} \rbrack}

\newcommand{\EExs}[1]{{\Exs{\!\Exs{#1}\!}}}
\newcommand{\AAll}[1]{{\All{\:\!\!\All{#1}\:\!\!}}}

\providecommand{\TTESet}[1][]{\mthset[#1]{TTE}}

\providecommand{\tteFun}[1][]{\mthfun[#1]{tte}}

\providecommand{\PTESet}[1][]{\mthset[#1]{PTE}}

\providecommand{\pteFun}[1][]{\mthfun[#1]{pte}}

% ...

\providecommand{\LabSet}[1][]{\mthset[#1]{\Sigma}}

\providecommand{\DirSet}[1][]{\mthset[#1]{\Delta}}

\providecommand{\TSet}[1][]{\mthset[#1]{T}}

\providecommand{\vFun}[1][]{\mthfun[#1]{v}}

\newcommand{\LTTuple}[4]
	{
	\ifx&#1&
		\ifx&#2&
			\tupleb{#3}{#4}
		\else
			\tuplec{#2}{#3}{#4}
		\fi
	\else
		\ifx&#2&
			\tuplec{#1}{#3}{#4}
		\else
			\tupled{#1}{#2}{#3}{#4}
		\fi
	\fi
	}

\newcommand{\LTDef}[3][]
	{
	\LTTuple {#2} {#3} {\TSet[#1]} {\vFun[#1]}
	}

\newcommand{\LTStruct}[1][]
	{
	\LTDef [#1] {} {}
	}

\newcommand{\TPG}{\txtname{Tpg}}

\newcommand{\TPA}{\txtname{Tpa}}

\providecommand{\NdESet}[1][]{\mthset[#1]{N_{e}}}

\providecommand{\NdOSet}[1][]{\mthset[#1]{N_{o}}}

\providecommand{\EdgRel}[1][]{\mthrel[#1]{E}}

\providecommand{\EdgERel}[1][]{\mthrel[#1]{E_{e}}}

\providecommand{\EdgORel}[1][]{\mthrel[#1]{E_{o}}}

\providecommand{\PosESet}[1][]{\mthset[#1]{Pos_{e}}}

\providecommand{\PosOSet}[1][]{\mthset[#1]{Pos_{o}}}

\providecommand{\posElm}[1][]{\mthelm[#1]{\varrho}}

\providecommand{\SchESet}[1][]{\mthset[#1]{Sch_{e}}}

\providecommand{\SchOSet}[1][]{\mthset[#1]{Sch_{o}}}

\providecommand{\scheFun}[1][]{\mthfun[#1]{s_{e}}}

\providecommand{\schoFun}[1][]{\mthfun[#1]{s_{o}}}

\providecommand{\MtcSet}[1][]{\mthset[#1]{Mtc}}

\providecommand{\mtcElm}[1][]{\mthelm[#1]{\varpi}}

\providecommand{\mtcFun}[1][]{\mthfun[#1]{mtc}}

\providecommand{\WinSet}[1][]{\mthset[#1]{Win}}

\newcommand{\TPATuple}[4]
	{
	\ifx&#4&
		\tuplec{#1}{#2}{#3}
	\else
		\tupled{#1}{#2}{#3}{#4}
	\fi
	}

\newcommand{\TPADef}[2][]
	{
	\TPATuple {\NdESet[#2]} {\NdOSet[#2]} {\EdgRel[#2]} {#1}
	}

\newcommand{\TPAStruct}[1][ {\nElm[0]} ]
	{
	\TPADef [#1] {}
	}

\newcommand{\TPGTuple}[2]
	{
	\tupleb{#1}{#2}
	}

\newcommand{\TPGDef}[1][]
	{
	\TPGTuple {\AName[#1]} {\WinSet[#1]}
	}

\newcommand{\TPGStruct}[1][]
	{
	\TPGDef [#1]
	}

\providecommand{\APSet}[1][]{\mthset[#1]{AP}}

\providecommand{\WSet}[1][]{\mthset[#1]{W}}

\providecommand{\RRel}[1][]{\mthrel[#1]{R}}

\providecommand{\LFun}[1][]{\mthfun[#1]{L}}

\newcommand{\KSTuple}[5]
	{
	\ifx&#5&
		\tupled{#1}{#2}{#3}{#4}
	\else
		\tuplee{#1}{#2}{#3}{#4}{#5}
	\fi
	}

\newcommand{\KSDef}[2][]
	{
	\KSTuple {\APSet} {\WSet[#2]} {\RRel[#2]} {\LFun[#2]} {#1}
	}

\newcommand{\KSStruct}[1][ {\wElm[0]} ]
	{
	\KSDef [#1] {}
	}

\providecommand{\PthSet}[1][]{\mthset[#1]{Pth}}

\providecommand{\pthSym}[1][]{\mthsym[#1]{\pi}}

\providecommand{\TrkSet}[1][]{\mthset[#1]{Trk}}

\providecommand{\trkSym}[1][]{\mthsym[#1]{\rho}}

\newcommand{\CGS}{\txtname{Cgs}}

\providecommand{\APSet}[1][]{\mthset[#1]{AP}}

\providecommand{\AgSet}[1][]{\mthset[#1]{Ag}}

\providecommand{\AcSet}[1][]{\mthset[#1]{Ac}}

\providecommand{\StSet}[1][]{\mthset[#1]{St}}

\providecommand{\labFun}[1][]{\mthfun[#1]{\lambda}}

\providecommand{\chcFun}[1][]{\mthfun[#1]{\xi}}

\providecommand{\trnFun}[1][]{\mthfun[#1]{\tau}}

\providecommand{\DecSet}[1][]{\mthset[#1]{Dc}}

\providecommand{\decFun}[1][]{\mthfun[#1]{d}}

\providecommand{\ownFun}[1][]{\mthfun[#1]{\eta}}

\newcommand{\CGSTuple}[7]
	{
	\ifx&#7&
		\tuplef{#1}{#2}{#3}{#4}{#5}{#6}
	\else
		\tupleg{#1}{#2}{#3}{#4}{#5}{#6}{#7}
	\fi
	}

\newcommand{\ECGSTuple}[8]
	{
	\ifx&#8&
		\tupleg{#1}{#2}{#3}{#4}{#5}{#6}{#7}
	\else
		\tupleh{#1}{#2}{#3}{#4}{#5}{#6}{#7}{#8}
	\fi
	}

\newcommand{\CGSDef}[2][]
	{
	\CGSTuple {\APSet} {\AgSet} {\AcSet[#2]} {\StSet[#2]} {\labFun[#2]}
		{\trnFun[#2]} {#1}
	}

\newcommand{\ECGSDef}[2][]
	{
	\ECGSTuple {\APSet} {\AgSet} {\AcSet[#2]} {\StSet[#2]} {\labFun[#2]}
		{\chcFun[#2]} {\trnFun[#2]} {#1}
	}

\newcommand{\CGSStruct}[1][ {\sElm[0]} ]
	{
	\CGSDef [#1] {}
	}

\newcommand{\ECGSStruct}[1][ {\sElm[0]} ]
	{
	\ECGSDef [#1] {}
	}

\providecommand{\StrSet}[1][]{\mthset[#1]{Str}}

\providecommand{\strFun}[1][]{\mthfun[#1]{f}}

\providecommand{\TrkSet}[1][]{\mthset[#1]{Trk}}

\providecommand{\trkElm}[1][]{\mthelm[#1]{\rho}}

\providecommand{\trkSym}[1][]{\mthsym[#1]{\rho}}

\providecommand{\PthSet}[1][]{\mthset[#1]{Pth}}

\providecommand{\pthElm}[1][]{\mthelm[#1]{\pi}}

\providecommand{\pthSym}[1][]{\mthsym[#1]{\pi}}

\providecommand{\playElm}[1][]{\mthelm[#1]{\pi}}

\providecommand{\playFun}[1][]{\mthfun[#1]{play}}

\newcommand{\QPSet}[1][]{\mthset[#1]{Qnt}}

\newcommand{\qpSym}[1][]{\mthsym[#1]{\wp}}

\newcommand{\qpElm}[1][]{\mthelm[#1]{\wp}}

\newcommand{\QPVSet}[1][]{\mthset[#1]{V}}

\newcommand{\QPEVSet}[2][]{\EExs{#2}}

\newcommand{\QPAVSet}[2][]{\AAll{#2}}

\newcommand{\qpordRel}[1][]{\mthrel[#1]{<}}

\newcommand{\qpdepRel}[1][]{\mthrel[#1]{\rightsquigarrow}}

\newcommand{\QPDepSet}[1][]{\mthset[#1]{Dep}}

\newcommand{\BPSet}[1][]{\mthset[#1]{Bnd}}

\newcommand{\bpFun}[1][]{\mthfun[#1]{\zeta}}

\newcommand{\bpSym}[1][]{\mthsym[#1]{\flat}}

\newcommand{\bpElm}[1][]{\mthelm[#1]{\flat}}

\providecommand{\ValSet}[1][]{\mthset[#1]{Val}}

\providecommand{\valFun}[1][]{\mthfun[#1]{v}}

\newcommand{\SpcSet}[1][]{\mthset[#1]{DM}}

\newcommand{\ESpcSet}[1][]{\mthset[#1]{EDM}}

\newcommand{\spcFun}[1][]{\mthfun[#1]{\theta}}

\providecommand{\DSet}[1][]{\mthset[#1]{D}}

\newcommand{\DSTuple}[4]
	{
	\ifx&#4&
		\tuplec{#1}{#2}{#3}
	\else
		\tupled{#1}{#2}{#3}{#4}
	\fi
	}

\newcommand{\PCPTuple}[4]
	{
	\ifx&#4&
		\tuplec{#1}{#2}{#3}
	\else
		\tupled{#1}{#2}{#3}{#4}
	\fi
	}

\newcommand{\NBW}{\txtname{Nbw}}

\newcommand{\UCW}{\txtname{Ucw}}

\providecommand{\SymSet}[1][]{\mthset[#1]{\Sigma}}

\providecommand{\QSet}[1][]{\mthset[#1]{Q}}

\providecommand{\PSet}[1][]{\mthset[#1]{P}}

\providecommand{\atFun}[1][]{\mthfun[#1]{\delta}}

\newcommand{\WATuple}[5]
	{
	\ifx&#5&
		\tupled{#1}{#2}{#3}{#4}
	\else
		\tuplee{#1}{#2}{#3}{#4}{#5}
	\fi
	}

\newcommand{\WMTuple}[7]
	{
	\ifx&#7&
		\tuplef{#1}{#2}{#3}{#4}{#5}{#6}
	\else
		\tupleg{#1}{#2}{#3}{#4}{#5}{#6}{#7}
	\fi
	}

\newcommand{\NTA}{\txtname{Nta}}
\newcommand{\UTA}{\txtname{Uta}}
\newcommand{\ATA}{\txtname{Ata}}

\newcommand{\UCT}{\txtname{Uct}}
\newcommand{\ACT}{\txtname{Act}}

\newcommand{\NPT}{\txtname{Npt}}
\newcommand{\UPT}{\txtname{Upt}}
\newcommand{\APT}{\txtname{Apt}}

\providecommand{\SymSet}[1][]{\mthset[#1]{\Sigma}}

\providecommand{\RDirSet}[1][]{\mthset[#1]{\Delta}}

\providecommand{\QSet}[1][]{\mthset[#1]{Q}}

\providecommand{\PSet}[1][]{\mthset[#1]{P}}

\providecommand{\atFun}[1][]{\mthfun[#1]{\delta}}

\newcommand{\TATuple}[6]
	{
	\ifx&#2&
		\ifx&#6&
			\tupled{#1}{#3}{#4}{#5}
		\else
			\tuplee{#1}{#3}{#4}{#5}{#6}
		\fi
	\else
		\ifx&#6&
			\tuplee{#1}{#2}{#3}{#4}{#5}
		\else
			\tuplef{#1}{#2}{#3}{#4}{#5}{#6}
		\fi
	\fi
	}

\newcommand{\TMTuple}[8]
	{
	\ifx&#3&
		\ifx&#8&
			\tuplef{#1}{#2}{#4}{#5}{#6}{#7}
		\else
			\tupleg{#1}{#2}{#4}{#5}{#6}{#7}{#8}
		\fi
	\else
		\ifx&#8&
			\tupleg{#1}{#2}{#3}{#4}{#5}{#6}{#7}
		\else
			\tupleh{#1}{#2}{#3}{#4}{#5}{#6}{#7}{#8}
		\fi
	\fi
	}

\newcommand{\TADef}[4][q_{0}]
	{
	\TATuple {\SymSet} {#2} {\QSet[#4]} {\atFun[#4]} {#1} {#3}
	}

\newcommand{\TAStruct}[2][q_{0}]
	{
	\TADef [#1] {#2} {\aleph} {}
	}

\newcommand{\ATAStruct}[1][q_{0}]
	{
	\TAStruct [#1] {\RDirSet}
	}

\providecommand{\BoolSet}[1][]{\BNot[#1]}

\providecommand{\PBoolSet}[1][]{\BNot[#1]^{+}}

\providecommand{\LangSet}[1][]{\mthset[#1]{L}}

\providecommand{\infFun}[1][]{\mthfun[#1]{inf}}

\newcommand{\PTime}{\txtname{PTime}}

\newcommand{\PTimeC}{\PTime-\CComp}

\newcommand{\ExpTime}{\txtname{ExpTime}}

\newcommand{\ExpTimeC}{\ExpTime-\CComp}

\newcommand{\ExpSpace}{\txtname{ExpSpace}}
\newcommand{\ExpSpaceH}{\ExpSpace-\HComp}

\newcommand{\NElmTime}{\txtname{NonElementaryTime}}

\newcommand{\NElmSpace}{\txtname{NonElementarySpace}}
\newcommand{\NElmSpaceH}{\NElmSpace-\HComp}

\newcommand{\HComp}{\txtname{hard}}

\newcommand{\CComp}{\txtname{complete}}

\renewcommand{\epsilon}{\varepsilon}

\usetikzlibrary{arrows,shapes,backgrounds}

\tikzstyle{every node} = [draw = black, fill = white, thin]
\tikzstyle{every edge} += [thick]

\tikzstyle{plain0} = [draw = none, fill = none]
\tikzstyle{plain1} = [draw = none]

\tikzstyle{player} = [circle]
\tikzstyle{player0} = [circle]
\tikzstyle{player1} = [regular polygon, regular polygon sides = 4]
\tikzstyle{player2} = [diamond]
\tikzstyle{player3} = [regular polygon, regular polygon sides = 5]

\newcommand{\toignore}[1]{}

\title
	{
	Reasoning About Strategies: \\
	On the Model-Checking Problem
	}

\author
	{
	FABIO MOGAVERO, ANIELLO MURANO, and GIUSEPPE PERELLI
	\affil{\\Universit\'a degli Studi di Napoli "Federico II", Napoli, Italy.}
	MOSHE Y. VARDI
	\affil{\\Rice University, Houston, Texas, USA.}
	}

\begin{abstract}

	In \emph{open systems verification}, to formally check for \emph{reliability},
	one needs an appropriate formalism to model the \emph{interaction} between
	\emph{agents} and express the \emph{correctness} of the system no matter how
	the \emph{environment} behaves.
	An important contribution in this context is given by \emph{modal logics} for
	\emph{strategic ability}, in the setting of \emph{multi-agent games}, such as
	\ATL, \ATLS, and the like.
	Recently, Chatterjee, Henzinger, and Piterman introduced \emph{Strategy
	Logic}, which we denote here by \CHPSL, with the aim of getting a powerful
	framework for reasoning explicitly about strategies.
	\CHPSL\ is obtained by using \emph{first-order quantifications} over
	strategies and has been investigated in the very specific setting of
	\emph{two-agents turned-based games}, where a non-elementary model-checking
	algorithm has been provided.
	While \CHPSL\ is a very expressive logic, we claim that it does not fully
	capture the strategic aspects of multi-agent systems.

	In this paper, we introduce and study a more general strategy logic, denoted
	\SL, for reasoning about strategies in \emph{multi-agent concurrent games}.
	We prove that \SL\ includes \CHPSL, while maintaining a decidable
	model-checking problem.
	In particular, the algorithm we propose is computationally not harder than the
	best one known for \CHPSL.
	Moreover, we prove that such a problem for \SL\ is \NElmSpaceH.
	This negative result has spurred us to investigate here syntactic fragments of
	\SL, strictly subsuming \ATLS, with the hope of obtaining an elementary
	model-checking problem.
	Among the others, we study the sublogics \NGSL, \BGSL, and \OGSL.
	They encompass formulas in a special \emph{prenex normal form} having,
	respectively, nested temporal goals, Boolean combinations of goals and, a
	single goal at a time.
	About these logics, we prove that the model-checking problem for \OGSL\ is
	2\ExpTimeC, thus not harder than the one for \ATLS.
	In contrast, \NGSL\ turns out to be \NElmSpaceH, strengthening the
	corresponding result for \SL.
	Finally, we observe that \BGSL\ includes \CHPSL, while its model-checking
	problem relies between \NElmTime\ and 2\ExpTime.

\end{abstract}





\acmformat{... .}

\begin{document}

	\flushbottom

	\begin{bottomstuff}
		This work is partially based on the paper~\cite{MMV10b}, which appeared in
		FSTTCS'10.
	\end{bottomstuff}

	\maketitle




\begin{section}{Introduction}
	\label{sec:introduction}

	In system design, \emph{model checking} is a well-established formal method
	that allows to automatically check for global system
	correctness~\cite{CE81,QS81,CGP02}.
	In such a framework, in order to check whether a system satisfies a required
	property, we describe its structure in a mathematical model (such as
	\emph{Kripke structures}~\cite{Kri63} or \emph{labeled transition
	systems}~\cite{Kel76}), specify the property with a formula of a temporal
	logic (such as \LTL~\cite{Pnu77}, \CTL~\cite{CE81}, or \CTLS~\cite{EH86}), and
	check formally that the model satisfies the formula.
	In the last decade, interest has arisen in analyzing the behavior of
	individual components or sets of them in systems with several entities.
	This interest has started in reactive systems, which are systems that interact
	continually with their environments.
	In \emph{module checking}~\cite{KVW01}, the system is modeled as a module that
	interacts with its environment and correctness means that a desired property
	holds with respect to all such interactions.
	\\ \indent
	Starting from the study of module checking, researchers have looked for logics
	focusing on the strategic behavior of agents in multi-agent
	systems~\cite{AHK02,Pau02,JH04}.
	One of the most important development in this field is \emph{Alternating-Time
	Temporal Logic} (\ATLS, for short), introduced by Alur, Henzinger, and
	Kupferman~\cite{AHK02}.
	\ATLS\ allows reasoning about strategies of agents with temporal goals.
	Formally, it is obtained as a generalization of \CTLS\ in which the path
	quantifiers, \emph{there exists} ``$\E$'' and \emph{for all} ``$\A$'', are
	replaced with \emph{strategic modalities} of the form ``$\EExs{\ASet}$'' and
	``$\AAll{\ASet}$'', where $\ASet$ is a set of \emph{agents} (a.k.a.
	\emph{players}).
	Strategic modalities over agent sets are used to express cooperation and
	competition among them in order to achieve certain goals.
	In particular, these modalities express selective quantifications over those
	paths that are the result of infinite games between a coalition and its
	complement.
	\\ \indent
	\ATLS\ formulas are interpreted over \emph{concurrent game structures} (\CGS,
	for short)~\cite{AHK02}, which model interacting processes.
	Given a \CGS\ $\GName$ and a set $\ASet$ of agents, the \ATLS\ formula
	$\EExs{\ASet} \psi$ is satisfied at a state $\sElm$ of $\GName$ if there is a
	set of strategies for agents in $\ASet$ such that, no matter strategies are
	executed by agents not in $\ASet$, the resulting outcome of the interaction in
	$\GName$ satisfies $\psi$ at $\sElm$.
	Thus, \ATLS\ can express properties related to the interaction among
	components, while \CTLS\ can only express property of the global system.
	As an example, consider	the property ``processes $\alpha$ and $\beta$
	cooperate to ensure that a system (having more than two processes) never
	enters a failure state''.
	This can be expressed by the \ATLS\ formula $\EExs{\{ \alpha, \beta \}} \G
	\neg \mathit{fail}$, where $\G$ is the classical \LTL\ temporal operators
	``\emph{globally}''.
	\CTLS, in contrast, cannot express this property~\cite{AHK02}.
	Indeed, it can only assert whether the set of all agents may or may not
	prevent the system from entering a failure state.
	\\ \indent
	The price that one has to pay for the greater expressiveness of \ATLS\ is the
	increased complexity of model checking.
	Indeed, both its model-checking and satisfiability problems are
	2\ExpTimeC~\cite{AHK02,Sch08}.
	\\ \indent
	Despite its powerful expressiveness, \ATLS\ suffers from a strong limitation,
	due to the fact that strategies are treated only implicitly, through
	modalities that refer to games between competing coalitions.
	To overcome this problem, Chatterjee, Henzinger, and Piterman introduced
	\emph{Strategy Logic} (\CHPSL, for short)~\cite{CHP07}, a logic that treats
	strategies in \emph{two-player turn-based games} as explicit \emph{first-order
	objects}.
	In \CHPSL, the \ATLS\ formula $\EExs{\{ \alpha \}} \psi$, for a system modeled
	by a \CGS\ with agents $\alpha$ and $\beta$, becomes $\exists \xSym. \forall
	\ySym. \psi(\xSym, \ySym)$, i.e., ``there exists a player-$\alpha$ strategy
	$\xSym$ such that for all player-$\beta$ strategies $\ySym$, the unique
	infinite path resulting from the two players following the strategies $\xSym$
	and $\ySym$ satisfies the property $\psi$''.
	The explicit treatment of strategies in this logic allows to state many
	properties not expressible in \ATLS.
	In particular, it is shown in~\cite{CHP07} that \ATLS, in the restricted case
	of two-agent turn-based games, corresponds to a proper one-alternation
	fragment of \CHPSL.
	The authors of that work have also shown that the model-checking problem for
	\CHPSL\ is decidable, although only a non-elementary algorithm for it, both in
	the size of system and formula, has been provided, leaving as open question
	whether an algorithm with a better complexity exists or not.
	The complementary question about the decidability of the satisfiability
	problem for \CHPSL\ was also left open and, as far as we known, it is not
	addressed in other papers apart our preliminary work~\cite{MMV10b}.
	\\ \indent
	While the basic idea exploited in~\cite{CHP07} to quantify over strategies and
	then to commit agents explicitly to certain of these strategies turns to be
	very powerful and useful~\cite{FKL10}, \CHPSL\ still presents severe
	limitations.
	Among the others, it needs to be extended to the more general concurrent
	multi-agent setting.
	Also, the specific syntax considered there allows only a weak kind of strategy
	commitment.
	For example, \CHPSL\ does not allow different players to share the same
	strategy, suggesting that strategies have yet to become first-class objects in
	this logic.
	Moreover, an agent cannot change his strategy during a play without forcing
	the other to do the same.
	\\ \indent
	These considerations, as well as all questions left open about decision
	problems, led us to introduce and investigate a new \emph{Strategy
	Logic}, denoted \SL, as a more general framework than \CHPSL, for explicit
	reasoning about strategies in \emph{multi-agent concurrent games}.
	Syntactically, \SL\ extends \LTL\ by means of two \emph{strategy quantifiers},
	the existential $\EExs{\xElm}$ and the universal $\AAll{\xElm}$, as well as
	\emph{agent binding} $(\aElm, \xElm)$, where $\aElm$ is an agent and $\xElm$ a
	variable.
	Intuitively, these elements can be respectively read as \emph{``there exists a
	strategy $\xElm$''}, \emph{``for all strategies $\xElm$''}, and \emph{``bind
	agent $\aElm$ to the strategy associated with $\xElm$''}.
	For example, in a \CGS\ with the three agents $\alpha$, $\beta$, $\gamma$, the
	previous \ATLS\ formula $\EExs{\{ \alpha, \beta \}} \G \neg \mathit{fail}$ can
	be translated in the \SL\ formula $\EExs{\xSym} \EExs{\ySym} \AAll{\zSym}
	(\alpha, \xSym) (\beta, \ySym) (\gamma, \zSym) (\G \neg \mathit{fail})$.
	The variables $\xSym$ and $\ySym$ are used to select two strategies for the
	agents $\alpha$ and $\beta$, respectively, while $\zSym$ is used to select one
	for the agent $\gamma$ such that their composition, after the binding, results
	in a play where $\mathit{fail}$ is never met.
	Note that we can also require, by means of an appropriate choice of agent
	bindings, that agents $\alpha$ and $\beta$ share the same strategy, using the
	formula $\EExs{\xSym} \AAll{\zSym} (\alpha, \xSym) (\beta, \xSym) (\gamma,
	\zSym) (\G \neg \mathit{fail})$.
	Furthermore, we may vary the structure of the game by changing the way the
	quantifiers alternate, as in the formula $\EExs{\xSym} \AAll{\zSym}
	\EExs{\ySym} (\alpha, \xSym) (\beta, \ySym) (\alpha, \zSym) (\G \neg
	\mathit{fail})$.
	In this case, $\xSym$ remains uniform w.r.t. $\zSym$, but $\ySym$ becomes
	dependent on it.
	Finally, we can change the strategy that one agent uses during the play
	without changing those of the other agents, by simply using nested bindings,
	as in the formula $\EExs{\xSym} \EExs{\ySym} \AAll{\zSym} \EExs{\wSym}
	(\alpha, \xSym) (\beta, \ySym) (\gamma, \zSym) (\G (\gamma, \wSym) \G \neg
	\mathit{fail})$.
	The last examples intuitively show that \SL\ is a extension of both \ATLS\ and
	\CHPSL.
	It is worth noting that the pattern of modal quantifications over strategies
	and binding to agents can be extended to other linear-time temporal logics
	than \LTL, such as the linear \MuCalculus~\cite{Var88}.
	In fact, the use of \LTL\ here is only a matter of simplicity in presenting
	our framework, and changing the embedded temporal logic only involves few
	side-changes in proofs and decision procedures.
	\\ \indent
	As one of the main results in this paper about \SL, we show that the
	model-checking problem is non-elementarily decidable.
	To gain this, we use an \emph{automata-theoretic approach}~\cite{KVW00}.
	Precisely, we reduce the decision problem for our logic to the emptiness
	problem of a suitable \emph{alternating parity tree automaton}, which is an
	\emph{alternating tree automaton} (see~\cite{GTW02}, for a survey) along with
	a \emph{parity acceptance condition}~\cite{MS95}.
	Due to the operations of projection required by the elimination of
	quantifications on strategies, which induce at any step an exponential
	blow-up, the overall size of the required automaton is non-elementary in the
	size of the formula, while it is only polynomial in the size of the model.
	Thus, together with the complexity of the automata-nonemptiness calculation,
	we obtain that the model checking problem is in \PTime, w.r.t. the size of the
	model, and \NElmTime, w.r.t. the size of the specification.
	Hence, the algorithm we propose is computationally not harder than the best
	one known for \CHPSL\ and even a non-elementary improvement with respect to
	the model.
	This fact allows for practical applications of \SL\ in the field of system
	verification just as those done for the monadic second-order logic on infinite
	objects~\cite{EKM98}.
	Moreover, we prove that our problem has a non-elementary lower bound.
	Specifically, it is $k$-\ExpSpaceH\ in the alternation number $k$ of
	quantifications in the specification.
	\\ \indent
	The contrast between the high complexity of the model-checking problem for our
	logic and the elementary one for \ATLS\ has spurred us to investigate
	syntactic fragments of \SL, strictly subsuming \ATLS, with a better
	complexity.
	In particular, by means of these sublogics, we would like to understand why
	\SL\ is computationally more difficult than \ATLS.
	\\ \indent
	The main fragments we study here are \emph{Nested-Goal}, \emph{Boolean-Goal},
	and \emph{One-Goal Strategy Logic}, respectively denoted by \NGSL, \BGSL, and
	\OGSL.
	Note that the last, differently from the first two, was introduced
	in~\cite{MMPV12}.
	They encompass formulas in a special prenex normal form having nested temporal
	goals, Boolean combinations of goals, and a single goal at a time,
	respectively.
	For goal we mean an \SL\ formula of the type $\bpElm \psi$, where $\bpElm$ is
	a binding prefix of the form $(\alpha_{1}, \xElm[1]), \ldots, (\alpha_{n},
	\xElm[n])$ containing all the involved agents and $\psi$ is an agent-full
	formula.
	With more detail, the idea behind \NGSL\ is that, when in $\psi$ there is a
	quantification over a variable, then there are quantifications of all free
	variables contained in the inner subformulas.
	So, a subgoal of $\psi$ that has a variable quantified in $\psi$ itself cannot
	use other variables quantified out of this formula.
	Thus, goals can be only nested or combined with Boolean and temporal
	operators.
	\BGSL\ and \OGSL\ further restrict the use of goals.
	In particular, in \OGSL, each temporal formula $\psi$ is prefixed by a
	quantification-binding prefix $\qpElm \bpElm$ that quantifies over a tuple of
	strategies and binds them to all agents.
	\\ \indent
	As main results about these fragments, we prove that the model-checking
	problem for \OGSL\ is 2\ExpTimeC, thus not harder than the one for \ATLS.
	On the contrary, for \NGSL, it is both \NElmTime\ and \NElmSpaceH\ and thus we
	enforce the corresponding result for \SL.
	Finally, we observe that \BGSL\ includes \CHPSL, while the relative
	model-checking problem relies between 2\ExpTime\ and \NElmTime.
	\\ \indent
	To achieve all positive results about \OGSL, we use a fundamental property of
	the semantics of this logic, called \emph{elementariness}, which allows us to
	strongly simplify the reasoning about strategies by reducing it to a set of
	reasonings about actions.
	This intrinsic characteristic of \OGSL, which unfortunately is not shared by
	the other fragments, asserts that, in a determined history of the play, the
	value of an existential quantified strategy depends only on the values of
	strategies, from which the first depends, on the same history.
	This means that, to choose an existential strategy, we do not need to know the
	entire structure of universal strategies, as for \SL, but only their values on
	the histories of interest.
	Technically, to describe this property, we make use of the machinery of
	\emph{dependence map}, which defines a Skolemization procedure for \SL,
	inspired by the one in first order logic.
	\\ \indent
	By means of elementariness, we can modify the \SL\ model-checking procedure
	via alternating tree automata in such a way that we avoid the projection
	operations by using a dedicated automaton that makes an action quantification
	for each node of the tree model.
	Consequently, the resulting automaton is only exponential in the size of the
	formula, independently from its alternation number.
	Thus, together with the complexity of the automata-nonemptiness calculation,
	we get that the model-checking procedure for \OGSL\ is 2\ExpTime.
	Clearly, the elementariness property also holds for \ATLS, as it is included
	in \OGSL.
	In particular, although it has not been explicitly stated, this property is
	crucial for most of the results achieved in literature about \ATLS\ by means
	of automata (see~\cite{Sch08}, as an example).
	Moreover, we believe that our proof techniques are of independent interest and
	applicable to other logics as well.
	\vspace{0.25cm}
	\\ \indent
	\emph{Related works.}
	%
		Several works have focused on extensions of \ATLS\ to incorporate more
		powerful strategic constructs.
		Among them, we recall \emph{Alternating-Time \MuCalculus} (\AMuCalculus, for
		short)~\cite{AHK02}, \emph{Game Logic} (\GL, for short)~\cite{AHK02},
		\emph{Quantified Decision Modality \MuCalculus} (\textsc{q}D$\mu$, for
		short)~\cite{Pin07}, \emph{Coordination Logic} (\CL, for short)~\cite{FS10},
		and some extensions of \ATLS\ considered in~\cite{BLLM09}.
		\AMuCalculus\ and \textsc{q}D$\mu$ are intrinsically different from \SL\ (as
		well as from \CHPSL\ and \ATLS) as they are obtained by extending the
		propositional $\mu$-calculus~\cite{Koz83} with strategic modalities.
		\CL\ is similar to \textsc{q}D$\mu$ but with \LTL\ temporal operators
		instead of explicit fixpoint constructors.
		\GL\ is strictly included in \CHPSL, in the case of two-player turn-based
		games, but it does not use any explicit treatment of strategies, neither it
		does the extensions of \ATLS\ introduced in~\cite{BLLM09}.
		In particular, the latter work consider restrictions on the memory for
		strategy quantifiers.
		Thus, all above logics are different from \SL, which we recall it aims
		to be a minimal but powerful logic to reason about strategic behavior in
		multi-agent systems.
		A very recent generalization of \ATLS, which results to be expressive but
		a proper sublogic of \SL, is also proposed in~\cite{CLM10}.
		In this logic, a quantification over strategies does not reset the
		strategies previously quantified but allows to maintain them in a particular
		context in order to be reused.
		This makes the logic much more expressive than \ATLS.
		On the other hand, as it does not allow agents to share the same strategy,
		it is not comparable with the fragments we have considered in this paper.
		Finally, we want to remark that our non-elementary hardness proof about the
		\SL\ model-checking problem is inspired by and improves a proof proposed for
		their logic and communicated to us~\cite{CLM10(PC)} by the authors
		of~\cite{CLM10}.
	%
	%
	\vspace{0.25cm}
	\\ \indent
	\emph{Note on~\cite{MMV10b}.}
	%
		Preliminary results on \SL\ appeared in~\cite{MMV10b}.
		We presented there a 2\ExpTime\ algorithm for the model-checking problem.
		The described procedure applies only to the \OGSL\ fragment, as model
		checking for full \SL\ is non-elementary.
	%
	%
	\vspace{0.25cm}
	\\ \indent
	\emph{Outline.}
	%
		The remaining part of this work is structured as follows.
		In Section~\ref{sec:sl}, we recall the semantic framework based on
		concurrent game structures and introduce syntax and semantics of \SL.
		Then, in Section~\ref{sec:modchkhrd}, we show the non-elementary lower bound
		for the model-checking problem.
		After this, in Section~\ref{sec:strqnt}, we start the study of few syntactic
		and semantic \SL\ fragments and introduce the concepts of dependence map and
		elementary satisfiability.
		Finally, in Section~\ref{sec:modchkprc}, we describe the model-checking
		automata-theoretic procedures for all \SL\ fragments.
		Note that, in the accompanying Appendix~\ref{app:mthnot}, we recall standard
		mathematical notation and some basic definitions that are used in the paper.
		However, for the sake of a simpler understanding of the technical part, we
		make a reminder, by means of footnotes, for each first use of a non trivial
		or immediate mathematical concept.
		The paper is self contained.
		All missing proofs in the main body of the work are reported in appendix.
	%

\end{section}





\newcommand{\figexmprs}
	{
	\begin{wrapfigure}[10]{r}{0.275\textwidth}
		\vspace{-15pt}
		\begin{center}
			\footnotesize
			\mbox{\scalebox{0.80}[0.80]{
			\begin{tikzpicture}
				[node distance = 2.5cm, bend angle = 25, shorten >= 2pt, shorten <= 2pt]
				\node [player]
							(SI)
							{$\stackrel{\sSym[i]}{\emptyset}$};
				\node [player]
							(SA)
							[below left of = SI]
							{$\stackrel{\sSym[\ASym]}{\wSym[\ASym]}$};
				\node [player]
							(SB)
							[below right of = SI]
							{$\stackrel{\sSym[\BSym]}{\wSym[\BSym]}$};
				\path[-stealth']
					(SI)	edge	[]
											node [plain1] {$\DSet[\ASym]$}
											(SA)
								edge	[]
											node [plain1] {$\DSet[\BSym]$}
											(SB)
								edge	[loop above]
											node [plain1] {$\DSet[i]$}
											()
					(SA)	edge	[loop below]
											node [plain1] {$**$}
											()
					(SB)	edge	[loop below]
											node [plain1] {$**$}
											()
					;
			\end{tikzpicture}
			}}
			\caption{\label{fig:exm:prs} The \CGS\ $\GName[P\!RS]$.}
		\end{center}
		\vspace{-15pt}
	\end{wrapfigure}
	}

\newcommand{\figexmpd}
	{
	\begin{wrapfigure}[11]{r}{0.275\textwidth}
		\vspace{-27.5pt}
		\begin{center}
			\footnotesize
			\mbox{\scalebox{0.80}[0.80]{
			\begin{tikzpicture}
				[node distance = 2.5cm, bend angle = 25, shorten >= 2pt, shorten <= 2pt]
				\node [player]
							(SI)
							{$\stackrel{\sSym[i]}{\fSym[ {\ASym[1]} ], \fSym[ {\ASym[2]} ]}$};
				\node [player]
							(SA)
							[below left of = SI]
							{$\stackrel{\sSym[ {\ASym[1]} ]}{\fSym[ {\ASym[1]} ]}$};
				\node [player]
							(SJ)
							[below of = SI]
							{$\stackrel{\sSym[j]}{\emptyset}$};
				\node [player]
							(SB)
							[below right of = SI]
							{$\stackrel{\sSym[ {\ASym[2]} ]}{\fSym[ {\ASym[2]} ]}$};
				\path[-stealth']
					(SI)	edge	[]
											node [plain1] {$\DSym\CSym$}
											(SA)
								edge	[]
											node [plain1] {$\DSym\DSym$}
											(SJ)
								edge	[]
											node [plain1] {$\CSym\DSym$}
											(SB)
								edge	[loop above]
											node [plain1] {$\CSym\CSym$}
											()
					(SA)	edge	[loop below]
											node [plain1] {$**$}
											()
					(SJ)	edge	[loop below]
											node [plain1] {$**$}
											()
					(SB)	edge	[loop below]
											node [plain1] {$**$}
											()
					;
			\end{tikzpicture}
			}}
			\caption{\label{fig:exm:pd} The \CGS\ $\GName[P\!D]$.}
		\end{center}
		\vspace{-15pt}
	\end{wrapfigure}
	}

\newcommand{\figexmsv}
	{
	\begin{wrapfigure}[10]{r}{0.300\textwidth}
		\vspace{-25pt}
		\begin{center}
			\footnotesize
			\mbox{\scalebox{0.80}[0.80]{
			\begin{tikzpicture}
				[node distance = 3cm, bend angle = 15, shorten >= 2pt, shorten <= 2pt]
				\node [player]
							(S0)
							{$\stackrel{\sSym[0]}{\emptyset}$};
				\node [player]
							(S1)
							[below left of = S0]
							{$\stackrel{\sSym[1]}{\pSym}$};
				\node [player]
							(S2)
							[below of = S0]
							{$\stackrel{\sSym[2]}{\pSym, \qSym}$};
				\node [player]
							(S3)
							[below right of = S0]
							{$\stackrel{\sSym[3]}{\qSym}$};
				\path[-stealth']
					(S0)	edge	[bend left]
											node [pos = 0.65, plain1] {$00$}
											(S1)
								edge	[bend left]
											node [plain1] {$01$}
											(S2)
								edge	[bend left]
											node [plain1] {$10$}
											(S3)
								edge	[loop above]
											node [plain1] {$11$}
											()
					(S1)	edge	[bend left]
											node [plain1] {$**$}
											(S0)
					(S2)	edge	[bend left]
											node [plain1] {$**$}
											(S0)
					(S3)	edge	[bend left]
											node [pos = 0.35, plain1] {$**$}
											(S0)
					;
			\end{tikzpicture}
			}}
			\caption{\label{fig:exm:sv} The \CGS\ $\GName[S\!V]$.}
		\end{center}
		\vspace{-15pt}
	\end{wrapfigure}
	}

\newcommand{\figlmmqptlrdc}
	{
	\begin{wrapfigure}[5]{r}{0.250\textwidth}
		\vspace{-10pt}
		\begin{center}
			\footnotesize
			\mbox{\scalebox{0.80}[0.80]{
			\begin{tikzpicture}
				[node distance = 3cm, bend angle = 15, shorten >= 2pt, shorten <= 2pt]
				\node [player]
							(S0)
							{$\stackrel{\sSym[0]}{\emptyset}$};
				\node [player]
							(S1)
							[right of = S0]
							{$\stackrel{\sSym[1]}{\pSym}$};
				\path[-stealth']
					(S0)	edge	[bend left]
											node [plain1] {$\Tt$}
											(S1)
								edge	[loop above]
											node [plain1] {$\Ff$}
											()
					(S1)	edge	[bend left]
											node [plain1] {$\Ff$}
											(S0)
								edge	[loop above]
											node [plain1] {$\Tt$}
											()
					;
			\end{tikzpicture}
			}}
			\caption{\label{fig:lmm:qptl(rdc)} The \CGS\ $\GName[Rdc]$.}
		\end{center}
		\vspace{-15pt}
	\end{wrapfigure}
	}

\newcommand{\figthmogslvsatlsexp}
	{
	\begin{figure}[htbp]
		\vspace{-0pt}
		\begin{center}
			\footnotesize
			\mbox{
			\subfigure[\CGS\ {$\GName[1]$}.]
				{
				\label{fig:thm:ogslvsatls(exp:1)}
				\scalebox{1.00}[0.80]{
				\begin{tikzpicture}
				[node distance = 2.5cm, bend angle = 25, shorten >= 2pt, shorten <= 2pt]
				\node [player]
							(S0)
							{$\stackrel{\sSym[0]}{\emptyset}$};
				\node [player]
							(S1)
							[below left of = S0]
							{$\stackrel{\sSym[1]}{\pSym}$};
				\node [player]
							(S2)
							[below right of = S0]
							{$\stackrel{\sSym[2]}{\emptyset}$};
				\path[-stealth']
					(S0)	edge	[]
											node [plain1] {$\DSet[1]$}
											(S1)
								edge	[]
											node [plain1]
											{$\DecSet[ {\GName[1]} ] \setminus \DSet[1]$}
											(S2)
					(S1)	edge	[loop below]
											node [plain1] {$***$}
											()
					(S2)	edge	[loop below]
											node [plain1] {$***$}
											()
					;
				\end{tikzpicture}
				}}
			\subfiggap\subfiggap\subfiggap\subfiggap
			\subfigure[\CGS\ {$\GName[2]$}.]
				{
				\label{fig:thm:ogslvsatls(exp:2)}
				\scalebox{1.00}[0.80]{
				\begin{tikzpicture}
				[node distance = 2.5cm, bend angle = 25, shorten >= 2pt, shorten <= 2pt]
				\node [player]
							(S0)
							{$\stackrel{\sSym[0]}{\emptyset}$};
				\node [player]
							(S1)
							[below left of = S0]
							{$\stackrel{\sSym[1]}{\pSym}$};
				\node [player]
							(S2)
							[below right of = S0]
							{$\stackrel{\sSym[2]}{\emptyset}$};
				\path[-stealth']
					(S0)	edge	[]
											node [plain1] {$\DSet[2]$}
											(S1)
								edge	[]
											node [plain1]
											{$\DecSet[ {\GName[2]} ] \setminus \DSet[2]$}
											(S2)
					(S1)	edge	[loop below]
											node [plain1] {$***$}
											()
					(S2)	edge	[loop below]
											node [plain1] {$***$}
											()
					;
				\end{tikzpicture}
				}}
			}
			\caption{\label{fig:thm:ogslvsatls(exp)} Alternation-$2$ non-equivalent
				\CGS s.}
		\end{center}
		\vspace{-10pt}
	\end{figure}
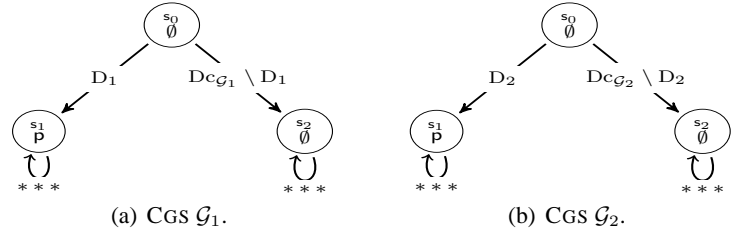
	}




\begin{section}{Strategy Logic}
	\label{sec:sl}

	In this section, we introduce \emph{Strategy Logic}, an extension of the
	classic linear-time temporal logic \LTL~\cite{Pnu77} along with the concepts
	of strategy quantifications and agent binding.
	Our aim is to define a formalism that allows to express strategic plans over
	temporal goals in a way that separates the part related to the strategic
	reasoning from that concerning the tactical one.
	This distinctive feature is achieved by decoupling the instantiation of
	strategies, done through the quantifications, from their application by means
	of bindings.
	Our proposal, on the line marked by its precursor \CHPSL~\cite{CHP07,CHP10}
	and differently from classical temporal logics~\cite{Eme90}, turns in a logic
	that is not simply propositional but predicative, since we treat strategies as
	a first order concept via the use of agents and variables as explicit
	syntactic elements.
	This fact let us to write Boolean combinations and nesting of complex
	predicates, linked together by some common strategic choice, which may
	represent each one a different temporal goal.
	However, it is worth noting that the technical approach we follow here is
	quite different from that used for the definition of \CHPSL, which is based,
	on the syntactic side, on the \CTLS\ formula framework~\cite{EH86} and,
	\mbox{on the semantic one, on the two-player turn-based game
	model~\cite{PP04}.}

	The section is organized as follows.
	In Subsection~\ref{subsec:undfrw}, we recall the definition of concurrent
	game structure used to interpret Strategy Logic, whose syntax is introduced in
	Subsection~\ref{subsec:syntax}.
	Then, in Subsection~\ref{subsec:bascnp}, we give, among the others, the
	notions of strategy and play, which are finally used, in
	Subsection~\ref{subsec:semantics}, to define the semantics of the logic.

	\begin{subsection}{Underlying framework}
		\label{subsec:undfrw}

		As semantic framework for our logic language, we use a \emph{graph-based
		model} for \emph{multi-player games} named \emph{concurrent game
		structure}~\cite{AHK02}.
		Intuitively, this mathematical formalism provides a generalization of
		\emph{Kripke structures}~\cite{Kri63} and \emph{labeled transition
		systems}~\cite{Kel76}, modeling \emph{multi-agent systems} viewed as games,
		in which players perform \emph{concurrent actions} chosen strategically as a
		function on the history of the play.
		\begin{definition}[Concurrent Game Structures]
			\label{def:cgs}
			A \emph{concurrent game structure} (\emph{\CGS}, for short) is a tuple
			$\GName \defeq \CGSStruct$, where $\APSet$ and $\AgSet$ are finite
			non-empty sets of \emph{atomic propositions} and \emph{agents}, $\AcSet$
			and $\StSet$ are enumerable non-empty sets of \emph{actions} and
			\emph{states}, $\sElm[0] \in \StSet$ is a designated \emph{initial state},
			and $\labFun : \StSet \to \pow{\APSet}$ is a \emph{labeling function} that
			maps each state to the set of atomic propositions true in that state.
			Let $\DecSet \defeq \AcSet^{\AgSet}$ be the set of \emph{decisions}, i.e.,
			functions from $\AgSet$ to $\AcSet$ representing the choices of an action
			for each agent.~\footnote{In the following, we use both $\XSet \to \YSet$
			and $\YSet^{\XSet}$ to denote the set of functions from the domain $\XSet$
			to the codomain $\YSet$.}
			Then, $\trnFun: \StSet \times \DecSet \to \StSet$ is a \emph{transition
			function} mapping a pair of a state and a decision to a state.
		\end{definition}
		Observe that elements in $\StSet$ are not global states of the system, but
		states of the environment in which the agents operate.
		Thus, they can be viewed as states of the game, which do not include the
		local states of the agents.
		From a practical point of view, this means that all agents have perfect
		information on the whole game, since local states are not taken into account
		in the choice of actions~\cite{FHMV95}.
		Observe also that, differently from other similar formalizations, each agent
		has the same set of possible executable actions, independently of the
		current state and of choices made by other agents.
		However, as already reported in literature~\cite{Pin07}, this simplifying
		choice does not result in a limitation of our semantics framework and allow
		us to give a simpler and clearer explanation of all formal definitions and
		techniques we work on.

		From now on, apart from the examples and if not differently stated, all \CGS
		s are defined on the same sets of atomic propositions $\APSet$ and agents
		$\AgSet$, so, when we introduce a new structure in our reasonings, we do not
		make explicit their definition anymore.
		In addition, we use the italic letters $\pElm$, $\aElm$, $\cElm$, and
		$\sElm$, possibly with indexes, as meta-variables on, respectively, the
		atomic propositions $\pSym, \qSym, \ldots$ in $\APSet$, the agents $\alpha,
		\beta, \gamma, \ldots$ in $\AgSet$, the actions $0, 1, \ldots$ in $\AcSet$,
		and the states $\sSym, \ldots$ in $\StSet$.
		Finally, we use the name of a \CGS\ as a subscript to extract the components
		from its tuple-structure.
		Accordingly, if $\GName = \CGSStruct$, we have that $\AcSet[\GName] =
		\AcSet$, $\labFun[\GName] = \labFun$, $\sElm[0\GName] = \sSym[0]$, and so
		on.
		Furthermore, we use the same notational concept to make explicit to which
		\CGS\ the set $\DecSet$ of decisions is related to.
		Note that, \mbox{we omit the subscripts if the structure can be
		unambiguously individuated from the context.}

		\par\figexmprs\
		Now, to get attitude to the introduced semantic framework, let us describe
		two running examples of simple concurrent games.
		In particular, we start by modeling the \emph{paper, rock, and scissor}
		game.
		\begin{example}[Paper, Rock, and Scissor]
			\label{exm:prs}
			Consider the classic two-player concurrent game \emph{paper, rock, and
			scissor} (PRS, for short) as represented in Figure~\vref{fig:exm:prs},
			where a play continues until one of the participants catches the move of
			the other.
			Vertexes are states of the game and labels on edges represent decisions of
			agents or sets of them, where the symbol $*$ is used in place of every
			possible action.
			In this specific case, since there are only two agents, the pair of
			symbols $**$ indicates the whole set $\DecSet$ of decisions.
			The agents ``Alice'' and ``Bob'' in $\AgSet \defeq \{ \ASym, \BSym \}$
			have as possible actions those in the set $\AcSet \defeq \{ \PSym, \RSym,
			\SSym \}$, which stand for ``paper'', ``rock'', and ``scissor'',
			respectively.
			During the play, the game can stay in one of the three states in $\StSet
			\defeq \{ \sSym[i], \sSym[\ASym], \sSym[\BSym] \}$, which represent,
			respectively, the waiting moment, named \emph{idle}, and the two winner
			positions.
			The latter ones are labeled with one of the atomic propositions in $\APSet
			\defeq \{ \wSym[\ASym], \wSym[\BSym] \}$, in order to represent who is the
			winner.
			The catch of one action over another is described by the relation $\CRel
			\defeq \{ (\PSym, \RSym), (\RSym, \SSym), (\SSym, \PSym) \} \subseteq
			\AcSet \times \AcSet$.
			We can now define the \CGS\ $\GName[P\!RS] \defeq \CGSStruct[ { \sSym[i] }
			]$ for the PRS game, with the labeling given by $\labFun(\sSym[i]) \defeq
			\emptyset$, $\labFun(\sSym[\ASym]) \defeq \{ \wSym[\ASym] \}$, and
			$\labFun(\sSym[\BSym]) \defeq \{ \wSym[\BSym] \}$ and the transition
			function set as follows, where $\DSet[\ASym] \defeq \set{ \decFun \in
			\DecSet[ { \GName[P\!RS] } ] }{ (\decFun(\ASym), \decFun(\BSym)) \in \CRel
			}$ and $\DSet[\BSym] \defeq \set{ \decFun \in \DecSet[ { \GName[P\!RS] } ]
			}{ (\decFun(\BSym), \decFun(\ASym)) \in \CRel }$ are the sets of winning
			decisions for the two agents: if $\sElm = \sSym[i]$ and $\decFun \in
			\DSet[\ASym]$ then $\trnFun(\sElm, \decFun) \defeq \sSym[\ASym]$, else if
			$\sElm = \sSym[i]$ and $\decFun \in \DSet[\BSym]$ then $\trnFun(\sElm,
			\decFun) \defeq \sSym[\BSym]$, otherwise $\trnFun(\sElm, \decFun) \defeq
			\sElm$.
			Note that, when none of the two agents catches the action of the other,
			i.e., the used decision is in $\DSet[i] \defeq \DecSet[ { \GName[P\!RS] }
			] \setminus (\DSet[\ASym] \cup \DSet[\BSym])$, the play remains in the
			idle state to allow another try, otherwise it is stuck in a winning
			position forever.
		\end{example}

		\par\figexmpd\
		We now describe a non-classic qualitative version of the well-known
		\emph{prisoner's dilemma}.
		\begin{example}[Prisoner's Dilemma]
			\label{exm:pd}
			In the \emph{prisoner's dilemma} (PD, for short), two accomplices are
			interrogated in separated rooms by the police, which offers them the same
			agreement.
			If one defects, i.e., testifies for the prosecution against the other,
			while the other cooperates, i.e., remains silent, the defector goes free
			and the silent accomplice goes to jail.
			If both cooperate, they remain free, but will be surely interrogated in
			the next future waiting for a defection.
			On the other hand, if every one defects, both go to jail.
			It is ensured that no one will know about the choice made by the other.
			This tricky situation can be modeled by the \CGS\ $\GName[P\!D] \defeq
			\CGSStruct[ {\sSym[i]} ]$ depicted in Figure~\vref{fig:exm:pd}, where the
			agents ``Accomplice-1'' and ``Accomplice-2'' in $\AgSet \defeq \{
			\ASym[1], \ASym[2] \}$ can chose an action in $\AcSet \defeq \{ \CSym,
			\DSym \}$, which stand for ``cooperation'' and ``defection'',
			respectively.
			There are four states in $\StSet \defeq \{ \sSym[i], \sSym[ {\ASym[1]} ],
			\sSym[ {\ASym[2]} ], \sSym[j] \}$.
			In the idle state $\sSym[i]$ the agents are waiting for the interrogation,
			while $\sSym[j]$ represents the jail for both of them.
			The remaining states $\sSym[ {\ASym[1]} ]$ and $\sSym[ {\ASym[2]} ]$
			indicate, instead, the situations in which only one of the agents become
			definitely free.
			To characterize the different meaning of these states, we use the atomic
			propositions in $\APSet \defeq \{ \fSym[ {\ASym[1]} ], \fSym[ {\ASym[2]} ]
			\}$, which denote who is ``free'', by defining the following labeling:
			$\labFun(\sSym[i]) \defeq \{ \fSym[ {\ASym[1]} ], \fSym[ {\ASym[2]} ] \}$,
			$\labFun(\sSym[ {\ASym[1]} ]) \defeq \{ \fSym[ {\ASym[1]} ] \}$,
			$\labFun(\sSym[ {\ASym[2]} ]) \defeq \{ \fSym[ {\ASym[2]} ] \}$, and
			$\labFun(\sSym[j]) \defeq \emptyset$.
			The transition function $\trnFun$ can be easily deduced by the figure.
		\end{example}

	\end{subsection}

	\begin{subsection}{Syntax}
		\label{subsec:syntax}

		\emph{Strategy Logic} (\emph{\SL}, for short) syntactically extends \LTL\ by
		means of two \emph{strategy quantifiers}, the existential $\EExs{\xElm}$ and
		the universal $\AAll{\xElm}$, and \emph{agent binding} $(\aElm, \xElm)$,
		where $\aElm$ is an agent and $\xElm$ a variable.
		Intuitively, these new elements can be respectively read as \emph{``there
		exists a strategy $\xElm$''}, \emph{``for all strategies $\xElm$''}, and
		\emph{``bind agent $\aElm$ to the strategy associated with the variable
		$\xElm$''}.
		The formal syntax of \SL\ follows.
		\begin{definition}[\SL\ Syntax]
			\label{def:sl(syntax)}
			\SL\ \emph{formulas} are built inductively from the sets of atomic
			propositions $\APSet$, variables $\VarSet$, and agents $\AgSet$, by using
			the following grammar, where $\pElm \in \APSet$, $\xElm \in \VarSet$, and
			$\aElm \in \AgSet$:
			\begin{center}
				$\varphi ::= \pElm \mid \neg \varphi \mid \varphi \wedge \varphi \mid
				\varphi \vee \varphi \mid \X \varphi \mid \varphi \:\U \varphi \mid
				\varphi \:\R \varphi \mid \EExs{\xElm} \varphi \mid \AAll{\xElm} \varphi
				\mid (\aElm, \xElm) \varphi$.
			\end{center}
			\SL\ denotes the infinite set of formulas generated by the above rules.
		\end{definition}
		Observe that, by construction, \LTL\ is a proper syntactic fragment of \SL,
		i.e., $\LTL \subset \SL$.
		In order to abbreviate the writing of formulas, we use the boolean values
		true $\Tt$ and false $\Ff$ and the well-known temporal operators future $\F
		\varphi \defeq \Tt\: \U \varphi$ and globally $\G \varphi \defeq \Ff\: \R
		\varphi$.
		Moreover, we use the italic letters $\xElm, \yElm, \zElm, \ldots$, possibly
		with indexes, as meta-variables on the variables $\xSym, \ySym, \zSym,
		\ldots$ in $\VarSet$.

		A first classic notation related to the \SL\ syntax that we need to
		introduce is that of \emph{subformula}, i.e., a syntactic expression that is
		part of an a priori given formula.
		By $\mthfun{sub} : \SL \to \pow{\SL}$ we formally denote the function
		returning the set of subformulas of an \SL\ formula.
		For instance, consider $\varphi = \EExs{\xSym} (\alpha, \xSym) (\F \pSym)$.
		Then, it is immediate to see that $\sub{\varphi} = \{ \varphi, (\alpha,
		\xSym) (\F \pSym), (\F \pSym), \pSym, \Tt \}$.

		Normally, predicative logics need the concepts of free and bound
		\emph{placeholders} in order to formally define the meaning of their
		formulas.
		The placeholders are used to represent particular positions in syntactic
		expressions that have to be highlighted, since they have a crucial role in
		the definition of the semantics.
		In first order logic, for instance, there is only one type of placeholders,
		which is represented by the variables.
		In \SL, instead, we have both agents and variables as placeholders, as it
		can be noted by its syntax, in order to distinguish between the
		quantification of a strategy and its application by an agent.
		Consequently, we need a way to differentiate if an agent has an associated
		strategy via a variable and if a variable is quantified.
		To do this, we use the set of \emph{free agents/variables} as the subset of
		$\AgSet \cup \VarSet$ containing \emph{(i)} all agents for which there is no
		binding after the occurrence of a temporal operator and \emph{(ii)} all
		variables for which there is a binding but no quantifications.
		\begin{definition}[\SL\ Free Agents/Variables]
			\label{def:sl(freeagvar)}
			The set of \emph{free agents/variables} of an \SL\ formula is given by the
			function $\mthfun{free} : \SL \to \pow{\AgSet \cup \VarSet}$ defined as
			follows:
			\begin{enumerate}[(i)]
				\setlength{\itemsep}{2pt}
				\item\label{def:sl(freeagvar:ap)}
					$\free{\pElm} \defeq \emptyset$, where $\pElm \in \APSet$;
				\item\label{def:sl(freeagvar:neg)}
					$\free{\neg \varphi} \defeq \free{\varphi}$;
				\item\label{def:sl(freeagvar:conjdisj)}
					$\free{\varphi_{1} \Opr \varphi_{2}} \defeq \free{\varphi_{1}} \cup
					\free{\varphi_{2}}$, where $\Opr\! \in \{ \wedge, \vee \}$;
				\item\label{def:sl(freeagvar:next)}
					$\free{\X \varphi} \defeq \AgSet \cup \free{\varphi}$;
				\item\label{def:sl(freeagvar:untilrelease)}
					$\free{\varphi_{1} \Opr \varphi_{2}} \defeq \AgSet \cup
					\free{\varphi_{1}} \cup \free{\varphi_{2}}$, where $\Opr\! \in \{
					\U\!\!, \R\! \}$;
				\item\label{def:sl(freeagvar:qnt)}
					$\free{\Qnt \varphi} \defeq \free{\varphi} \setminus \{ \xElm \}$,
					where $\Qnt\! \in \set{ \EExs{\xElm}, \AAll{\xElm} }{ \xElm \in
					\VarSet }$;
				\item\label{def:sl(freeagvar:bndprs)}
					$\free{(\aElm, \xElm) \varphi} \defeq \free{\varphi}$, if $\aElm
					\not\in \free{\varphi}$, where $\aElm \in \AgSet$ and $\xElm \in
					\VarSet$;
				\item\label{def:sl(freeagvar:bndrem)}
					$\free{(\aElm, \xElm) \varphi} \defeq (\free{\varphi} \setminus \{
					\aElm \}) \cup \{ \xElm \}$, if $\aElm \in \free{\varphi}$, where
					$\aElm \in \AgSet$ and $\xElm \in \VarSet$.
			\end{enumerate}
			A formula $\varphi$ without free agents (resp., variables), i.e., with
			$\free{\varphi} \cap \AgSet = \emptyset$ (resp., $\free{\varphi} \cap
			\VarSet = \emptyset$), is named \emph{agent-closed} (resp.,
			\emph{variable-closed}).
			If $\varphi$ is both agent- and variable-closed, it is referred to as a
			\emph{sentence}.
			The function $\mthfun{snt} : \SL \to \pow{\SL}$ returns the set of
			\emph{subsentences} $\snt{\varphi} \defeq \set{ \phi \in \sub{\varphi} }{
			\free{\phi} = \emptyset }$ for each \SL\ formula $\varphi$.
		\end{definition}
		Observe that, on one hand, free agents are introduced in
		Items~\ref{def:sl(freeagvar:next)} and~\ref{def:sl(freeagvar:untilrelease)}
		and removed in Item~\ref{def:sl(freeagvar:bndrem)}.
		On the other hand, free variables are introduced in
		Item~\ref{def:sl(freeagvar:bndrem)} and removed in
		Item~\ref{def:sl(freeagvar:qnt)}.
		As an example, let $\varphi = \EExs{\xSym} (\alpha, \xSym) (\beta, \ySym)
		(\F \pSym)$ be a formula on the agents $\AgSet = \{ \alpha, \beta, \gamma
		\}$.
		Then, we have $\free{\varphi} = \{ \gamma, \ySym \}$, since $\gamma$ is an
		agent without any binding after $\F \pSym$ and $\ySym$ has no quantification
		at all.
		Consider also the formulas $(\alpha, \zSym) \varphi$ and $(\gamma, \zSym)
		\varphi$, where the subformula $\varphi$ is the same as above.
		Then, we have $\free{(\alpha, \zSym) \varphi} = \free{\varphi}$ and
		$\free{(\gamma, \zSym) \varphi} = \{ \ySym, \zSym \}$, since $\alpha$ is not
		free in $\varphi$ but $\gamma$ is, i.e., $\alpha \notin \free{\varphi}$ and
		$\gamma \in \free{\varphi}$.
		So, $(\gamma, \zSym) \varphi$ is agent-closed while $(\alpha, \zSym)
		\varphi$ is not.

		Similarly to the case of first order logic, another important concept that
		characterizes the syntax of \SL\ is that of the \emph{alternation number} of
		quantifiers, i.e., the maximum number of quantifier switches $\EExs{\cdot}
		\AAll{\cdot}$, $\AAll{\cdot} \EExs{\cdot}$, $\EExs{\cdot} \neg
		\EExs{\cdot}$, or $\AAll{\cdot} \neg \AAll{\cdot}$ that bind a variable in a
		subformula that is not a sentence.
		The constraint on the kind of subformulas that are considered here means
		that, when we evaluate the number of such switches, we consider each
		possible subsentence as an atomic proposition, hence, its quantifiers are
		not taken into account.
		Moreover, it is important to observe that vacuous quantifications, i.e.,
		quantifications on variable that are not free in the immediate inner
		subformula, need to be not considered at all in the counting of quantifier
		switches.
		This value is crucial when we want to analyze the complexity of the decision
		problems of fragments of our logic, since higher alternation can usually
		mean higher complexity.
		By $\mthfun{alt} : \SL \to \SetN$ we formally denote the function returning
		the alternation number of an \SL\ formula.
		Furthermore, the fragment $\SL[k-\text{alt}] \defeq \set{ \varphi \in \SL }{
		\forall \varphi' \in \sub{\varphi} \:.\: \alt{\varphi'} \leq k }$ of \SL,
		for $k \in \SetN$, denotes the subset of formulas having all subformulas
		with alternation number bounded by $k$.
		For instance, consider the sentence $\varphi = \AAll{\xSym} \EExs{\ySym}
		(\alpha, \xSym) (\beta, \ySym) (\F \varphi')$ with $\varphi' = \AAll{\xSym}
		\EExs{\ySym} (\alpha, \xSym) (\beta, \ySym) (\X \pSym)$, on the set of
		agents $\AgSet = \{ \alpha, \beta \}$.
		Then, the alternation number $\alt{\varphi}$ is $1$ and not $3$, as one can
		think at a first glance, since $\varphi'$ is a sentence.
		Moreover, it holds that $\alt{\varphi'} = 1$.
		Hence, $\varphi \in \SL[1-\text{alt}]$.
		On the other hand, if we substitute $\varphi'$ with $\varphi'' =
		\AAll{\xSym} (\alpha, \xSym) (\X \pSym)$, we have that $\alt{\varphi} = 2$,
		since $\varphi''$ is not a sentence.
		Thus, it holds that $\varphi \not\in \SL[1-\text{alt}]$ but $\varphi \in
		\SL[2-\text{alt}]$.

		At this point, in order to practice with the syntax of our logic by
		expressing game-theoretic concepts through formulas, we describe two
		examples of important properties that are possible to write in \SL, but
		neither in \ATLS~\cite{AHK02} nor in \CHPSL.
		This is clarified later in the paper.
		The first concept we introduce is the well-known deterministic concurrent
		multi-player \emph{Nash equilibrium} for Boolean valued payoffs.
		\begin{example}[Nash Equilibrium]
			\label{exm:ne}
			Consider the $n$ agents $\alpha_{1}, \ldots, \alpha_{n}$ of a game, each
			of them having, respectively, a possibly different temporal goal described
			by one of the \LTL\ formulas $\psi_{1}, \!\ldots\!, \psi_{n}$.
			Then, we can express the existence of a strategy profile $(\xSym[1],
			\ldots, \xSym[n])$ that is a \emph{Nash equilibrium} (NE, for short) for
			$\alpha_{1}, \ldots, \alpha_{n}$ w.r.t.\ $\psi_{1}, \ldots, \psi_{n}$ by
			using the \SL[$1$-alt] sentence $\varphi_{\!N\!\!E} \!\defeq\!
			\EExs{\xSym[1]} (\alpha_{1}, \xSym[1]) \cdots \EExs{\xSym[n]} (\alpha_{n},
			\xSym[n]) \: \psi_{\!N\!\!E}$, where $\psi_{\!N\!\!E} \!\defeq\!
			\bigwedge_{i = 1}^{n} (\EExs{\ySym} (\alpha_{i}, \ySym) \psi_{i})
			\rightarrow \psi_{i}$ is a variable-closed formula.
			Informally, this asserts that every agent $\alpha_{i}$ has $\xSym[i]$ as
			one of the best strategy w.r.t.\ the goal $\psi_{i}$, once all the other
			strategies of the remaining agents $\alpha_{j}$, with $j \neq i$, have
			been fixed to $\xSym[j]$.
			Note that here we are only considering equilibria under deterministic
			strategies.
		\end{example}

		As in physics, also in game theory an equilibrium is not always stable.
		Indeed, there are games like the PD of Example~\vref{exm:pd} having Nash
		equilibria that are instable.
		One of the simplest concepts of stability that is possible to think is
		called \emph{stability profile}.
		\begin{example}[Stability Profile]
			\label{exm:sp}
			Think about the same situation of the above example on NE.
			Then, a \emph{stability profile} (SP, for short) is a strategy profile
			$(\xSym[1], \ldots, \xSym[n])$ for $\alpha_{1}, \ldots, \alpha_{n}$
			w.r.t.\ $\psi_{1}, \ldots, \psi_{n}$ such that there is no agent
			$\alpha_{i}$ that can choose a different strategy from $\xSym[i]$ without
			changing its own payoff and penalizing the payoff of another agent
			$\alpha_{j}$, with $j \neq i$.
			To represent the existence of such a profile, we can use the \SL[$1$-alt]
			sentence $\varphi_{\!S\!P} \defeq \EExs{\xSym[1]} (\alpha_{1}, \xSym[1])
			\cdots \EExs{\xSym[n]} (\alpha_{n}, \xSym[n]) \: \psi_{\!S\!P}$, where
			$\psi_{\!S\!P} \defeq \bigwedge_{i,j = 1, i \neq j}^{n} \psi_{j}
			\rightarrow \AAll{\ySym} ((\psi_{i} \leftrightarrow (\alpha_{i}, \ySym)
			\psi_{i}) \rightarrow (\alpha_{i}, \ySym) \psi_{j})$.
			Informally, with the $\psi_{\!S\!P}$ subformula, we assert that, if
			$\alpha_{j}$ is able to achieve his goal $\psi_{j}$, all strategies
			$\ySym$ of $\alpha_{i}$ that left unchanged the payoff related to
			$\psi_{i}$, also let $\alpha_{j}$ to maintain his achieved goal.
			At this point, it is very easy to ensure the existence of an NE that is
			also an SP, by using the \SL[$1$-alt] sentence $\varphi_{\!S\!N\!\!E}
			\defeq \EExs{\xSym[1]} (\alpha_{1}, \xSym[1]) \cdots \EExs{\xSym[n]}
			(\alpha_{n}, \xSym[n]) \: \psi_{\!S\!P} \wedge \psi_{\!N\!\!E}$.
		\end{example}

	\end{subsection}

	\begin{subsection}{Basic concepts}
		\label{subsec:bascnp}

		Before continuing with the description of our logic, we have to introduce
		some basic concepts, regarding a generic \CGS, that are at the base of the
		semantics formalization.
		Remind that a description of used mathematical notation is reported in
		Appendix~\ref{app:mthnot}.

		We start with the notions of \emph{track} and \emph{path}.
		Intuitively, tracks and paths of a \CGS\ $\GName$ are legal sequences of
		reachable states in $\GName$ that can be respectively seen as partial and
		complete descriptions of possible outcomes of the game modeled by $\GName$
		itself.
		\begin{definition}[Tracks and Paths]
			\label{def:trkpth}
			A \emph{track} (resp., \emph{path}) in a \CGS\ $\GName$ is a finite
			(resp., an infinite) sequence of states $\trkElm \in \StSet^{*}$ (resp.,
			$\pthElm \in \StSet^{\omega}$) such that, for all $i \in
			\numco{0}{\card{\trkElm} - 1}$ (resp., $i \in \SetN$), there exists a
			decision $\decFun \in \DecSet$ such that $(\trkElm)_{i + 1} =
			\trnFun((\trkElm)_{i}, \decFun)$ (resp., $(\pthElm)_{i + 1} =
			\trnFun((\pthElm)_{i}, \decFun)$).~\footnote{The notation $(\wElm)_{i} \in
			\Sigma$ indicates the \emph{element} of index $i \in
			\numco{0}{\card{\wElm}}$ of a non-empty sequence $\wElm \in
			\Sigma^{\infty}$.}
			A track $\trkElm$ is \emph{non-trivial} if it has non-zero length, i.e.,
			$\card{\trkElm} > 0$ that is $\trkElm \neq \epsilon$.~\footnote{The
			Greek letter $\epsilon$ stands for the \emph{empty sequence}.}
			The set $\TrkSet \subseteq \StSet^{+}$ (resp., $\PthSet \subseteq
			\StSet^{\omega}$) contains all non-trivial tracks (resp., paths).
			Moreover, $\TrkSet(\sElm) \defeq \set{ \trkElm \in \TrkSet }{
			\fst{\trkElm} = \sElm }$ (resp., $\PthSet(\sElm) \defeq \set{ \pthElm \in
			\PthSet }{ \fst{\pthElm} = \sElm }$) indicates the subsets of tracks
			(resp., paths) starting at a state $\sElm \in \StSet$.~\footnote{By
			$\fst{\wElm} \defeq (\wElm)_{0}$ it is denoted the \emph{first element} of
			a non-empty sequence $\wElm \in \Sigma^{\infty}$.}
		\end{definition}
		For instance, consider the PRS game of Example~\vref{exm:prs}.
		Then, $\trkSym = \sSym[i] \cdot \sSym[\ASym] \in \StSet^{+}$ and $\pthSym =
		\sSym[i]^{\omega} \in \StSet^{\omega}$ are, respectively, a track and a path
		in the \CGS\ $\GName[P\!RS]$.
		Moreover, it holds that $\TrkSet = \sSym[i]^{+} + \sSym[i]^{*} \cdot
		(\sSym[\ASym]^{+} + \sSym[\BSym]^{+})$ and $\PthSet = \sSym[i]^{\omega} +
		\sSym[i]^{*} \cdot (\sSym[\ASym]^{\omega} + \sSym[\BSym]^{\omega})$.

		At this point, we can define the concept of \emph{strategy}.
		Intuitively, a strategy is a scheme for an agent that contains all choices
		of actions as a function of the history of the current outcome.
		However, observe that here we do not set an a priori connection between a
		strategy and an agent, since the same strategy can be used by more than one
		agent at the same time.
		\begin{definition}[Strategies]
			\label{def:str}
			A \emph{strategy} in a \CGS\ $\GName$ is a partial function $\strFun :
			\TrkSet \pto \AcSet$ that maps each non-trivial track in its domain to an
			action.
			For a state $\sElm \in \StSet$, a strategy $\strFun$ is said
			\emph{$\sElm$-total} if it is defined on all tracks starting in $\sElm$,
			i.e., $\dom{\strFun} = \TrkSet(\sElm)$.
			The set $\StrSet \defeq \TrkSet \pto \AcSet$ (resp., $\StrSet(\sElm)
			\defeq \TrkSet(\sElm) \to \AcSet$) contains all (resp., $\sElm$-total)
			strategies.
		\end{definition}
		An example of strategy in the \CGS\ $\GName[P\!RS]$ is the function
		$\strFun[1] \in \StrSet(\sSym[i])$ that maps each track having length
		multiple of $3$ to the action $\PSym$, the tracks whose remainder of length
		modulo $3$ is $1$ to the action $\RSym$, and the remaining tracks to the
		action $\SSym$.
		A different strategy is given by the function $\strFun[2] \in
		\StrSet(\sSym[i])$ that returns the action $\PSym$, if the tracks ends in
		$\sSym[\ASym]$ or $\sSym[\BSym]$ or if its length is neither a second nor a
		third power of a positive number, the action $\RSym$, if the length is a
		square power, and the action $\SSym$, otherwise.

		An important operation on strategies is that of \emph{translation} along a
		given track, which is used to determine which part of a strategy has yet to
		be used in the game.
		\begin{definition}[Strategy Translation]
			\label{def:strtrn}
			Let $\strFun \in \StrSet$ be a strategy and $\trkElm \in \dom{\strFun}$ a
			track in its domain.
			Then, $(\strFun)_{\trkElm} \in \StrSet$ denotes the \emph{translation} of
			$\strFun$ along $\trkElm$, i.e., the strategy with
			$\dom{(\strFun)_{\trkElm}} \defeq \set{ \trkElm' \in
			\TrkSet(\lst{\trkElm}) }{ \trkElm \cdot \trkElm'_{\geq 1} \in
			\dom{\strFun} }$ such that $(\strFun)_{\trkElm}(\trkElm') \defeq
			\strFun(\trkElm \cdot \trkElm'_{\geq 1})$, for all $\trkElm' \in
			\dom{(\strFun)_{\trkElm}}$.~\footnote{By $\lst{\wElm} \defeq
			(\wElm)_{\card{\wElm} - 1}$ it is denoted the \emph{last element} of a
			finite non-empty sequence $\wElm \in \Sigma^{*}$.}~\footnote{The notation
			$(\wElm)_{\geq i} \in \Sigma^{\infty}$ indicates the \emph{suffix} from
			index $i \in \numcc{0}{\card{\wElm}}$ inwards of a non-empty sequence
			$\wElm \in \Sigma^{\infty}$.}
		\end{definition}
		Intuitively, the translation $(\strFun)_{\trkElm}$ is the update of the
		strategy $\strFun$, once the history of the game becomes $\trkElm$.
		It is important to observe that, if $\strFun$ is a $\fst{\trkElm}$-total
		strategy then $(\strFun)_{\trkElm}$ is $\lst{\trkElm}$-total.
		For instance, consider the two tracks $\trkElm[1] = \sSym[i]^{4} \in
		\TrkSet(\sSym[i])$ and $\trkElm[2] = \sSym[i]^{4} \cdot \sSym[\ASym]^{2} \in
		\TrkSet(\sSym[i])$ in the \CGS\ $\GName[P\!RS]$ and the strategy $\strFun[1]
		\in \StrSet(\sSym[i])$ previously described.
		Then, we have that $(\strFun[1])_{\trkElm[1]} = \strFun[1]$, while
		$(\strFun[1])_{\trkElm[2]} \in \StrSet(\sSym[\ASym])$ maps each track having
		length multiple of $3$ to the action $\SSym$, each track whose remainder of
		length modulo $3$ is $1$ to the action $\PSym$, and the remaining tracks to
		the action $\RSym$.

		We now introduce the notion of \emph{assignment}.
		Intuitively, an assignment gives a valuation of variables with strategies,
		where the latter are used to determine the behavior of agents in the game.
		With more detail, as in the case of first order logic, we use this concept
		as a technical tool to quantify over strategies associated with variables,
		independently of agents to which they are related to.
		So, assignments are \mbox{used precisely as a way to define a correspondence
		between variables and agents via strategies.}
		\begin{definition}[Assignments]
			\label{def:asg}
			An \emph{assignment} in a \CGS\ $\GName$ is a partial function $\asgFun :
			\VarSet \cup \AgSet \pto \StrSet$ mapping variables and agents in its
			domain to a strategy.
			An assignment $\asgFun$ is \emph{complete} if it is defined on all agents,
			i.e., $\AgSet \subseteq \dom{\asgFun}$.
			For a state $\sElm \in \StSet$, it is said that $\asgFun$ is
			\emph{$\sElm$-total} if all strategies $\asgFun(\lElm)$ are $\sElm$-total,
			for $\lElm \in \dom{\asgFun}$.
			The set $\AsgSet \defeq \VarSet \cup \AgSet \pto \StrSet$ (resp.,
			$\AsgSet(\sElm) \defeq \VarSet \cup \AgSet \pto \StrSet(\sElm)$) contains
			all (resp., $\sElm$-total) assignments.
			Moreover, $\AsgSet(\XSet) \defeq \XSet \to \StrSet$ (resp.,
			$\AsgSet(\XSet, \sElm) \defeq \XSet \to \StrSet(\sElm)$) indicates the
			subset of \emph{$\XSet$-defined} (resp., $\sElm$-total) assignments, i.e.,
			(resp., $\sElm$-total) assignments defined on the set $\XSet \subseteq
			\VarSet \cup \AgSet$.
		\end{definition}
		As an example of assignment, let us consider the function $\asgFun[1]
		\in \AsgSet$ in the \CGS\ $\GName[P\!RS]$, defined on the set $\{ \ASym,
		\xSym \}$, whose values are $\strFun[1]$ on $\ASym$ and $\strFun[2]$ on
		$\xSym$, where the strategies $\strFun[1], \strFun[2] \in \StrSet(\sSym[i])$
		are those described above.
		Another examples is given by the assignment $\asgFun[2] \in \AsgSet$,
		defined on the set $\{ \ASym, \BSym \}$, such that $\asgFun[2](\ASym) =
		\asgFun[1](\xSym)$ and $\asgFun[2](\BSym) = \asgFun[1](\ASym)$.
		Note that both are $\sSym[i]$-total and the latter is also complete while
		the former is not.

		As in the case of strategies, it is useful to define the operation of
		\emph{translation} along a given track for assignments too.
		\begin{definition}[Assignment Translation]
			\label{def:asgtrn}
			For a given state $\sElm \in \StSet$, let $\asgFun \in \AsgSet(\sElm)$ be
			an $\sElm$-total assignment and $\trkElm \in \TrkSet(\sElm)$ a track.
			Then, $(\asgFun)_{\trkElm} \in \AsgSet(\lst{\trkElm})$ denotes the
			\emph{translation} of $\asgFun$ along $\trkElm$, i.e., the
			$\lst{\trkElm}$-total assignment, with $\dom{(\asgFun)_{\trkElm}} \defeq
			\dom{\asgFun}$, such that $(\asgFun)_{\trkElm}(\lElm) \defeq
			(\asgFun(\lElm))_{\trkElm}$, for all $\lElm \in \dom{\asgFun}$.
		\end{definition}
		Intuitively, the translation $(\asgFun)_{\trkElm}$ is the simultaneous
		update of all strategies $\asgFun(\lElm)$ defined by the assignment
		$\asgFun$, once the history of the game becomes $\trkElm$.

		Given an assignment $\asgFun$, an agent or variable $\lElm$, and a strategy
		$\strFun$, it is important to define a notation to represent the
		\emph{redefinition} of $\asgFun$, i.e., a new assignment equal to the first
		on all elements of its domain but $\lElm$, on which it assumes the value
		$\strFun$.
		\begin{definition}[Assignment Redefinition]
			\label{def:asgrdf}
			Let $\asgFun \in \AsgSet$ be an assignment, $\strFun \in \StrSet$ a
			strategy and $\lElm \in \VarSet \cup \AgSet$ either an agent or a
			variable.
			Then, $\asgFun[][\lElm \mapsto \strFun] \in \AsgSet$ denotes the new
			assignment defined on $\dom{\asgFun[][\lElm \mapsto \strFun]} \defeq
			\dom{\asgFun} \cup \{ \lElm \}$ that returns $\strFun$ on $\lElm$ and is
			equal to $\asgFun$ on the remaining part of its domain, i.e.,
			$\asgFun[][\lElm \mapsto \strFun](\lElm) \defeq \strFun$ and
			$\asgFun[][\lElm \mapsto \strFun](\lElm') \defeq \asgFun(\lElm')$, for all
			$\lElm' \in \dom{\asgFun} \setminus \{ \lElm \}$.
		\end{definition}
		Intuitively, if we have to add or update a strategy that needs to be bound
		by an agent or variable, we can simply take the old assignment and redefine
		it by using the above notation.
		It is worth to observe that, if $\asgFun$ and $\strFun$ are $\sElm$-total
		then $\asgFun[][\lElm \mapsto \strFun]$ is $\sElm$-total too.

		Now, we can introduce the concept of \emph{play} in a game.
		Intuitively, a play is the unique outcome of the game determined by all
		agent strategies participating to it.
		\begin{definition}[Plays]
			\label{def:play}
			A path $\playElm \in \PthSet(\sElm)$ starting at a state $\sElm \in
			\StSet$ is a \emph{play} w.r.t.\ a complete $\sElm$-total assignment
			$\asgFun \in \AsgSet(\sElm)$ (\emph{$(\asgFun, \sElm)$-play}, for short)
			if, for all $i \in \SetN$, it holds that $(\playElm)_{i + 1} =
			\trnFun((\playElm)_{i}, \decFun)$, where $\decFun(\aElm) \defeq
			\asgFun(\aElm)((\playElm)_{\leq i})$, for each $\aElm \in
			\AgSet$.~\footnote{The notation $(\wElm)_{\leq i} \in \Sigma^{*}$
			indicates the \emph{prefix} up to index $i \in \numcc{0}{\card{\wElm}}$ of
			a non-empty sequence $\wElm \in \Sigma^{\infty}$.}
			The partial function $\playFun : \AsgSet \times \StSet \pto \PthSet$, with
			$\dom{\playFun} \defeq \set{ (\asgFun, \sElm) }{ \AgSet \subseteq
			\dom{\asgFun} \land \asgFun \in \AsgSet(\sElm) \land \sElm \in \StSet }$,
			returns the $(\asgFun, \sElm)$-play $\playFun(\asgFun, \sElm) \in
			\PthSet(\sElm)$, for all pairs $(\asgFun, \sElm)$ in its domain.
		\end{definition}
		As a last example, consider again the complete $\sSym[i]$-total assignment
		$\asgFun[2]$ previously described for the \CGS\ $\GName[P\!RS]$, which
		returns the strategies $\strFun[2]$ and $\strFun[1]$ on the agents $\ASym$
		and $\BSym$, respectively.
		Then, we have that $\playFun(\asgFun[2], \sSym[i]) = \sSym[i]^{3} \cdot
		\sSym[\BSym]^{\omega}$.
		This means that the play is won by the agent $\BSym$.

		Finally, we give the definition of global translation of a complete
		assignment together with a related state, which is used to calculate, at a
		certain step of the play, what is the current state and its updated
		assignment.
		\begin{definition}[Global Translation]
			\label{def:gbltrn}
			For a given state $\sElm \in \StSet$ and a complete $\sElm$-total
			assignment $\asgFun \in \AsgSet(\sElm)$, the \emph{$i$-th global
			translation} of $(\asgFun, \sElm)$, with $i \in \SetN$, is the pair of a
			complete assignment and a state $(\asgFun, \sElm)^{i} \defeq
			((\asgFun)_{(\playElm)_{\leq i}}, (\playElm)_{i})$, where $\playElm =
			\playFun(\asgFun, \sElm)$.
		\end{definition}

		In order to avoid any ambiguity of interpretation of the described notions,
		we may use the name of a \CGS\ as a subscript of the sets and functions just
		introduced to clarify to which structure they are related to, as in the case
		of components in the tuple-structure of the \CGS\ itself.

	\end{subsection}

	\begin{subsection}{Semantics}
		\label{subsec:semantics}

		As already reported at the beginning of this section, just like \ATLS\ and
		differently from \CHPSL, the semantics of \SL\ is defined w.r.t.\ concurrent
		game structures.
		For a \CGS\ $\GName$, one of its states $\sElm$, and an $\sElm$-total
		assignment $\asgFun$ with $\free{\varphi} \subseteq \dom{\asgFun}$, we write
		$\GName, \asgFun, \sElm \models \varphi$ to indicate that the formula
		$\varphi$ holds at $\sElm$ in $\GName$ under $\asgFun$.
		The semantics of \SL\ formulas involving the atomic propositions, the
		Boolean connectives $\neg$, $\wedge$, and $\vee$, as well as the temporal
		operators $\X\!$, $\U\!$, and $\R\!$ is defined as usual in \LTL.
		The novel part resides in the formalization of the meaning of strategy
		quantifications $\EExs{\xElm}$ and $\AAll{\xElm}$ and agent binding $(\aElm,
		\xElm)$.
		\begin{definition}[\SL\ Semantics]
			\label{def:sl(semantics)}
			Given a \CGS\ $\GName$, for all \SL\ formulas $\varphi$, states $\sElm \in
			\StSet$, and $\sElm$-total assignments $\asgFun \in \AsgSet(\sElm)$ with
			$\free{\varphi} \subseteq \dom{\asgFun}$, the modeling relation $\GName,
			\asgFun, \sElm \models \varphi$ is inductively defined as follows.
			\begin{enumerate}
				\item\label{def:sl(semantics:ap)}
					$\GName, \asgFun, \sElm \models \pElm$ if $\pElm \in \labFun(\sElm)$,
					with $\pElm \in \APSet$.
				\item\label{def:sl(semantics:bool)}
					For all formulas $\varphi$, $\varphi_{1}$, and $\varphi_{2}$, it holds
					that:
					\begin{enumerate}
						\item\label{def:sl(semantics:neg)}
							$\GName, \asgFun, \sElm \models \neg \varphi$ if not $\GName,
							\asgFun, \sElm \models \varphi$, that is $\GName, \asgFun, \sElm
							\not\models \varphi$;
						\item\label{def:sl(semantics:conj)}
							$\GName, \asgFun, \sElm \models \varphi_{1} \wedge \varphi_{2}$ if
							$\GName, \asgFun, \sElm \models \varphi_{1}$ and $\GName, \asgFun,
							\sElm \models \varphi_{2}$;
						\item\label{def:sl(semantics:disj)}
							$\GName, \asgFun, \sElm \models \varphi_{1} \vee \varphi_{2}$ if
							$\GName, \asgFun, \sElm \models \varphi_{1}$ or $\GName, \asgFun,
							\sElm \models \varphi_{2}$.
					\end{enumerate}
				\item\label{def:sl(semantics:qnt)}
					For a variable $\xElm \in \VarSet$ and a formula $\varphi$, it holds
					that:
					\begin{enumerate}
						\item\label{def:sl(semantics:eqnt)}
							$\GName, \asgFun, \sElm \models \EExs{\xElm} \varphi$ if there
							exists an $\sElm$-total strategy $\strFun \in \StrSet(\sElm)$ such
							that $\GName, \asgFun[][\xElm \mapsto \strFun], \sElm \models
							\varphi$;
						\item\label{def:sl(semantics:aqnt)}
							$\GName, \asgFun, \sElm \models \AAll{\xElm} \varphi$ if for all
							$\sElm$-total strategies $\strFun \in \StrSet(\sElm)$ it holds
							that $\GName, \asgFun[][\xElm \mapsto \strFun], \sElm \models
							\varphi$.
					\end{enumerate}
				\item\label{def:sl(semantics:bnd)}
					For an agent $\aElm \in \AgSet$, a variable $\xElm \in \VarSet$, and a
					formula $\varphi$, it holds that $\GName, \asgFun, \sElm \models
					(\aElm, \xElm) \varphi$ if $\GName, \asgFun[][\aElm \mapsto
					\asgFun(\xElm)], \sElm \models \varphi$.
				\item\label{def:sl(semantics:path)}
					Finally, if the assignment $\asgFun$ is also complete, for all
					formulas $\varphi$, $\varphi_{1}$, and $\varphi_{2}$, it holds that:
					\begin{enumerate}
						\item\label{def:sl(semantics:next)}
							$\GName, \asgFun, \sElm \models \X \varphi$ if $\GName, (\asgFun,
							\sElm)^{1} \models \varphi$;
						\item\label{def:sl(semantics:until)}
							$\GName, \asgFun, \sElm \models \varphi_{1} \U \varphi_{2}$ if
							there is an index $i \in \SetN$ with $k \leq i$ such that $\GName,
							(\asgFun, \sElm)^{i} \models \varphi_{2}$ and, for all indexes $j
							\in \SetN$ with $k \leq j < i$, it holds that $\GName, (\asgFun,
							\sElm)^{j} \models \varphi_{1}$;
						\item\label{def:sl(semantics:release)}
							$\GName, \asgFun, \sElm \models \varphi_{1} \R \varphi_{2}$ if,
							for all indexes $i \in \SetN$ with $k \leq i$, it holds that
							$\GName, (\asgFun, \sElm)^{i} \models \varphi_{2}$ or there is an
							index $j \in \SetN$ with $k \leq j < i$ such that $\GName,
							(\asgFun, \sElm)^{j} \models \varphi_{1}$.
					\end{enumerate}
			\end{enumerate}
		\end{definition}
		Intuitively, at Items~\ref{def:sl(semantics:eqnt)}
		and~\ref{def:sl(semantics:aqnt)}, respectively, we evaluate the existential
		$\EExs{\xElm}$ and universal $\AAll{\xElm}$ quantifiers over strategies, by
		associating them to the variable $\xElm$.
		Moreover, at Item~\ref{def:sl(semantics:bnd)}, by means of an agent binding
		$(\aElm, \xElm)$, we commit the agent $\aElm$ to a strategy associated with
		the variable $\xElm$.
		It is evident that, due to Items~\ref{def:sl(semantics:next)},
		\ref{def:sl(semantics:until)}, and~\ref{def:sl(semantics:release)}, the
		\LTL\ semantics is simply embedded into the \SL\ one.

		In order to complete the description of the semantics, we now give the
		classic notions of \emph{model} and \emph{satisfiability} of an \SL\
		sentence.
		\begin{definition}[\SL\ Satisfiability]
			\label{def:sl(sat)}
			We say that a \CGS\ $\GName$ is a \emph{model} of an \SL\ sentence
			$\varphi$, in symbols $\GName \models \varphi$, if $\GName, \emptyfun,
			\sElm[0] \models \varphi$.~\footnote{The symbol $\emptyfun$ stands for the
			empty function.}
			In general, we also say that $\GName$ is a \emph{model} for $\varphi$ on
			$\sElm \in \StSet$, in symbols $\GName, \sElm \models \varphi$, if
			$\GName, \emptyfun, \sElm \models \varphi$.
			An \SL\ sentence $\varphi$ is \emph{satisfiable} if there is a model for
			it.
		\end{definition}

		It remains to introduce the concepts of \emph{implication} and
		\emph{equivalence} between \SL\ formulas, which are useful to describe
		transformations preserving the meaning of a specification.
		\begin{definition}[\SL\ Implication and Equivalence]
			\label{def:sl(impeqv)}
			\ Given two \SL\ formulas $\varphi_{1}$ and $\varphi_{2}$ with
			$\free{\varphi_{1}} = \free{\varphi_{2}}$, we say that $\varphi_{1}$
			\emph{implies} $\varphi_{2}$, in symbols $\varphi_{1} \implies
			\varphi_{2}$, if, for all \CGS s $\GName$, states $\sElm \in \StSet$, and
			$\free{\varphi_{1}}$-defined $\sElm$-total assignments $\asgFun \in
			\AsgSet(\free{\varphi_{1}}, \sElm)$, it holds that if $\GName, \asgFun,
			\sElm \models \varphi_{1}$ then $\GName, \asgFun, \sElm \models
			\varphi_{2}$.
			Accordingly, we say that $\varphi_{1}$ is \emph{equivalent} to
			$\varphi_{2}$, in symbols $\varphi_{1} \equiv \varphi_{2}$, if both
			$\varphi_{1} \implies \varphi_{2}$ and $\varphi_{2} \implies \varphi_{1}$
			hold.
		\end{definition}

		In the rest of the paper, especially when we describe a decision procedure,
		we may consider formulas in \emph{existential normal form} (\emph{\enf}, for
		short) and \emph{positive normal form} (\emph{\pnf}, for short), i.e.,
		formulas in which only existential quantifiers appear or in which the
		negation is applied only to atomic propositions.
		In fact, it is to this aim that we have considered in the syntax of \SL\
		both the Boolean connectives $\wedge$ and $\vee$, the temporal operators
		$\U\!\!$, and $\R\!\!$, and the strategy quantifiers $\EExs{ \cdot }$ and
		$\AAll{ \cdot }$.
		Indeed, all formulas can be linearly translated in \pnf\ by using De
		Morgan's laws together with the following equivalences, which directly
		follow from the semantics of the logic: $\neg \X \varphi \equiv \X \neg
		\varphi$, $\neg (\varphi_{1} \U \varphi_{2}) \equiv (\neg \varphi_{1}) \R
		(\neg \varphi_{2})$, $\neg \EExs{x} \varphi \equiv \AAll{x} \neg \varphi$,
		and $\neg (\aElm, \xElm) \varphi \equiv (\aElm, \xElm) \neg \varphi$.

		\par\figexmsv\
		At this point, in order to better understand the meaning of our logic, we
		discuss two examples in which we describe the evaluation of the semantics of
		some formula w.r.t.\ the a priori given \CGS s.
		We start by explaining how a strategy can be shared by different agents.
		\begin{example}[Shared Variable]
			\label{exm:sv}
			Consider the \SL[$2$-alt] sentence $\varphi = \EExs{\xSym} \AAll{\ySym}
			\EExs{\zSym} ((\alpha, \xSym) (\beta, \ySym) (\X \pSym) \wedge (\alpha,
			\ySym) (\beta, \zSym) (\X \qSym))$.
			It is immediate to note that both agents $\alpha$ and $\beta$ use the
			strategy associated with $\ySym$ to achieve simultaneously the \LTL\
			temporal goals $\X \pSym$ and $\X \qSym$.
			A model for $\varphi$ is given by the \CGS\ $\GName[S\!V] \defeq \CGSTuple
			{\{ \pSym, \qSym \}} {\{ \alpha, \beta\}} {\{ 0, 1 \}} {\{ \sSym[0],
			\sSym[1], \sSym[2], \sSym[3] \}} {\labFun} {\trnFun} {\sSym[0]}$, where
			$\labFun(\sSym[0]) \defeq \emptyset$, $\labFun(\sSym[1]) \defeq \{ \pSym
			\}$, $\labFun(\sSym[2]) \defeq \{ \pSym, \qSym \}$, $\labFun(\sSym[3])
			\defeq \{ \qSym \}$, $\trnFun(\sSym[0], (0, 0)) \defeq \sSym[1]$,
			$\trnFun(\sSym[0], (0, 1)) \defeq \sSym[2]$, $\trnFun(\sSym[0], (1, 0))
			\defeq \sSym[3]$, and all the remaining transitions (with any decision) go
			to $\sSym[0]$.
			In Figure~\vref{fig:exm:sv}, we report a graphical representation of the
			structure.
			Clearly, $\GName[S\!V] \models \varphi$ by letting, on $\sSym[0]$, the
			variables $\xSym$ to chose action $0$ (the goal $(\alpha, \xSym) (\beta,
			\ySym) (\X \pSym)$ is satisfied for any choice of $\ySym$, since we can
			move from $\sSym[0]$ to either $\sSym[1]$ or $\sSym[2]$, both labeled with
			$\pSym$) and $\zSym$ to choose action $1$ when $\ySym$ has action $0$ and,
			vice versa, $0$ when $\ySym$ has $1$ (in both cases, the goal
			$(\alpha, \ySym) (\beta, \zSym) (\X \qSym)$ is satisfied, since one can
			move from $\sSym[0]$ to either $\sSym[2]$ or $\sSym[3]$, both labeled with
			$\qSym$).
		\end{example}

		We now discuss an application of the concepts of Nash equilibrium and
		stability profile to both the prisoner's dilemma and the paper, rock, and
		scissor game.
		\begin{example}[Equilibrium Profiles]
			\label{exm:ep}
			Let us first to consider the \CGS\ $\GName[P\!D]$ of the prisoner's
			dilemma described in the Example~\vref{exm:pd}.
			Intuitively, each of the two accomplices $\ASym[1]$ and $\ASym[2]$ want to
			avoid the prison.
			These goals can be, respectively, represented by the \LTL\ formulas
			$\psi_{\ASym[1]} \defeq \G \fSym[ {\ASym[1]} ]$ and $\psi_{\ASym[2]}
			\defeq \G \fSym[ {\ASym[2]} ]$.
			The existence of a Nash equilibrium in $\GName[P\!D]$ for the two
			accomplices w.r.t.\ the above goals can be written as $\phi_{\!N\!E}
			\defeq \EExs{\xSym[1]} (\ASym[1], \xSym[1]) \EExs{\xSym[2]} (\ASym[2],
			\xSym[2]) \: \psi_{\!N\!E}$, where $\psi_{\!N\!E} \defeq ((\EExs{\ySym}
			(\ASym[1], \ySym) \psi_{\ASym[1]}) \rightarrow \psi_{\ASym[1]}) \wedge
			((\EExs{\ySym} (\ASym[2], \ySym) \psi_{\ASym[2]}) \rightarrow
			\psi_{\ASym[2]})$, which results to be an instantiation of the general
			sentence $\varphi_{\!N\!E}$ of Example~\vref{exm:ne}.
			In the same way, the existence of a stable Nash equilibrium can be
			represented with the sentence $\phi_{\!S\!N\!\!E} \defeq \EExs{\xSym[1]}
			(\ASym[1], \xSym[1]) \EExs{\xSym[2]} (\ASym[2], \xSym[2]) \: \psi_{\!N\!E}
			\wedge \psi_{\!S\!P}$, where $\psi_{\!S\!P} \defeq (\psi_{1} \rightarrow
			\AAll{\ySym} ((\psi_{2} \leftrightarrow (\ASym[2], \ySym) \psi_{2})
			\rightarrow (\ASym[2], \ySym) \psi_{1})) \wedge (\psi_{2} \rightarrow
			\AAll{\ySym} ((\psi_{1} \leftrightarrow (\ASym[1], \ySym) \psi_{1})
			\rightarrow (\ASym[1], \ySym) \psi_{2}))$, which is a particular case of
			the sentence $\varphi_{\!S\!N\!\!E}$ of Example~\vref{exm:sp}.
			Now, it is easy to see that $\GName[P\!D] \models \phi_{\!S\!N\!E}$ and,
			so, $\GName[P\!D] \models \phi_{\!N\!E}$.
			Indeed, an assignment $\asgFun \in \AsgSet[ {\GName[P\!D]} ](\AgSet,
			\sSym[i])$, for which $\asgFun(\ASym[1])(\sSym[i]) =
			\asgFun(\ASym[2])(\sSym[i]) = \DSym$, is a stable equilibrium profile,
			i.e., it is such that $\GName[P\!D], \asgFun, \sSym[i] \models
			\psi_{\!N\!E} \wedge \psi_{\!S\!P}$.
			This is due to the fact that, if an agent $\ASym[k]$, for $k \in \{ 1, 2
			\}$, choses another strategy $\strFun \in \StrSet[ {\GName[P\!D]}
			](\sSym[i])$, he is still unable to achieve his goal $\psi_{k}$, i.e.,
			$\GName[P\!D], \asgFun[][\ASym[k] \mapsto \strFun], \sSym[i] \not\models
			\psi_{k}$, so, he cannot improve his payoff.
			Moreover, this equilibrium is stable, since the payoff of an agent cannot
			be made worse by the changing of the strategy of the other agent.
			However, it is interesting to note that there are instable equilibria too.
			One of these is represented by the assignment $\asgFun' \in \AsgSet[
			{\GName[P\!D]} ](\AgSet, \sSym[i])$, for which
			$\asgFun'(\ASym[1])(\sSym[i]^{j}) = \asgFun'(\ASym[2])(\sSym[i]^{j}) =
			\CSym$, for all $j \in \SetN$.
			Indeed, we have that $\GName[P\!D], \asgFun', \sSym[i] \models
			\psi_{\!N\!E}$, since $\GName[P\!D], \asgFun', \sSym[i] \models \psi_{1}$
			and $\GName[P\!D], \asgFun', \sSym[i] \models \psi_{2}$, but
			$\GName[P\!D], \asgFun', \sSym[i] \not\models \psi_{\!S\!P}$.
			The latter property holds because, if one of the agents $\ASym[k]$, for $k
			\in \{ 1, 2 \}$, choses a different strategy $\strFun' \in \StrSet[
			{\GName[P\!D]} ](\sSym[i])$ for which there is a $j \in \SetN$ such that
			$\strFun'(\sSym[i]^{j}) = \DSym$, he cannot improve his payoff but makes
			surely worse the payoff of the other agent, i.e., $\GName[P\!D],
			\asgFun'[\ASym[k] \mapsto \strFun'], \sSym[i] \models \psi_{k}$ but
			$\GName[P\!D], \asgFun'[\ASym[k] \mapsto \strFun'], \sSym[i] \not\models
			\psi_{3 - k}$.
			Finally, consider the \CGS\ $\GName[P\!RS]$ of the paper, rock, and
			scissor game described in the Example~\vref{exm:prs} together with the
			associated formula for the Nash equilibrium $\phi_{\!N\!E} \defeq
			\EExs{\xSym[1]} (\ASym, \xSym[1]) \EExs{\xSym[2]} (\BSym, \xSym[2]) \:
			\psi_{\!N\!E}$, where $\psi_{\!N\!E} \defeq ((\EExs{\ySym} (\ASym, \ySym)
			\psi_{\ASym}) \rightarrow \psi_{\ASym}) \wedge ((\EExs{\ySym} (\BSym,
			\ySym) \psi_{\BSym}) \rightarrow \psi_{\BSym})$ with $\psi_{\ASym} \defeq
			\F \wSym[\ASym]$ and $\psi_{\BSym} \defeq \F \wSym[\BSym]$ representing
			the \LTL\ temporal goals for Alice and Bob, respectively.
			Then, it is not hard to see that $\GName[P\!RS] \not\models
			\phi_{\!N\!E}$, i.e., there are no Nash equilibria in this game, since
			there is necessarily an agent that can improve his/her payoff by changing
			his/her strategy.
		\end{example}

		Finally, we want to remark that our semantics framework, based on concurrent
		game structures, is enough expressive to describe turn-based features in the
		multi-agent case too.
		This is possible by simply allowing the transition function to depend only
		on the choice of actions of an a priori given agent for each state.
		\begin{definition}[Turn-Based Game Structures]
			\label{def:tbgs}
			A \CGS\ $\GName$ is \emph{turn-based} if there exists a function $\ownFun
			: \StSet \to \AgSet$, named \emph{owner function}, such that, for all
			states $\sElm \in \StSet$ and decisions $\decFun[1], \decFun[2] \in
			\DecSet$, it holds that if $\decFun[1](\ownFun(\sElm)) =
			\decFun[2](\ownFun(\sElm))$ then $\trnFun(\sElm, \decFun[1]) =
			\trnFun(\sElm, \decFun[2])$.
		\end{definition}
		Intuitively, a \CGS\ is turn-based if it is possible to associate with each
		state an agent, i.e., the owner of the state, which is responsible for the
		choice of the successor of that state.
		It is immediate to observe that $\ownFun$ introduces a partitioning of the
		set of states into $\card{\rng{\ownFun}}$ components, each one ruled by a
		single agent.
		Moreover, observe that a \CGS\ having just one agent is trivially
		turn-based, since this agent is the only possible owner of all states.

		In the following, as one can expect, we also consider the case in which \SL\
		has its semantics defined on turn-based \CGS\ only.
		In such an eventuality, we name the resulting semantic fragment
		\emph{Turn-based Strategy Logic} (\TBSL, for short) and refer to the related
		satisfiability concept as \emph{turn-based satisfiability}.

	\end{subsection}

\end{section}




\begin{section}{Model-Checking Hardness}
	\label{sec:modchkhrd}

	In this section, we show the non-elementary lower bound for the model-checking
	problem of \SL.
	Precisely, we prove that, for sentences having alternation number $k$, this
	problem is $k$-\ExpSpaceH.
	To this aim, in Subsection~\ref{subsec:qptl}, we first recall syntax and
	semantics of \QPTL~\cite{Sis83}.
	Then, in Subsection~\ref{subsec:hrdrdc}, we give a reduction from the
	satisfiability problem for this logic to the model-checking problem for \SL.

	\begin{subsection}{Quantified propositional temporal logic}
		\label{subsec:qptl}

		\emph{Quantified Propositional Temporal Logic} (\emph{\QPTL}, for short)
		syntactically extends the old-style temporal logic with the \emph{future}
		$\F\!$ and \emph{global} $\G\!$ operators by means of two \emph{proposition
		quantifiers}, the existential $\exists \qElm .$ and the universal $\forall
		\qElm .$, where $\qElm$ is an atomic proposition.
		Intuitively, these elements can be respectively read as \emph{``there exists
		an evaluation of $\qElm$''} and \emph{``for all evaluations of $\qElm$''}.
		The formal syntax of \QPTL\ follows.
		\begin{definition}[\QPTL\ Syntax]
			\label{def:qptl(syntax)}
			\QPTL\ \emph{formulas} are built inductively from the sets of atomic
			propositions $\APSet$, by using the following grammar, where $\pElm \in
			\APSet$:
			\begin{center}
				$\varphi ::= \pElm \mid \neg \varphi \mid \varphi \wedge \varphi \mid
				\varphi \vee \varphi \mid \X \varphi \mid \F \varphi \mid \G \varphi
				\mid \exists \pElm . \varphi \mid \forall \pElm . \varphi$.
			\end{center}
			\QPTL\ denotes the infinite set of formulas generated by the above
			grammar.
		\end{definition}

		Similarly to \SL, we use the concepts of subformula, free atomic
		proposition, sentence, and alternation number, together with the \QPTL\
		syntactic fragment of bounded alternation \QPTL[$k$-alt], with $k \in
		\SetN$.

		In order to define the semantics of \QPTL, we have first to introduce the
		concepts of truth evaluations used to interpret the meaning of atomic
		propositions at the passing of time.
		\begin{definition}[Truth Evaluations]
			\label{def:trhevl}
			A \emph{temporal truth evaluation} is a function $\tteFun : \SetN \to \{
			\Ff, \Tt \}$ that maps each natural number to a Boolean value.
			Moreover, a \emph{propositional truth evaluation} is a partial function
			$\pteFun : \APSet \pto \TTESet$ mapping every atomic proposition in its
			domain to a temporal truth evaluation.
			The sets $\TTESet \defeq \SetN \to \{ \Ff, \Tt \}$ and $\PTESet \defeq
			\APSet \pto \TTESet$ contain, respectively, all temporal and propositional
			truth evaluations.
		\end{definition}

		At this point, we have the tool to define the interpretation of \QPTL\
		formulas.
		For a propositional truth evaluation $\pteFun$ with $\free{\varphi}
		\subseteq \dom{\pteFun}$ and a number $k$, we write $\pteFun, k \models
		\varphi$ to indicate that the formula $\varphi$ holds at the $k$-th position
		of the $\pteFun$.
		\begin{definition}[\QPTL\ Semantics]
			\label{def:qptl(semantics)}
			For all \QPTL\ formulas $\varphi$, propositional truth evaluation $\pteFun
			\in \PTESet$ with $\free{\varphi} \subseteq \dom{\pteFun}$, and numbers $k
			\in \SetN$, the modeling relation $\pteFun, k \models \varphi$ is
			inductively defined as follows.
			\begin{enumerate}
				\item\label{def:qptl(semantics:ap)}
					$\pteFun, k \models \pElm$ iff $\pteFun(\pElm)(k) = \Tt$, with
					$\pElm \in \APSet$.
				\item\label{def:qptl(semantics:bool)}
					For all formulas $\varphi$, $\varphi_{1}$, and $\varphi_{2}$, it holds
					that:
					\begin{enumerate}
						\item\label{def:qptl(semantics:neg)}
							$\pteFun, k \models \neg \varphi$ iff not $\pteFun, k \models
							\varphi$, that is $\pteFun, k \not\models \varphi$;
						\item\label{def:qptl(semantics:conj)}
							$\pteFun, k \models \varphi_{1} \wedge \varphi_{2}$ iff
							$\pteFun, k \models \varphi_{1}$ and $\pteFun, k \models
							\varphi_{2}$;
						\item\label{def:qptl(semantics:disj)}
							$\pteFun, k \models \varphi_{1} \vee \varphi_{2}$ iff $\pteFun,
							k \models \varphi_{1}$ or $\pteFun, k \models \varphi_{2}$;
						\item\label{def:qptl(semantics:next)}
							$\pteFun, k \models \X \varphi$ iff $\pteFun, k + 1 \models
							\varphi$;
						\item\label{def:qptl(semantics:future)}
							$\pteFun, k \models \F \varphi$ iff there is an index $i \in
							\SetN$ with $k \leq i$ such that $\pteFun, i \models \varphi$;
						\item\label{def:qptl(semantics:globally)}
							$\pteFun, k \models \G \varphi$ iff, for all indexes $i \in \SetN$
							with $k \leq i$, it holds that $\pteFun, i \models \varphi$.
					\end{enumerate}
				\item\label{def:qptl(semantics:qnt)}
					For an atomic proposition $\qElm \in \APSet$ and a formula $\varphi$,
					it holds that:
					\begin{enumerate}
						\item\label{def:qptl(semantics:eqnt)}
							$\pteFun, k \models \exists \qElm . \varphi$ iff there exists a
							temporal truth evaluation $\tteFun \in \TTESet$ such that
							$\pteFun[][\qElm \mapsto \tteFun], k \models \varphi$;
						\item\label{def:qptl(semantics:aqnt)}
							$\pteFun, k \models \forall \qElm . \varphi$ iff for all temporal
							truth evaluations $\tteFun \in \TTESet$ it holds that
							$\pteFun[][\qElm \mapsto \tteFun], k \models \varphi$.
					\end{enumerate}
			\end{enumerate}
		\end{definition}
		Obviously, a \QPTL\ sentence $\varphi$ is \emph{satisfiable} if $\emptyfun,
		0 \models \varphi$.
		Observe that the described semantics is slightly different but completely
		equivalent to that proposed and used in~\cite{SVW87} to prove the
		non-elementary hardness result for the satisfiability problem.

	\end{subsection}

	\begin{subsection}{Non-elementary lower-bound}
		\label{subsec:hrdrdc}

		We can show how the solution of \QPTL\ satisfiability problem can be reduced
		to that of the model-checking problem for \SL, over a turn-based constant
		size \CGS\ with a unique atomic proposition.

		\par\figlmmqptlrdc\
		In order to do this, we first prove the following auxiliary lemma, which
		actually represents the main step of the above mentioned reduction.
		\begin{lemma}[\QPTL\ Reduction]
			\label{lmm:qptl(rdc)}
			There is a one-agent \CGS\ $\GName[Rdc]$ such that, for each
			\QPTL[$k$-alt] sentence $\varphi$, with $k \in \SetN$, there exists an
			\TBSL[$k$-alt] variable-closed formula $\trn{\varphi}$ such that $\varphi$
			is satisfiable iff $\GName[Rdc], \asgFun, \sElm[0] \models \trn{\varphi}$,
			for all complete assignments $\asgFun \in \AsgSet(\AgSet, \sElm[0])$.
		\end{lemma}
		\begin{proof}
			Consider the one-agent \CGS\ $\GName[Rdc] \defeq \CGSTuple { \{ \pSym \} }
			{ \{ \alpha \} } { \{ \Ff, \Tt \} } { \{ \sSym[0], \sSym[1] \} } {\labFun}
			{\trnFun} {\sSym[0]}$ depicted in Figure~\vref{fig:lmm:qptl(rdc)}, where
			the two actions are the Boolean values false and true and where the
			labeling and transition functions $\labFun$ and $\trnFun$ are set as
			follows: $\labFun(\sSym[0]) \defeq \emptyset$, $\labFun(\sSym[1]) \defeq
			\{ \pSym \}$, and $\trnFun(\sElm, \decFun) = \sSym[0]$ iff
			$\decFun(\alpha) = \Ff$, for all $\sElm \in \StSet$ and $\decFun \in
			\DecSet$.
			It is evident that $\GName[Rdc]$ is a turn-based \CGS.
			Moreover, consider the transformation function $\trn{\cdot} : \QPTL \to
			\SL$ inductively defined as follows:
			\begin{itemize}
				\item
					$\trn{\qElm} \defeq (\alpha, \xSym[\qElm]) \X \pSym$, for $\qElm \in
					\APSet$;
				\item
					$\trn{\exists \qElm . \varphi} \defeq \EExs{\xSym[\qElm]}
					\trn{\varphi}$;
				\item
					$\trn{\forall \qElm . \varphi} \defeq \AAll{\xSym[\qElm]}
					\trn{\varphi}$;
				\item
					$\trn{\Opr \varphi} \defeq \Opr \trn{\varphi}$, where $\Opr \in \{
					\neg, \X\!, \F\!, \G\! \}$;
				\item
					$\trn{\varphi_{1} \Opr \varphi_{2}} \defeq \trn{\varphi_{1}} \Opr
					\trn{\varphi_{2}}$, where $\Opr \in \{ \wedge, \vee \}$.
			\end{itemize}
			It is not hard to see that a \QPTL\ formula $\varphi$ is a sentence iff
			$\trn{\varphi}$ is variable-closed.
			Furthermore, we have that $\alt{\trn{\varphi}} = \alt{\varphi}$.

			At this point, it remains to prove that, a \QPTL\ sentence $\varphi$ is
			satisfiable iff $\GName[Rdc], \asgFun, \sSym[0] \! \models \trn{\varphi}$,
			for all total assignments $\asgFun \in \AsgSet(\{ \alpha \}, \sSym[0])$.
			To do this by induction on the structure of $\varphi$, we actually show a
			stronger result asserting that, for all subformulas $\psi \in
			\sub{\varphi}$, propositional truth evaluations $\pteFun \in \PTESet$, and
			$i \in \SetN$, it holds that $\pteFun, i \models \psi$ iff $\GName[Rdc],
			(\asgFun, \sSym[0])^{i} \models \trn{\psi}$, for each total assignment
			$\asgFun \in \AsgSet(\{ \alpha \} \cup \set{ \xSym[\qElm] \in \VarSet }{
			\qElm \in \free{\psi} }, \sSym[0])$ such that
			$\asgFun(\xSym[\qElm])((\playElm)_{\leq n}) = \pteFun(\qElm)(n)$, where
			$\playElm \defeq \playFun(\asgFun, \sSym[0])$, for all $\qElm \in
			\free{\psi}$ and $n \in \numco{i}{\omega}$.

			Here, we only show the base case of atomic propositions and the two
			inductive cases regarding the proposition quantifiers.
			The remaining cases of Boolean connectives and temporal operators are
			straightforward and left to the reader as a simple exercise.

			\begin{itemize}
				\item
					$\psi = \qElm$.

					\hspace{0.75em}
					By Item~\ref{def:qptl(semantics:ap)} of
					Definition~\ref{def:qptl(semantics)} of \QPTL\ semantics, we have that
					$\pteFun, i \models \qElm$ iff $\pteFun(\qElm)(i) = \Tt$.
					Thus, due to the above constraint on the assignment, it follows that
					$\pteFun, i \models \qElm$ iff $\asgFun(\xSym[\qElm])((\playElm)_{\leq
					i}) = \Tt$.
					Now, by applying Items~\ref{def:sl(semantics:bnd)}
					and~\ref{def:sl(semantics:next)} of Definition~\ref{def:sl(semantics)}
					of \SL\ semantics, we have that $\GName[Rdc], (\asgFun, \sSym[0])^{i}
					\models (\alpha, \xSym[\qElm]) \X \pSym$ iff $\GName[Rdc],
					(\asgFun'[\alpha \mapsto \asgFun'(\xSym[\qElm])], \sElm')^{1} \models
					\pSym$, where $(\asgFun', \sElm') = (\asgFun, \sSym[0])^{i}$.
					At this point, due to the particular structure of the \CGS\
					$\GName[Rdc]$, we have that $\GName[Rdc], (\asgFun'[\alpha \mapsto
					\asgFun'(\xSym[\qElm])], \sElm')^{1} \models \pSym$ iff
					$(\playElm')_{1} = \sSym[1]$, where $\playElm' \defeq
					\playFun(\asgFun'[\alpha \mapsto \asgFun'(\xSym[\qElm])], \sElm')$,
					which in turn is equivalent to
					$\asgFun'(\xSym[\qElm])((\playElm')_{\leq 0}) = \Tt$.
					So, $\GName[Rdc], (\asgFun, \sSym[0])^{i} \models (\alpha,
					\xSym[\qElm]) \X \pSym$ iff $\asgFun'(\xSym[\qElm])((\playElm')_{\leq
					0}) = \Tt$.
					Now, by observing that $(\playElm')_{\leq 0} = (\playElm)_{i}$ and
					using the above definition of $\asgFun'$, we obtain that
					$\asgFun'(\xSym[\qElm])((\playElm')_{\leq 0}) =
					\asgFun(\xSym[\qElm])((\playElm)_{\leq i})$.
					Hence, $\pteFun, i \models \qElm$ iff $\pteFun(\qElm)(i) =
					\asgFun(\xSym[\qElm])((\playElm)_{\leq i}) = \Tt =
					\asgFun'(\xSym[\qElm])((\playElm')_{\leq 0})$ iff $\GName[Rdc],
					(\asgFun, \sSym[0])^{i} \models (\alpha, \xSym[\qElm]) \X \pSym$.
				\item
					$\psi = \exists \qElm. \psi'$.

					\hspace{0.75em}
					\emph{[Only if].}
					If $\pteFun, i \models \exists \qElm. \psi'$, by
					Item~\ref{def:qptl(semantics:eqnt)} of
					Definition~\ref{def:qptl(semantics)}, there exists a temporal truth
					evaluation $\tteFun \in \TTESet$ such that $\pteFun[][\qElm \mapsto
					\tteFun], i \models \psi'$.
					Now, consider a strategy $\strFun \in \StrSet(\sSym[0])$ such that
					$\strFun((\playElm)_{\leq n}) = \tteFun(n)$, for all $n \in
					\numco{i}{\omega}$.
					Then, it is evident that $\asgFun[][\xSym[\qElm] \mapsto
					\strFun](\xSym[\qElm'])((\playElm)_{\leq n}) = \pteFun[][\qElm \mapsto
					\tteFun](\qElm')(n)$, for all $\qElm' \in \free{\psi}$ and $n \in
					\numco{i}{\omega}$.
					So, by the inductive hypothesis, it follows that $\GName[Rdc],
					(\asgFun[][\xSym[\qElm] \mapsto \strFun], \sSym[0])^{i} \models
					\trn{\psi'}$.
					Thus, we have that $\GName[Rdc], (\asgFun, \sSym[0])^{i} \models
					\EExs{\xSym[\qElm]} \trn{\psi'}$.

					\hspace{0.75em}
					\emph{[If].}
					If $\GName[Rdc], (\asgFun, \sSym[0])^{i} \models \EExs{\xSym[\qElm]}
					\trn{\psi'}$, there exists a strategy $\strFun \in \StrSet(\sSym[0])$
					such that $\GName[Rdc], (\asgFun[][\xSym[\qElm] \mapsto \strFun],
					\sSym[0])^{i} \models \trn{\psi'}$.
					Now, consider a temporal truth evaluation $\tteFun \in \TTESet$ such
					that $\tteFun(n) = \strFun((\playElm)_{\leq n})$, for all $n \in
					\numco{i}{\omega}$.
					Then, it is evident that $\asgFun[][\xSym[\qElm] \mapsto
					\strFun](\xSym[\qElm'])((\playElm)_{\leq n}) = \pteFun[][\qElm \mapsto
					\tteFun](\qElm')(n)$, for all $\qElm' \in \free{\psi}$ and $n \in
					\numco{i}{\omega}$.
					So, by the inductive hypothesis, it follows that $\pteFun[][\qElm
					\mapsto \tteFun], i \models \psi'$.
					Thus, by Item~\ref{def:qptl(semantics:eqnt)} of
					Definition~\ref{def:qptl(semantics)}, we have that $\pteFun, i \models
					\exists \qElm. \psi'$.
				\item
					$\psi = \forall \qElm. \psi'$.

					\hspace{0.75em}
					\emph{[Only if].}
					For each strategy $\strFun \in \StrSet(\sSym[0])$, consider a temporal
					truth evaluation $\tteFun \in \TTESet$ such that $\tteFun(n) =
					\strFun((\playElm)_{\leq n})$, for all $n \in \numco{i}{\omega}$.
					It is evident that $\asgFun[][\xSym[\qElm] \mapsto
					\strFun](\xSym[\qElm'])((\playElm)_{\leq n}) = \pteFun[][\qElm \mapsto
					\tteFun](\qElm')(n)$, for all $\qElm' \in \free{\psi}$ and $n \in
					\numco{i}{\omega}$.
					Now, since $\pteFun, i \models \forall \qElm. \psi'$, by
					Item~\ref{def:qptl(semantics:aqnt)} of
					Definition~\ref{def:qptl(semantics)}, it follows that $\pteFun[][\qElm
					\mapsto \tteFun], i \models \psi'$.
					So, by the inductive hypothesis, for each strategy $\strFun \in
					\StrSet(\sSym[0])$, it holds that $\GName[Rdc],
					(\asgFun[][\xSym[\qElm] \mapsto \strFun], \sSym[0])^{i} \models
					\trn{\psi'}$.
					Thus, we have that $\GName[Rdc], (\asgFun, \sSym[0])^{i} \models
					\AAll{\xSym[\qElm]} \trn{\psi'}$.

					\hspace{0.75em}
					\emph{[If].}
					For each temporal truth evaluation $\tteFun \in \TTESet$, consider a
					strategy $\strFun \in \StrSet(\sSym[0])$ such that
					$\strFun((\playElm)_{\leq n}) = \tteFun(n)$, for all $n \in
					\numco{i}{\omega}$.
					It is evident that $\asgFun[][\xSym[\qElm] \mapsto
					\strFun](\xSym[\qElm'])((\playElm)_{\leq n}) =
					\pteFun[][\qElm \mapsto \tteFun](\qElm')(n)$, for all $\qElm' \in
					\free{\psi}$ and $n \in \numco{i}{\omega}$.
					Now, since $\GName[Rdc], (\asgFun, \sSym[0])^{i} \models
					\AAll{\xSym[\qElm]} \trn{\psi'}$, it follows that $\GName[Rdc],
					(\asgFun[][\xSym[\qElm] \mapsto \strFun], \sSym[0])^{i} \models
					\trn{\psi'}$.
					So, by the inductive hypothesis, for each temporal truth evaluation
					$\tteFun \in \TTESet$, it holds that $\pteFun[][\qElm \mapsto
					\tteFun], i \models \psi'$.
					Thus, by Item~\ref{def:qptl(semantics:aqnt)} of
					Definition~\ref{def:qptl(semantics)}, we have that $\pteFun, i \models
					\forall \qElm. \psi'$.
			\end{itemize}
			Thus, we are done with the proof.
		\end{proof}

		Now, we can show the full reduction that allows us to state the existence of
		a non-elementary lower-bound for the model-checking problem of \TBSL\ and,
		thus, of \SL.
		\begin{theorem}[\TBSL\ Model-Checking Hardness]
			\label{thm:tbsl(modchkhrd)}
			The model-checking problem for \TBSL[$k$-alt] is $k$-\ExpSpaceH.
		\end{theorem}
		\begin{proof}
			Let $\varphi$ be a \QPTL[$k$-alt] sentence, $\trn{\varphi}$ the related
			\TBSL[$k$-alt] variable-closed formula, and $\GName[Rdc]$ the turn-based
			\CGS\ of Lemma~\ref{lmm:qptl(rdc)} of \QPTL\ reduction.
			Then, by applying the previous mentioned lemma, it is easy to see that
			$\varphi$ is satisfiable iff $\GName[Rdc] \models \AAll{\xSym} (\alpha,
			\xSym) \trn{\varphi}$ iff $\GName[Rdc] \models \EExs{\xSym} (\alpha,
			\xSym) \trn{\varphi}$.
			Thus, the satisfiability problem for \QPTL\ can be reduced to the
			model-checking problem for \TBSL.
			Now, since the satisfiability problem for \QPTL[$k$-alt] is
			$k$-\ExpSpaceH~\cite{SVW87}, we have that the model-checking problem for
			\TBSL[$k$-alt] is $k$-\ExpSpaceH\ as well.
		\end{proof}

		The following corollary is an immediate consequence of the previous theorem.
		\begin{corollary}[\SL\ Model-Checking Hardness]
			\label{cor:sl(modchkhrd)}
			The model-checking problem for \SL[$k$-alt] is $k$-\ExpSpaceH.
		\end{corollary}

	\end{subsection}

\end{section}




\begin{section}{Strategy Quantifications}
	\label{sec:strqnt}

	Since model checking for \SL\ is non-elementary hard while the same problem
	for \ATLS\ is only 2\ExpTimeC, a question that naturally arises is whether
	there are proper fragments of \SL\ of practical interest, still strictly
	subsuming \ATLS, that reside in such a complexity gap.
	In this section, we answer positively to this question and go even further.
	Precisely, we enlighten a fundamental property that, if satisfied, allows to
	retain a 2\ExpTimeC\ model-checking problem.
	We refer to such a property as \emph{elementariness}.
	To formally introduce this concept, we use the notion of \emph{dependence map}
	as a machinery.

	The remaining part of this section is organized as follows.
	In Subsection~\ref{subsec:synfrg}, we describe three syntactic fragments of
	\SL, named \NGSL, \BGSL, and \OGSL, having the peculiarity to use strategy
	quantifications grouped in atomic blocks.
	Then, in Subsection~\ref{subsec:qntspc}, we define the notion of dependence
	map, which is used, in Subsection~\ref{subsec:elmqnt}, to introduce the
	concept of elementariness.
	Finally, in Subsection~\ref{subsec:elmprp}, we prove a fundamental result,
	which is at the base of our elementary model-checking procedure for \OGSL.

	\begin{subsection}{Syntactic fragments}
		\label{subsec:synfrg}

		In order to formalize the syntactic fragments of \SL\ we want to
		investigate, we first need to define the concepts of \emph{quantification}
		and \emph{binding prefixes}.
		\begin{definition}[Prefixes]
			\label{def:prf}
			A \emph{quantification prefix} over a set $\VSet \subseteq \VarSet$ of
			variables is a finite word $\qpElm \in \set{ \EExs{\xElm}, \AAll{\xElm} }{
			\xElm \in \VSet }^{\card{\VSet}}$ of length $\card{\VSet}$ such that each
			variable $\xElm \in \VSet$ occurs just once in $\qpElm$, i.e., there is
			exactly one index $i \in \numco{0}{\card{\VSet}}$ such that $(\qpElm)_{i}
			\in \{ \EExs{\xElm}, \AAll{\xElm} \}$.
			A \emph{binding prefix} over a set of variables $\VSet \subseteq \VarSet$
			is a finite word $\bpElm \in \set{ (\aElm, \xElm) }{ \aElm \in \AgSet
			\land \xElm \in \VSet }^{\card{\AgSet}}$ of length $\card{\AgSet}$ such
			that each agent $\aElm \in \AgSet$ occurs just once in $\bpElm$, i.e.,
			there is exactly one index $i \in \numco{0}{\card{\AgSet}}$ for which
			$(\bpElm)_{i} \in \set{ (\aElm, \xElm) }{ \xElm \in \VSet }$.
			Finally, $\QPSet(\VSet) \subseteq \set{ \EExs{\xElm}, \AAll{\xElm} }{
			\xElm \in \VSet }^{\card{\VSet}}$ and $\BPSet(\VSet) \subseteq \set{
			(\aElm, \xElm) }{ \aElm \in \AgSet \land \xElm \in \VSet
			}^{\card{\AgSet}}$ denote, respectively, the sets of all quantification
			and binding prefixes over variables in $\VSet$.
		\end{definition}

		We now have all tools to define the syntactic fragments we want to analyze,
		which we name, respectively, \emph{Nested-Goal}, \emph{Boolean-Goal}, and
		\emph{One-Goal Strategy Logic} (\NGSL, \BGSL, and \OGSL, for short).
		For \emph{goal} we mean an \SL\ agent-closed formula of the kind $\bpElm
		\varphi$, with $\AgSet \subseteq \free{\varphi}$, being $\bpElm \in
		\BndSet(\VarSet)$ a binding prefix.
		The idea behind \NGSL\ is that, when there is a quantification over a
		variable used in a goal, we are forced to quantify over all free variables
		of the inner subformula containing the goal itself, by using a
		quantification prefix.
		In this way, the subformula is build only by nesting and Boolean
		combinations of goals.
		In addition, with \BGSL\ we avoid nested goals sharing the variables of a
		same quantification prefix, but allow their Boolean combinations.
		Finally, \OGSL\ forces the use of a different quantification prefix for each
		single goal in the formula.
		The formal syntax of \NGSL, \BGSL, and \OGSL\ follows.
		\begin{definition}[\NGSL, \BGSL, and \OGSL\ Syntax]
			\label{def:xgsl(syntax)}
			\NGSL\ formulas are built inductively from the sets of atomic propositions
			$\APSet$, quantification prefixes $\QPSet(\VSet)$ for any $\VSet \subseteq
			\VarSet$, and binding prefixes $\BPSet(\VarSet)$, by using the following
			grammar, with $\pElm \in \APSet$, $\qpElm \in \cup_{\VSet \subseteq
			\VarSet} \QPSet(\VSet)$, and $\bpElm \in \BPSet(\VarSet)$:
			\begin{center}
				$\varphi ::= \pElm \mid \neg \varphi \mid \varphi \wedge \varphi \mid
				\varphi \vee \varphi \mid \X \varphi \mid \varphi \:\U \varphi \mid
				\varphi \:\R \varphi \mid \qpElm \varphi \mid \bpElm \varphi$,
			\end{center}
			where in the formation rule $\qpElm \varphi$ it is ensured that $\varphi$
			is agent-closed and $\qpElm \in \QPSet(\free{\varphi})$.\\
			In addition, \BGSL\ formulas are determined by splitting the above
			syntactic class in two different parts, of which the second is dedicated
			to build the Boolean combinations of goals avoiding their nesting:
			\begin{center}
				$\varphi ::= \pElm \mid \neg \varphi \mid \varphi \wedge \varphi \mid
				\varphi \vee \varphi \mid \X \varphi \mid \varphi \:\U \varphi \mid
				\varphi \:\R \varphi \mid \qpElm \psi$,\\
				$\psi ::= \bpElm \varphi \mid \neg \psi \mid \psi \wedge \psi \mid \psi
				\vee \psi$,
			\end{center}
			where in the formation rule $\qpElm \psi$ it is ensured that $\qpElm \in
			\QPSet(\free{\psi})$.\\
			Finally, the simpler \OGSL\ formulas are obtained by forcing each goal to
			be coupled with a quantification prefix:
			\begin{center}
				$\varphi ::= \pElm \mid \neg \varphi \mid \varphi \wedge \varphi \mid
				\varphi \vee \varphi \mid \X \varphi \mid \varphi \:\U \varphi \mid
				\varphi \:\R \varphi \mid \qpElm \bpElm \varphi$,
			\end{center}
			where in the formation rule $\qpElm \bpElm \varphi$ it is ensured that
			$\qpElm \in \QPSet(\free{\bpElm \varphi})$.\\
			$\SL \supset \NGSL \supset \BGSL \supset \OGSL$ denotes the syntactic
			chain of infinite sets of formulas generated by the respective grammars
			with the associated constraints on free variables of goals.
		\end{definition}
		Intuitively, in \NGSL, \BGSL, and \OGSL, we force the writing of formulas to
		use atomic blocks of quantifications and bindings, where the related free
		variables are strictly coupled with those that are effectively quantified in
		the prefix just before the binding.
		In a nutshell, we can only write formulas by using sentences of the form
		$\qpElm \psi$ belonging to a kind of \emph{prenex normal form} in which the
		quantifications contained into the \emph{matrix} $\psi$ only belong to the
		prefixes $\qpElm'$ of some inner subsentence $\qpElm' \psi' \in \snt{\qpElm
		\psi}$.

		An \NGSL\ sentence $\phi$ is \emph{principal} if it is of the form $\phi =
		\qpElm \psi$, where $\psi$ is agent-closed and $\qpElm \in
		\QPSet(\free{\psi})$.
		By $\psnt{\varphi} \subseteq \snt{\varphi}$ we denote the set of all
		principal subsentences of the formula $\varphi$.

		We now introduce other two general restrictions in which the numbers
		$\card{\AgSet}$ of agents and $\card{\VarSet}$ of variables that are used
		to write a formula are fixed to the a priori values $n, m \in
		\numco{1}{\omega}$, respectively.
		Moreover, we can also forbid the sharing of variables, i.e., each variable
		is binded to one agent only, so, we cannot force two agents to use the same
		strategy.
		We name these three fragments \SL[$n$-ag], \SL[$m$-var], and \SL[fvs],
		respectively.
		Note that, in the one agent fragment, the restriction on the sharing of
		variables between agents, naturally, does not act, i.e., $\SL[$1$-ag, fvs] =
		\SL[$1$-ag]$.

		To start to practice with the above fragments, consider again the sentence
		$\varphi$ of Example~\vref{exm:sv}.
		It is easy to see that it actually belongs to \BGSL[$2$-ag, $3$-var,
		$2$-alt], and so, to \NGSL, but not to \OGSL, since it is of the form
		$\qpSym (\bpSym[1] \X \pSym \wedge \bpSym[2] \X \qSym)$, where the
		quantification prefix is $\qpSym = \EExs{\xSym} \AAll{\ySym} \EExs{\zSym}$
		and the binding prefixes of the two goals are $\bpSym[1] = (\alpha, \xSym)
		(\beta, \ySym)$ and $\bpSym[2] = (\alpha, \ySym) (\beta, \zSym)$.

		Along the paper, sometimes we assert that a given formula $\varphi$ belongs
		to an \SL\ syntactic fragment also if its syntax does not precisely
		correspond to what is described by the relative grammar.
		We do this in order to make easier the reading and interpretation of the
		formula $\varphi$ itself and only in the case that it is simple to translate
		it into an equivalent formula that effectively belongs to the intended
		logic, by means of a simple generalization of classic rules used to put a
		formula of first order logic in the prenex normal form.
		For example, consider the sentence $\varphi_{\!N\!\!E}$ of
		Example~\vref{exm:ne} representing the existence of a Nash equilibrium.
		This formula is considered to belong to \BGSL[$n$-ag, $2n$-var, fvs,
		$1$-alt], since it can be easily translated in the form $\phi_{\!N\!\!E} =
		\qpSym \bigwedge_{i = 1}^{n} \bpSym[i] \psi_{i} \rightarrow \bpSym
		\psi_{i}$, where $\qpSym = \EExs{\xSym[1]} \cdots \EExs{\xSym[n]}
		\AAll{\ySym[1]} \cdots \AAll{\ySym[n]}$, $\bpSym = (\alpha_{1}, \xSym[1])
		\cdots (\alpha_{n}, \xSym[n])$, $\bpSym[i] = (\alpha_{1}, \xSym[1]) \cdots
		(\alpha_{i - 1}, \xSym[i - 1]) (\alpha_{i}, \ySym[i]) (\alpha_{i + 1},
		\xSym[i + 1]) \cdots (\alpha_{n}, \xSym[n])$, and $\free{\psi_{i}} =
		\AgSet$.
		As another example, consider the sentence $\varphi_{\!S\!P}$ of
		Example~\vref{exm:sp} representing the existence of a stability profile.
		Also this formula is considered to belong to \BGSL[$n$-ag, $2n$-var, fvs,
		$1$-alt], since it is equivalent to $\phi_{\!S\!P} = \qpSym \bigwedge_{i, j
		= 1, i \neq j}^{n} \bpSym \psi_{j} \rightarrow ((\bpSym \psi_{i}
		\leftrightarrow \bpSym[i] \psi_{i}) \rightarrow \bpSym[i] \psi_{j})$.
		Note that both $\phi_{\!N\!\!E}$ and $\phi_{\!S\!P}$ are principal
		sentences.

		Now, it is interesting to observe that \CTLS\ and \ATLS\ are exactly
		equivalent to \OGSL[fvs, $0$-alt] and \OGSL[fvs, $1$-alt], respectively.
		Moreover, \GL~\cite{AHK02} is the very simple fragment of \BGSL[fvs,
		$1$-alt] that forces all goals in a formula to have a common part containing
		all variables quantified before the unique possible alternation of the
		quantification prefix.
		Finally, we have that \CHPSL\ is the \TBBGSL[$2$-ag, fvs] fragment.

		\begin{remark}[\TBNGSL\ Model-Checking Hardness]
			\label{rmk:tbngsl(modchkhrd)}
			It is well-known that the non-elementary hardness result for the
			satisfiability problem of \QPTL~\cite{SVW87} already holds for formulas in
			prenex normal form.
			Now, it is not hard to see that the transformation described in
			Lemma~\ref{lmm:qptl(rdc)} of \QPTL\ reduction puts \QPTL[$k$-alt]
			sentences $\varphi$ in prenex normal form into \TBNGSL[$1$-ag, $k$-alt]
			variable-closed formulas $\trn{\varphi} = \qpElm \psi$.
			Moreover, the derived \TBSL[$1$-ag, $k$-alt] sentence $\EExs{\xSym}
			(\alpha, \xSym) \qpElm \psi$ used in Theorem~\ref{thm:tbsl(modchkhrd)} of
			\TBSL\ model-checking hardness is equivalent to the \TBNGSL[$1$-ag,
			$k$-alt] principal sentence $\EExs{\xSym} \qpElm (\alpha, \xSym) \psi$,
			since $\xSym$ is not used in the quantification prefix $\qpElm$.
			Thus, the hardness result for the model-checking problem holds for
			\TBNGSL[$1$-ag, $k$-alt] and, consequently, for \NGSL[$1$-ag, $k$-alt] as
			well.
			However, it is important to observe that, unfortunately, it is not know
			if such an hardness result holds for \TBBGSL\ or \BGSL\ and, in
			particular, for \CHPSL.
			We leave this problem open here.
		\end{remark}

		\figthmogslvsatlsexp
		At this point, we prove that \ATLS\ is strictly less expressive than \OGSL\
		and, consequently, than \BGSL\ and \NGSL.
		To do this, we show the existence of two structures that result to be
		equivalent only w.r.t.\ sentences having alternation number bounded by $1$.
		It can be interesting to note that, we use an ad-hoc technique based on a
		brute-force check to verify that all \ATLS\ formulas cannot distinguish
		between the two structures.
		A possible future line of research is to study variants of the
		Ehrenfeucht-Fra\"iss\'e game~\cite{EF95,Hod93} for \SL, which allow to
		determine whether two structures are or not equivalent w.r.t.\ a particular
		\SL\ fragment.
		\begin{theorem}[\OGSL\ vs \ATLS\ Expressiveness]
			\label{thm:ogslvsatls(exp)}
			There exists an \OGSL[$3$-ag, fvs, $2$-alt] sentence having no \ATLS\
			equivalent.
		\end{theorem}
		\begin{proof}
			Consider the two \CGS s $\GName[1] \defeq \CGSTuple {\{ \pSym \}} {\{
			\alpha, \beta, \gamma \}} {\{ 0, 1 \}} {\{ \sSym[0], \sSym[1], \sSym[2]
			\}} {\labFun} {\trnFun[1]} {\sSym[0]}$ and $\GName[2] \defeq \CGSTuple {\{
			\pSym \}} {\{ \alpha, \beta, \gamma \}} {\{ 0, 1, 2 \}} {\{ \sSym[0],
			\sSym[1], \sSym[2] \}} {\labFun} {\trnFun[2]} {\sSym[0]}$ depicted in
			Figure~\vref{fig:thm:ogslvsatls(exp)}, where $\labFun(\sSym[0]) =
			\labFun(\sSym[2]) \defeq \emptyset$, $\labFun(\sSym[1]) \defeq \{ \pSym
			\}$, $\DSet[1] \defeq \{ 00*, 11* \}$, and $\DSet[2] \defeq \{ 00*, 11*,
			12*, 200, 202, 211 \}$.
			Moreover, consider the \OGSL[$3$-ag, fvs, $2$-alt] sentence $\varphi^{*}
			\defeq \qpSym^{*} \bpSym^{*} \X \pSym$, where $\qpSym^{*} \defeq
			\AAll{\xSym} \EExs{\ySym} \AAll{\zElm}$ and $\bpSym^{*} \defeq (\alpha,
			\xSym) (\beta, \ySym) (\gamma, \zSym)$.
			Then, it is easy to see that $\GName[1] \models \varphi^{*}$ but
			$\GName[2] \not\models \varphi^{*}$.
			Indeed, $\GName[1], \asgFun[1], \sSym[0] \models \bpSym^{*} \X \pSym$, for
			all $\asgFun[1] \in \AsgSet[ {\GName[1]} ](\{ \xSym, \ySym, \zSym \},
			\sSym[0])$ such that $\asgFun[1](\ySym)(\sSym[0]) =
			\asgFun[1](\xSym)(\sSym[0])$, and $\GName[2], \asgFun[2], \sSym[0] \models
			\bpSym^{*} \X \neg \pSym$, for all $\asgFun[2] \in \AsgSet[ {\GName[2]}
			](\{ \xSym, \ySym, \zSym \}, \sSym[0])$ such that
			$\asgFun[2](\xSym)(\sSym[0]) = 2$ and $\asgFun[2](\zSym)(\sSym[0]) =
			(\asgFun[2](\ySym)(\sSym[0]) + 1) \bmod 3$.

			Now, due to the particular structure of the \CGS s $\GName[i]$ under exam,
			with $i \in \{ 1, 2 \}$, for each path $\pthElm \in \PthSet[ {\GName[i]}
			](\sSym[0])$, we have that either $\labFun((\pthElm)_{j}) = \{ \pSym \}$
			or $\labFun((\pthElm)_{j}) = \emptyset$, for all $j \in
			\numco{1}{\omega}$, i.e., apart from the initial state, the path is
			completely labeled either with $\{ \pSym \}$ or with $\emptyset$.
			Thus, it is easy to see that, for each \ATLS\ formula $\qpElm \bpElm
			\psi$, there is a literal $\lElm[\psi] \in \{ \pSym, \neg \pSym \}$ such
			that $\GName[i] \models \qpElm \bpElm \psi$ iff $\GName[i] \models \qpElm
			\bpElm \X \!\lElm[\psi]$, for all $i \in \{ 1, 2 \}$.
			W.l.o.g., we can suppose that $\bpElm = \bpSym^{*}$, since we are always
			able to uniformly rename the variables of the quantification and binding
			prefixes without changing the meaning of the sentence.

			At this point, it is easy to see that there exists an index $k \in \{ 1,
			2, 3 \}$ for which it holds that either $\qpSym[k] \bpSym^{*} \X
			\!\lElm[\psi] \implies \qpElm \bpSym^{*} \X \!\lElm[\psi]$ or $\qpElm
			\bpSym^{*} \X \!\lElm[\psi] \implies \dual{\qpElm[k]} \bpSym^{*} \X
			\!\lElm[\psi]$, where $\qpSym[1] \defeq \AAll{\xSym} \AAll{\zElm}
			\EExs{\ySym}$, $\qpSym[2] \defeq \EExs{\xSym} \EExs{\yElm} \AAll{\zSym}$,
			and $\qpSym[3] \defeq \AAll{\ySym} \AAll{\zElm} \EExs{\xSym}$.
			Thus, to prove that every \ATLS\ formula cannot distinguish between
			$\GName[1]$ and $\GName[2]$, we can simply show that the sentences
			$\qpSym[k] \bpSym^{*} \X \!\lElm$, with $k \in \{ 1, 2, 3 \}$ and $\lElm
			\in \{ \pSym, \neg \pSym \}$, do the same.
			In fact, it holds that $\GName[i] \models \qpSym[k] \bpSym^{*} \X
			\!\lElm$, for all $i \in \{ 1, 2 \}$, $k \in \{ 1, 2, 3 \}$, and $\lElm
			\in \{ \pSym, \neg \pSym \}$.
			Hence, the thesis holds.
			The check of the latter fact is trivial and left to the reader as an
			exercise.
		\end{proof}

	\end{subsection}

	\begin{subsection}{Dependence Maps}
		\label{subsec:qntspc}

		We now introduce the concept of dependence map of a quantification and show
		how any quantification prefix contained into an \SL\ formula can be
		represented by an adequate choice of a dependence map over strategies.
		The main idea here is inspired by what Skolem proposed for the first order
		logic in order to eliminate each existential quantification over variables,
		by substituting them with second order existential quantifications over
		functions, whose choice is uniform w.r.t.\ the universal variables.

		First, we introduce some notation regarding quantification prefixes.
		Let $\qpElm \in \QPSet(\VSet)$ be a quantification prefix over a set
		$\QPVSet(\qpElm) \defeq \VSet \subseteq \VarSet$ of variables.
		By $\QPEVSet{\qpElm} \defeq \set{ \xElm \in \QPVSet(\qpElm) }{ \exists i \in
		\numco{0}{\card{\qpElm}} .\: (\qpElm)_{i} = \EExs{\xElm} }$ and
		$\QPAVSet{\qpElm} \defeq \QPVSet(\qpElm) \setminus \QPEVSet{\qpElm}$ we
		denote, respectively, the sets of \emph{existential} and \emph{universal
		variables} quantified in $\qpElm$.
		For two variables $\xElm, \yElm \in \QPVSet(\qpElm)$, we say that $\xElm$
		\emph{precedes} $\yElm$ in $\qpElm$, in symbols $\xElm \qpordRel[\qpElm]
		\yElm$, if $\xElm$ occurs before $\yElm$ in $\qpElm$, i.e., there are two
		indexes $i, j \in \numco{0}{\card{\qpElm}}\!$, with $i < j$, such that
		$(\qpElm)_{i} \in \{ \EExs{\xElm}, \AAll{\xElm} \}$ and $(\qpElm)_{j} \in \{
		\EExs{\yElm}, \AAll{\yElm} \}$.
		Moreover, we say that $\yElm$ is \emph{functional dependent} on $\xElm$, in
		symbols $\xElm \qpdepRel[\qpElm] \yElm$, if $\xElm \in \QPAVSet{\qpElm}$,
		$\yElm \in \QPEVSet{\qpElm}$, and $\xElm \qpordRel_{\qpElm} \yElm$, i.e.,
		$\yElm$ is existentially quantified after that $\xElm$ is universally
		quantified, so, there may be a dependence between a value chosen by $\xElm$
		and that chosen by $\yElm$.
		This definition induces the set $\QPDepSet(\qpElm) \defeq \set{ (\xElm,
		\yElm) \in \QPVSet(\qpElm) \times \QPVSet(\qpElm) }{ \xElm \qpdepRel[\qpElm]
		\yElm }$ of \emph{dependence pairs} and its derived version
		$\QPDepSet(\qpElm, \yElm) \defeq \set{ \xElm \in \QPVSet(\qpElm) }{ \xElm
		\qpdepRel[\qpElm] \yElm }$ containing all variables from which $\yElm$
		depends.
		Finally, we use $\dual{\qpElm} \in \QPSet(\QPVSet(\qpElm))$ to indicate the
		quantification derived from $\qpElm$ by \emph{dualizing} each quantifier
		contained in it, i.e., for all indexes $i \in \numco{0}{\card{\qpElm}}\!$,
		it holds that $(\dual{\qpElm})_{i} = \EExs{\xElm}$ iff $(\qpElm)_{i} =
		\AAll{\xElm}$, with $\xElm \in \QPVSet(\qpElm)$.
		It is evident that $\QPEVSet{\dual{\qpElm}} = \QPAVSet{\qpElm}$ and
		$\QPAVSet{\dual{\qpElm}} = \QPEVSet{\qpElm}$.
		As an example, let $\qpSym = \AAll{\xSym} \EExs{\ySym} \EExs{\zSym}
		\AAll{\wSym} \EExs{\vSym}$.
		Then, we have $\QPEVSet{\qpSym} = \{ \ySym, \zSym, \vSym \}$,
		$\QPAVSet{\qpSym} = \{ \xSym, \wSym \}$, $\QPDepSet(\qpSym, \xSym) =
		\QPDepSet(\qpSym, \wSym) = \emptyset$, \mbox{$\QPDepSet(\qpSym, \ySym) =
		\QPDepSet(\qpSym, \zSym) = \{ \xSym \}$, $\QPDepSet(\qpSym, \vSym) = \{
		\xSym, \wSym \}$, and $\dual{\qpSym} = \EExs{\xSym} \AAll{\ySym}
		\AAll{\zSym} \EExs{\wSym} \AAll{\vSym}$.}

		Finally, we define the notion of \emph{valuation} of variables over a
		generic set $\DSet$, called \emph{domain}, i.e., a partial function $\valFun
		: \VarSet \pto \DSet$ mapping every variable in its domain to an element in
		$\DSet$.
		By $\ValSet[\DSet](\VSet) \defeq \VSet \to \DSet$ we denote the set of all
		valuation functions over $\DSet$ defined on $\VSet \subseteq \VarSet$.

		At this point, we give a general high-level semantics for the quantification
		prefixes by means of the following main definition of \emph{dependence map}.
		\begin{definition}[Dependence Maps]
			\label{def:qntspc}
			Let $\qpElm \in \QPSet(\VSet)$ be a quantification prefix over a set
			$\VSet \subseteq \VarSet$ of variables, and $\DSet$ a set.
			Then, a \emph{dependence map} for $\qpElm$ over $\DSet$ is a function
			$\spcFun : \ValSet[\DSet](\QPAVSet{\qpElm}) \to \ValSet[\DSet](\VSet)$
			satisfying the following properties:
			\begin{enumerate}
				\item
					\label{def:qntspc(aqnt)}
					$\spcFun(\valFun)_{\rst \QPAVSet{\qpElm}} \!=\! \valFun$, for all
					$\valFun \in \ValSet[\DSet](\QPAVSet{\qpElm})$;~\footnote{By
					$\gFun_{\rst \ZSet} : (\XSet \cap \ZSet) \to \YSet$ we denote the
					\emph{restriction} of a function $\gFun : \XSet \to \YSet$ to the
					elements in the set $\ZSet$.}
				\item
					\label{def:qntspc(eqnt)}
					$\spcFun(\valFun[1])(\xElm) \!=\! \spcFun(\valFun[2])(\xElm)$, for all
					$\valFun[1], \valFun[2] \!\in\! \ValSet[\DSet](\QPAVSet{\qpElm})$ and
					$\xElm \!\in\! \QPEVSet{\qpElm}$ such that $\valFun[1]_{\rst
					\QPDepSet(\qpElm, \xElm)} \!=\! \valFun[2]_{\rst \QPDepSet(\qpElm,
					\xElm)}$.
			\end{enumerate}
			$\SpcSet[\DSet](\qpElm)$ denotes the set of all dependence maps for
			$\qpElm$ over $\DSet$.
		\end{definition}
		Intuitively, Item~\ref{def:qntspc(aqnt)} asserts that $\spcFun$ takes the
		same values of its argument w.r.t.\ the universal variables in $\qpElm$ and
		Item~\ref{def:qntspc(eqnt)} ensures that the value of $\spcFun$ w.r.t.\ an
		existential variable $\xElm$ in $\qpElm$ does not depend on variables not in
		$\QPDepSet(\qpElm, \xElm)$.
		To get a better insight into this definition, a dependence map $\spcFun$ for
		$\qpElm$ can be considered as a set of \emph{Skolem functions} that, given a
		value for each variable in $\QPVSet(\qpElm)$ that is universally quantified
		in $\qpElm$, returns a possible value for all the existential variables in
		$\qpElm$, in a way that is consistent w.r.t.\ the order of quantifications.
		Observe that, each $\spcFun \in \SpcSet[\DSet](\qpElm)$ is injective, so,
		$\card{\rng{\spcFun}} = \card{\dom{\spcFun}} =
		\card{\DSet}^{\card{\QPAVSet{\qpElm}}}$.
		Moreover, $\card{\SpcSet[\DSet](\qpElm)} = \prod_{\xElm \in
		\QPEVSet{\qpElm}} \card{\DSet}^{\card{\DSet}^{\card{\QPDepSet(\qpElm,
		\xElm)}}}$.
		As an example, let $\DSet = \{ 0, 1 \}$ and $\qpSym = \AAll{\xSym}
		\EExs{\ySym} \AAll{\zSym} \in \QPSet(\VSet)$ be a quantification prefix over
		$\VSet = \{ \xSym, \ySym, \zSym \}$.
		Then, we have that $\card{\SpcSet[\DSet](\qpSym)} = 4$ and
		$\card{\SpcSet[\DSet](\dual{\qpSym})} = 8$.
		Moreover, the dependence maps $\spcFun[i] \in \SpcSet[\DSet](\qpSym)$ with
		$i \in \numcc{0}{3}$ and $\dual[i]{\spcFun} \in
		\SpcSet[\DSet](\dual{\qpSym})$ with $i \in \numcc{0}{7}$, for a particular
		fixed order, are such that $\spcFun[0](\valFun)(\ySym) = 0$,
		$\spcFun[1](\valFun)(\ySym) = \valFun(\xSym)$, $\spcFun[2](\valFun)(\ySym) =
		1 - \valFun(\xSym)$, and $\spcFun[3](\valFun)(\ySym) = 1$, for all $\valFun
		\in \ValSet[\DSet](\QPAVSet{\qpSym})$, and
		$\dual[i]{\spcFun}(\dual{\valFun})(\xSym) = 0$ with $i \in \numcc{0}{3}$,
		$\dual[i]{\spcFun}(\dual{\valFun})(\xSym) = 1$ with $i \in \numcc{4}{7}$,
		$\dual[0]{\spcFun}(\dual{\valFun})(\zSym) =
		\dual[4]{\spcFun}(\dual{\valFun})(\zSym) = 0$,
		$\dual[1]{\spcFun}(\dual{\valFun})(\zSym) =
		\dual[5]{\spcFun}(\dual{\valFun})(\zSym) = \dual{\valFun}(\ySym)$,
		$\dual[2]{\spcFun}(\dual{\valFun})(\zSym) =
		\dual[6]{\spcFun}(\dual{\valFun})(\zSym) = 1 - \dual{\valFun}(\ySym)$, and
		$\dual[3]{\spcFun}(\dual{\valFun})(\zSym) =
		\dual[7]{\spcFun}(\dual{\valFun})(\zSym) = 1$, for all $\dual{\valFun} \in
		\ValSet[\DSet](\QPAVSet{\dual{\qpSym}})$.

		We now prove the following fundamental theorem that describes how to
		eliminate the strategy quantifications of an \SL\ formula via a choice of a
		suitable dependence map over strategies.
		This procedure can be seen as the equivalent of \emph{Skolemization} in
		first order logic (see~\cite{Hod93}, for more details).
		\begin{theorem}[\SL\ Strategy Quantification]
			\label{thm:sl(strqnt)}
			Let $\GName$ be a \CGS\ and $\varphi = \qpElm \psi$ an \SL\ formula, being
			$\qpElm \in \QPSet(\VSet)$ a quantification prefix over a set $\VSet
			\subseteq \free{\psi} \cap \VarSet$ of variables.
			Then, for all assignments $\asgFun \in \AsgSet(\free{\varphi}, \sElm[0])$,
			the following holds: $\GName, \asgFun, \sElm[0] \models \varphi$ iff there
			exists a dependence map $\spcFun \in \SpcSet[ {\StrSet(\sElm[0])}
			](\qpElm)$ such that $\GName, \asgFun \umrg \spcFun(\asgFun'), \sElm[0]
			\models \psi$, for all $\asgFun' \in \AsgSet(\QPAVSet{\qpElm},
			\sElm[0])$.~\footnote{By $\gFun[1] \umrg \gFun[2] : (\XSet[1] \cup
			\XSet[2]) \to (\YSet[1] \cup \YSet[2])$ we denote the operation of
			\emph{union} of two functions $\gFun[1] : \XSet[1] \to \YSet[1]$ and
			$\gFun[2] : \XSet[2] \to \YSet[2]$ defined on disjoint domains, i.e.,
			$\XSet[1] \cap \XSet[2] = \emptyset$.}
		\end{theorem}
		\begin{proof}
			The proof proceeds by induction on the length of the quantification prefix
			$\qpElm$.
			For the base case $\card{\qpElm} = 0$, the thesis immediately follows,
			since $\QPAVSet{\qpElm} = \emptyset$ and, consequently, both
			$\SpcSet[\StrSet( {\sElm[0]} )](\qpElm)$ and $\AsgSet(\QPAVSet{\qpElm},
			\sElm[0])$ contain the empty function only (we are assuming, by
			convention, that $\emptyfun(\emptyfun) \defeq \emptyfun$).

			We now prove, separately, the two directions of the inductive case.

			\emph{[Only if].}
			Suppose that $\GName, \asgFun, \sElm[0] \models \varphi$, where $\qpElm =
			\Qnt \cdot \qpElm'$.
			Then, two possible cases arise: either $\Qnt = \EExs{\xElm}$ or $\Qnt =
			\AAll{\xElm}$.
			\begin{itemize}
				\item
					$\Qnt = \EExs{\xElm}$.

					By Item~\ref{def:sl(semantics:eqnt)} of
					Definition~\ref{def:sl(semantics)} of \SL\ semantics, there is a
					strategy $\strFun \in \StrSet(\sElm[0])$ such that $\GName,
					\asgFun[][\xElm \mapsto \strFun], \sElm[0] \models \qpElm' \psi$.
					Note that $\QPAVSet{\qpElm} = \QPAVSet{\qpElm'}$.
					By the inductive hypothesis, we have that there exists a dependence
					map $\spcFun \in \SpcSet[\StrSet( {\sElm[0]} )](\qpElm')$ such that
					$\GName, \asgFun[][\xElm \mapsto \strFun] \umrg \spcFun(\asgFun'),
					\sElm[0] \models \psi$, for all $\asgFun' \in
					\AsgSet(\QPAVSet{\qpElm'}, \sElm[0])$.
					Now, consider the function $\spcFun' : \AsgSet(\QPAVSet{\qpElm},
					\sElm[0]) \to \AsgSet(\VSet, \sElm[0])$ defined by $\spcFun'(\asgFun')
					\defeq \spcFun(\asgFun')[\xElm \mapsto \strFun]$, for all $\asgFun'
					\in \AsgSet(\QPAVSet{\qpElm}, \sElm[0])$.
					It is easy to check that $\spcFun'$ is a dependence map for $\qpElm$
					over $\StrSet(\sElm[0])$, i.e., $\spcFun' \in \SpcSet[\StrSet(
					{\sElm[0]} )](\qpElm)$.
					Moreover, $\asgFun[][\xElm \mapsto \strFun] \umrg \spcFun(\asgFun') =
					\asgFun \umrg \spcFun(\asgFun')[\xElm \mapsto \strFun] = \asgFun \umrg
					\spcFun'(\asgFun')$, for $\asgFun' \in \AsgSet(\QPAVSet{\qpElm},
					\sElm[0])$.
					Hence, $\GName, \asgFun \umrg \spcFun'(\asgFun'), \sElm[0] \models
					\psi$, for all $\asgFun' \in \AsgSet(\QPAVSet{\qpElm}, \sElm[0])$.
				\item
					$\Qnt = \AAll{\xElm}$.

					By Item~\ref{def:sl(semantics:aqnt)} of
					Definition~\ref{def:sl(semantics)}, we have that, for all strategies
					$\strFun \in \StrSet(\sElm[0])$, it holds that $\GName,
					\asgFun[][\xElm \mapsto \strFun], \sElm[0] \models \qpElm' \psi$.
					Note that $\QPAVSet{\qpElm} = \QPAVSet{\qpElm'} \cup \{ \xElm \}$.
					By the inductive hypothesis, we derive that, for each $\strFun \in
					\StrSet(\sElm[0])$, there exists a dependence map $\spcFun[\strFun]
					\in \SpcSet[\StrSet( {\sElm[0]} )](\qpElm')$ such that $\GName,
					\asgFun[][\xElm \mapsto \strFun] \umrg \spcFun[\strFun](\asgFun'),
					\sElm[0] \models \psi$, for all $\asgFun' \in
					\AsgSet(\QPAVSet{\qpElm'}, \sElm[0])$.
					Now, consider the function $\spcFun' : \AsgSet(\QPAVSet{\qpElm},
					\sElm[0]) \to \AsgSet(\VSet, \sElm[0])$ defined by $\spcFun'(\asgFun')
					\defeq \spcFun[\asgFun'(\xElm)](\asgFun[|']_{\rst
					\QPAVSet{\qpElm'}})[\xElm \mapsto \asgFun'(\xElm)]$, for all $\asgFun'
					\in \AsgSet(\QPAVSet{\qpElm}, \sElm[0])$.
					It is evident that $\spcFun'$ is a dependence map for $\qpElm$ over
					$\StrSet(\sElm[0])$, i.e., $\spcFun' \in \SpcSet[\StrSet( {\sElm[0]}
					)](\qpElm)$.
					Moreover, $\asgFun[][\xElm \mapsto \strFun] \umrg
					\spcFun[\strFun](\asgFun') = \asgFun \umrg
					\spcFun[\strFun](\asgFun')[\xElm \mapsto \strFun] = \asgFun \umrg
					\spcFun'(\asgFun'[\xElm \mapsto \strFun])$, for $\strFun \in
					\StrSet(\sElm[0])$ and $\asgFun' \in \AsgSet(\QPAVSet{\qpElm'},
					\sElm[0])$.
					Hence, $\GName, \asgFun \umrg \spcFun'(\asgFun'), \sElm[0] \models
					\psi$, for all $\asgFun' \in \AsgSet(\QPAVSet{\qpElm}, \sElm[0])$.
			\end{itemize}

			\emph{[If].}
			Suppose that there exists a dependence map $\spcFun \in \SpcSet[\StrSet(
			{\sElm[0]} )](\qpElm)$ such that $\GName, \asgFun \umrg \spcFun(\asgFun'),
			\sElm[0] \models \psi$, for all $\asgFun' \in \AsgSet(\QPAVSet{\qpElm},
			\sElm[0])$, where $\qpElm = \Qnt \cdot \qpElm'$.
			Then, two possible cases arise: either $\Qnt = \EExs{\xElm}$ or $\Qnt =
			\AAll{\xElm}$.
			\begin{itemize}
				\item
					$\Qnt = \EExs{\xElm}$.

					There is a strategy $\strFun \in \StrSet(\sElm[0])$ such that $\strFun
					= \spcFun(\asgFun')(\xElm)$, for all $\asgFun' \in
					\AsgSet(\QPAVSet{\qpElm}, \sElm[0])$.
					Note that $\QPAVSet{\qpElm} = \QPAVSet{\qpElm'}$.
					Consider the function $\spcFun' : \AsgSet(\QPAVSet{\qpElm'}, \sElm[0])
					\to \AsgSet(\VSet \setminus \{ \xElm \}, \sElm[0])$ defined by
					$\spcFun'(\asgFun') \defeq \spcFun(\asgFun')_{\rst (\VSet \setminus \{
					\xElm \})}$, for all $\asgFun' \in \AsgSet(\QPAVSet{\qpElm'},
					\sElm[0])$.
					It is easy to check that $\spcFun'$ is a dependence map for $\qpElm'$
					over $\StrSet(\sElm[0])$, i.e., $\spcFun' \in \SpcSet[\StrSet(
					{\sElm[0]} )](\qpElm')$.
					Moreover, $\asgFun \umrg \spcFun(\asgFun') = \asgFun \umrg
					\spcFun'(\asgFun')[\xElm \mapsto \strFun] = \asgFun[][\xElm \mapsto
					\strFun] \umrg \spcFun'(\asgFun')$, for $\asgFun' \in
					\AsgSet(\QPAVSet{\qpElm'}, \sElm[0])$.
					Then, it is evident that $\GName, \asgFun[][\xElm \to \strFun] \umrg
					\spcFun'(\asgFun'), \sElm[0] \models \psi$, for all $\asgFun' \in
					\AsgSet(\QPAVSet{\qpElm'}, \sElm[0])$.
					By the inductive hypothesis, we derive that $\GName, \asgFun[][\xElm
					\mapsto \strFun], \sElm[0] \models \qpElm' \psi$, which means that
					$\GName, \asgFun, \sElm[0] \models \varphi$, by
					Item~\ref{def:sl(semantics:eqnt)} of
					Definition~\ref{def:sl(semantics)} of \SL\ semantics.
				\item
					$\Qnt = \AAll{\xElm}$.

					First note that $\QPAVSet{\qpElm} = \QPAVSet{\qpElm'} \cup \{ \xElm
					\}$.
					Also, consider the functions $\spcFun[\strFun|'] :
					\AsgSet(\QPAVSet{\qpElm'}, \sElm[0]) \to \AsgSet(\VSet \setminus \{
					\xElm \}, \sElm[0])$ defined by $\spcFun[\strFun|'](\asgFun') \defeq
					\spcFun(\asgFun'[\xElm \mapsto \strFun])_{\rst (\VSet \setminus \{
					\xElm \})}$, for each $\strFun \in \StrSet(\sElm[0])$ and $\asgFun'
					\in \AsgSet(\QPAVSet{\qpElm'}, \sElm[0])$.
					It is easy to see that every $\spcFun[\strFun|']$ is a dependence map
					for $\qpElm'$ over $\StrSet(\sElm[0])$, i.e.,
					$\spcFun[\strFun|'] \in \SpcSet[\StrSet( {\sElm[0]} )](\qpElm')$.
					Moreover, $\asgFun \umrg \spcFun(\asgFun') = \asgFun \umrg
					\spcFun[\asgFun'(\xElm)|'](\asgFun'_{\rst \QPAVSet{\qpElm'}})[\xElm
					\mapsto \asgFun'(\xElm)] = \asgFun[][\xElm \mapsto \asgFun'(\xElm)]
					\umrg \spcFun[\asgFun'(\xElm)|'](\asgFun'_{\rst \QPAVSet{\qpElm'}})$,
					for $\asgFun' \in \AsgSet(\QPAVSet{\qpElm}, \sElm[0])$.
					Then, it is evident that $\GName, \asgFun[][\xElm \to \strFun] \umrg
					\spcFun[\strFun|'](\asgFun'), \sElm[0] \models \psi$, for all $\strFun
					\in \StrSet(\sElm[0])$ and $\asgFun' \in \AsgSet(\QPAVSet{\qpElm'},
					\sElm[0])$.
					By the inductive hypothesis, we derive that $\GName, \asgFun[][\xElm
					\mapsto \strFun], \sElm[0] \models \qpElm' \psi$, for all $\strFun \in
					\StrSet(\sElm[0])$, which means that $\GName, \asgFun, \sElm[0]
					\models \varphi$, by Item~\ref{def:sl(semantics:aqnt)} of
					Definition~\ref{def:sl(semantics)}.
			\end{itemize}
			Thus, the thesis of the theorem holds.
		\end{proof}

		As an immediate consequence of the previous result, we derive the following
		corollary.
		\begin{corollary}[\SL\ Strategy Quantification]
			\label{cor:sl(strqnt)}
			Let $\GName$ be a \CGS\ and $\varphi = \qpElm \psi$ an \SL\ sentence,
			where $\psi$ is agent-closed and $\qpElm \in \QPSet(\free{\psi})$.
			Then, $\GName \models \varphi$ iff there exists a dependence map $\spcFun
			\in \SpcSet[ {\StrSet(\sElm[0])} ](\qpElm)$ such that $\GName,
			\spcFun(\asgFun), \sElm[0] \models \psi$, for all $\asgFun \in
			\AsgSet(\QPAVSet{\qpElm}, \sElm[0])$.
		\end{corollary}

	\end{subsection}

	\begin{subsection}{Elementary quantifications}
		\label{subsec:elmqnt}

		We now have all tools we need to introduce the property of elementariness
		for a particular class of dependence maps.
		Intuitively, a dependence map over functions from a set $\TSet$ to a set
		$\DSet$ is elementary if it can be split into a set of dependence maps over
		$\DSet$, one for each element of $\TSet$.
		This idea allows us to enormously simplify the reasoning about strategy
		quantifications, since we can reduce them to a set of quantifications over
		actions, one for each track in their domains.
		This means that, under certain conditions, we can transform a dependence map
		$\spcFun \in \SpcSet[\StrSet(\sElm)](\qpElm)$ over strategies in a function
		$\adj{\spcFun} : \TrkSet(\sElm) \to \SpcSet[\AcSet](\qpElm)$ that associates
		with each track a dependence map over actions.

		To formally develop the above idea, we have first to introduce the generic
		concept of adjoint function and state an auxiliary lemma.
		\begin{definition}[Adjoint Functions]
			\label{def:adjfun}
			Let $\DSet$, $\TSet$, $\USet$, and $\VSet$ be four sets, and $\mFun :
			(\TSet \to \DSet)^{\USet} \to (\TSet \to \DSet)^{\VSet}$ and $\adj{\mFun}
			: \TSet \to (\DSet^{\USet} \to \DSet^{\VSet})$ two functions.
			Then, $\adj{\mFun}$ is the \emph{adjoint} of $\mFun$ if
			$\adj{\mFun}(\tElm)(\flip{\gFun}(\tElm))(\xElm) =
			\mFun(\gFun)(\xElm)(\tElm)$, for all $\gFun \in (\TSet \to
			\DSet)^{\USet}$, $\xElm \in \VSet$, and $\tElm \in \TSet$~\footnote{By
			$\flip{\gFun} : \YSet \to \XSet \to \ZSet$ we denote the operation of
			\emph{flipping} of a function $\gFun : \XSet \to \YSet \to \ZSet$.}
		\end{definition}
		Intuitively, $\adj{\mFun}$ is the adjoint of $\mFun$ if the dependence from
		the set $\TSet$ in both domain and codomain of the latter can be extracted
		and put as a common factor of the functor given by the former.
		This means also that, for every pair of functions  $\gFun[1], \gFun[2] \in
		(\TSet \to \DSet)^{\USet}$ such that $\flip{\gFun[1]}(\tElm) =
		\flip{\gFun[2]}(\tElm)$ for some $\tElm \in \TSet$, it holds that
		$\mFun(\gFun[1])(\xElm)(\tElm) = \mFun(\gFun[2])(\xElm)(\tElm)$, for all
		$\xElm \in \VSet$.
		It is immediate to observe that if a function has an adjoint then this
		adjoint is unique.
		At the same way, if one has an adjoint function then it is possible to
		determine the original function without any ambiguity.
		Thus, it is established a one-to-one correspondence between functions
		admitting an adjoint and the adjoint itself.

		Next lemma formally states the property briefly described above, i.e., that
		each dependence map over a set $\TSet \to \DSet$, admitting an adjoint
		function, can be represented as a function, with $\TSet$ as domain, which
		returns dependence maps over $\DSet$ as values.
		\begin{lemma}[Adjoint Dependence Maps]
			\label{lmm:adjspc}
			Let $\qpElm \in \QPSet(\VSet)$ be a quantification prefix over a set
			$\VSet \subseteq \VarSet$ of variables, $\DSet$ and $\TSet$ two sets, and
			$\spcFun : \ValSet[\TSet \to \DSet](\QPAVSet{\qpElm}) \to \ValSet[\TSet
			\to \DSet](\VSet)$ and $\adj{\spcFun} : \TSet \to
			(\ValSet[\DSet](\QPAVSet{\qpElm}) \to \ValSet[\DSet](\VSet))$ two
			functions such that $\adj{\spcFun}$ is the adjoint of $\spcFun$.
			Then, $\spcFun \in \SpcSet[\TSet \to \DSet](\qpElm)$ iff, for all $t \in
			\TSet$, it holds that $\adj{\spcFun}(t) \in \SpcSet[\DSet](\qpElm)$.
		\end{lemma}

		We now define the formal meaning of the elementariness of a dependence map
		over functions.
		\begin{definition}[Elementary Dependence Maps]
			\label{def:elmspc}
			Let $\qpElm \in \QPSet(\VSet)$ be a quantification prefix over a set
			$\VSet \subseteq \VarSet$ of variables, $\DSet$ and $\TSet$ two sets, and
			$\spcFun \in \SpcSet[\TSet \to \DSet](\qpElm)$ a dependence map for
			$\qpElm$ over $\TSet \to \DSet$.
			Then, $\spcFun$ is \emph{elementary} if it admits an adjoint function.
			$\ESpcSet[\TSet \to \DSet](\qpElm)$ \mbox{denotes the set of all
			elementary dependence maps for $\qpElm$ over $\TSet \to \DSet$.}
		\end{definition}
		It is important to observe that, unfortunately, there are dependence maps
		that are not elementary.
		To easily understand why this is actually the case, it is enough to count
		both the number of dependence maps $\SpcSet[\TSet \to \DSet](\qpElm)$ and of
		adjoint functions $\TSet \to \SpcSet[\DSet](\qpElm)$, where $\card{\DSet} >
		1$, $\card{\TSet} > 1$ and $\qpElm$ is such that there is an $\xElm \in
		\QPEVSet{\qpElm}$ for which $\QPDepSet(\qpElm, \xElm) \neq \emptyset$.
		Indeed, it holds that $\card{\SpcSet[\TSet \to \DSet](\qpElm)} =
		\prod_{\xElm \in \QPEVSet{\qpElm}} \card{\DSet}^{\card{\TSet} \cdot
		\card{\DSet}^{\card{\TSet} \cdot \card{\QPDepSet(\qpElm, \xElm)}}} >
		\prod_{\xElm \in \QPEVSet{\qpElm}} \card{\DSet}^{\card{\TSet} \cdot
		\card{\DSet}^{\card{\QPDepSet(\qpElm, \xElm)}}} = \card{\TSet \to
		\SpcSet[\DSet](\qpElm)}$.
		So, there are much more dependence maps, a number double exponential in
		$\card{\TSet}$, than possible adjoint functions, whose number is only
		exponential in this value.
		Furthermore, observe that the simple set
		$\QPSet[\exists^{*}\forall^{*}](\VSet) \defeq \set{ \qpElm \in \QPSet(\VSet)
		}{ \exists i \in \numcc{0}{\card{\qpElm}} \:.\: \QPAVSet{(\qpElm)_{< i}} =
		\emptyset \wedge \QPEVSet{(\qpElm)_{\geq i}} = \emptyset }$, for $\VSet
		\subseteq \VarSet$, is the maximal class of quantification prefixes that
		admits only elementary dependence maps over $\TSet \to \DSet$, i.e., it is
		such that each $\spcFun \in \SpcSet[\TSet \to \DSet](\qpElm)$ is elementary,
		for all $\qpElm \in \QPSet[\exists^{*}\forall^{*}](\VSet)$.
		This is due to the fact that there are no functional dependences between
		variables, i.e., for each $\xElm \in \QPEVSet{\qpElm}$, it holds that
		$\QPDepSet(\qpElm, \xElm) = \emptyset$.

		Finally, we can introduce a new very important semantics for \SL\ syntactic
		fragments, which is based on the concept of elementary dependence map over
		strategies, and we refer to the related satisfiability concept as
		\emph{elementary satisfiability}, in symbols $\emodels$.
		Intuitively, such a semantics has the peculiarity that a strategy, used in
		an existential quantification in order to satisfy a formula, is only chosen
		between those that are elementary w.r.t.\ the universal quantifications.
		In this way, when we have to decide what is its value $\cElm$ on a given
		track $\trkElm$, we do it only in dependence of the values on the same track
		of the strategies so far quantified, but not on their whole structure, as it
		is the case instead of the classic semantics.
		This means that $\cElm$ does not depend on the values of the other
		strategies on tracks $\trkElm'$ that extend $\trkElm$, i.e., it does not
		depend on future choices made on $\trkElm'$.
		In addition, we have that $\cElm$ does not depend on values on parallel
		tracks $\trkElm'$ that only share a prefix with $\trkElm$, i.e., it is
		independent on choices made on the possibly alternative futures $\trkElm'$.
		The elementary semantics of \NGSL\ formulas involving atomic propositions,
		Boolean connectives, temporal operators, and agent bindings is defined as
		for the classic one, where the modeling relation $\models$ is substituted
		with $\emodels$, and we omit to report it here.
		In the following definition, we only describe the part concerning the
		quantification prefixes.
		\begin{definition}[\NGSL\ Elementary Semantics]
			\label{def:ngsl(elementarysemantics)}
			Let $\GName$ be a \CGS, $\sElm \in \StSet$ one of its states, and $\qpElm
			\psi$ an \NGSL\ formula, where $\psi$ is agent-closed and $\qpElm \in
			\QPSet(\free{\psi})$.
			Then $\GName, \emptyfun, \sElm \emodels \qpElm \psi$ if there is an
			elementary dependence map $\spcFun \in \ESpcSet[\StrSet(\sElm)](\qpElm)$
			for $\qpElm$ over $\StrSet(\sElm)$ such that $\GName, \spcFun(\asgFun),
			\sElm \emodels \psi$, for all $\asgFun \in \AsgSet(\QPAVSet{\qpElm},
			\sElm)$.
		\end{definition}
		It is immediate to see a strong similarity between the statement of
		Corollary~\ref{cor:sl(strqnt)} of \SL\ strategy quantification and the
		previous definition.
		The only crucial difference resides in the choice of the kind of dependence
		map.
		Moreover, observe that, differently from the classic semantics, the
		quantifications in the prefix are not treated individually but as an atomic
		block.
		This is due to the necessity of having a strict correlation between the
		point-wise structure of the quantified strategies.

		\begin{remark}[\SL\ Elementary Semantics]
			\label{rmk:sl(elmentarysemantics)}
			It can be interesting to know that we do not define an elementary
			semantics for the whole \SL, since we are not able, at the moment, to
			easily use the concept of elementary dependence map, when the
			quantifications are not necessarily grouped in prefixes, i.e., when the
			formula is not in prenex normal form.
			In fact, this may represent a challenging problem, whose solution is left
			to future works.
		\end{remark}

		Due to the new semantics of \NGSL, we have to redefine the related concepts
		of model and satisfiability, in order to differentiate between the classic
		relation $\models$ and the elementary one $\emodels$.
		Indeed, as we show later, there are sentences that are satisfiable but not
		elementary satisfiable and vice versa.
		\begin{definition}[\NGSL\ Elementary Satisfiability]
			\label{def:ngsl(elmsat)}
			We say that a \CGS\ $\GName$ is an \emph{elementary model} of an \NGSL\
			sentence $\varphi$, in symbols $\GName \emodels \varphi$, if $\GName,
			\emptyfun, \sElm[0] \emodels \varphi$.
			In general, we also say that $\GName$ is a \emph{elementary model} for
			$\varphi$ on $\sElm \in \StSet$, in symbols $\GName, \sElm \emodels
			\varphi$, if $\GName, \emptyfun, \sElm \emodels \varphi$.
			An \NGSL\ sentence $\varphi$ is \emph{elementarily satisfiable} if there
			is an elementary model for it.
		\end{definition}

		We have to modify the concepts of implication and equivalence, as well.
		Indeed, also in this case we can have pairs of equivalent formulas that are
		not elementarily equivalent, and vice versa.
		Thus, we have to be careful when we use natural transformation between
		formulas, since it can be the case that they preserve the meaning only under
		the classic semantics.
		An example of this problem can arise when one want to put a formula in \pnf.
		\begin{definition}[\NGSL\ Elementary Implication and Equivalence]
			\label{def:ngsl(elmimpeqv)}
			Given two \NGSL\ formulas $\varphi_{1}$ and $\varphi_{2}$ with
			$\free{\varphi_{1}} = \free{\varphi_{2}}$, we say that $\varphi_{1}$
			\emph{elementarily implies} $\varphi_{2}$, in symbols $\varphi_{1}
			\eimplies \varphi_{2}$, if, for all \CGS s $\GName$, states $\sElm \in
			\StSet$, and $\free{\varphi_{1}}$-defined $\sElm$-total assignments
			$\asgFun \in \AsgSet(\free{\varphi_{1}}, \sElm)$, it holds that if
			$\GName, \asgFun, \sElm \emodels \varphi_{1}$ then $\GName, \asgFun, \sElm
			\emodels \varphi_{2}$.
			Accordingly, we say that $\varphi_{1}$ is \emph{elementarily equivalent}
			to $\varphi_{2}$, in symbols $\varphi_{1} \!\eequiv\! \varphi_{2}$, if
			both $\varphi_{1} \!\eimplies\! \varphi_{2}$ and $\varphi_{2}
			\!\eimplies\! \varphi_{1}$ hold.
		\end{definition}

	\end{subsection}

	\begin{subsection}{Elementariness and non-elementariness}
		\label{subsec:elmprp}

		Finally, we show that the introduced concept of elementary satisfiability is
		relevant to the context of our logic, as its applicability represents a
		demarcation line between ``easy'' and ``hard'' fragments of \SL.
		Moreover, we believe that it is because of this fundamental property that
		several well-known temporal logics are so robustly decidable~\cite{Var96}.

		\begin{remark}[{\NGSL[$0$-alt]} Elementariness]
			\label{rmk:ngsl(elm)}
			It is interesting to observe that, for every \CGS\ $\GName$ and
			\NGSL[$0$-alt] sentence $\varphi$, it holds that $\GName \models \varphi$
			iff $\GName \emodels \varphi$.
			This is an immediate consequence of the fact that all quantification
			prefixes $\qpElm$ used in $\varphi$ belong to
			$\QPSet[\exists^{*}\forall^{*}](\VSet)$, for  a given set $\VSet \subseteq
			\VarSet$ of variables.
			Thus, as already mentioned, the related dependence maps on strategies
			$\spcFun \in \SpcSet[\StrSet( {\sElm[0]} )](\qpElm)$ are necessarily
			elementary.
		\end{remark}

		By Corollary~\ref{cor:sl(strqnt)} of \SL\ strategy quantification, it is
		easy to see that the following coherence property about the elementariness
		of the \NGSL\ satisfiability holds.
		Intuitively, it asserts that every elementarily satisfiable sentence in
		\pnf\ is satisfiable too.
		\begin{theorem}[\NGSL\ Elementary Coherence]
			\label{thm:ngsl(elmchr)}
			Let $\GName$ be a \CGS, $\sElm \in \StSet$ one of its states, $\varphi$ an
			\NGSL\ formula in \pnf, and $\asgFun \in \AsgSet(\sElm)$ an $\sElm$-total
			assignment with $\free{\varphi} \subseteq \dom{\asgFun}$.
			Then, it holds that $\GName, \asgFun, \sElm \emodels \varphi$ implies
			$\GName, \asgFun, \sElm \models \varphi$.
		\end{theorem}
		\begin{proof}
			The proof proceeds by induction on the structure of the formula.
			For the sake of succinctness, we only show the crucial case of principal
			subsentences $\phi \in \psnt{\varphi}$, i.e., when $\phi$ is of the form
			$\qpSym \psi$, where $\qpElm \in \QPSet(\free{\psi})$ is a quantification
			prefix, and $\psi$ is an agent-closed formula.

			Suppose that $\GName, \emptyfun, \sElm \emodels \qpSym \psi$.
			Then, by Definition~\ref{def:ngsl(elementarysemantics)} of \NGSL\
			elementary semantics, there is an elementary dependence map $\spcFun \in
			\ESpcSet[ {\StrSet(\sElm)} ](\qpElm)$ such that, for all assignments
			$\asgFun \in \AsgSet(\QPAVSet{\qpElm}, \sElm)$, it holds that $\GName,
			\spcFun(\asgFun), \sElm \emodels \psi$.
			Now, by the inductive hypothesis, there is a dependence map $\spcFun \in
			\SpcSet[ {\StrSet(\sElm)} ](\qpElm)$ such that, for all assignments
			$\asgFun \in \AsgSet(\QPAVSet{\qpElm}, \sElm)$, it holds that $\GName,
			\spcFun(\asgFun), \sElm \models \psi$.
			Hence, by Corollary~\ref{cor:sl(strqnt)} of \SL\ strategy quantification,
			we have that $\GName, \emptyfun, \sElm \models \qpSym \psi$.
		\end{proof}

		However, it is worth noting that the converse property may not hold, as we
		show in the next theorem, i.e., there are sentences in \pnf\ that are
		satisfiable but not elementarily satisfiable.
		Note that the following results already holds for \CHPSL.
		\begin{theorem}[\TBBGSL\ Non-Elementariness]
			\label{thm:tbbgsl(nelm)}
			There exists a satisfiable \TBBGSL[$1$-ag, $2$-var, $1$-alt] sentence in
			\pnf\ that is not elementarily satisfiable.
		\end{theorem}
		\begin{proof}
			Consider the \TBBGSL[$1$-ag, $2$-var, $1$-alt] sentence $\varphi \defeq
			\varphi_{1} \wedge \varphi_{2}$ in \pnf\, where $\varphi_{1} \defeq \qpSym
			(\psi_{1} \wedge \psi_{2})$, with $\qpSym \defeq \AAll{\xSym}
			\EExs{\ySym}$, $\psi_{1} \defeq (\alpha, \xSym) \X \pSym \leftrightarrow
			(\alpha,\ySym) \X \neg \pSym$, and $\psi_{2} \defeq (\alpha, \xSym) \X \X
			\pSym \leftrightarrow (\alpha, \ySym) \X \X \pSym$, and $\varphi_{2}
			\defeq \AAll{\xSym} (\alpha, \xSym) \X ((\EExs{\xSym} (\alpha, \xSym) \X
			\pSym) \wedge (\EExs{\xSym} (\alpha, \xSym) \X \neg \pSym))$.
			Moreover, note that the \TBOGSL[$1$-ag, $1$-var, $0$-alt] sentence
			$\varphi_{2}$ is equivalent to the \CTL\ formula $\A \X ((\E \X \pSym)
			\wedge (\E \X \neg \pSym))$.
			Then, it is easy to see that the turn-based \CGS\ $\GName[Rdc]$ of
			Figure~\vref{fig:lmm:qptl(rdc)} satisfies $\varphi$.
			Indeed, $\GName[Rdc], \spcFun(\asgFun), \sSym[0] \models \psi_{1} \wedge
			\psi_{2}$, for all assignments $\asgFun \in \AsgSet(\{ \xSym \},
			\sSym[0])$, where the non-elementary dependence map $\spcFun \in \SpcSet[
			{\StrSet(\sSym[0])} ](\qpSym)$ is such that
			$\spcFun(\asgFun)(\ySym)(\sSym[0]) = \neg \asgFun(\xSym)(\sSym[0])$ and
			$\spcFun(\asgFun)(\ySym)(\sSym[0] \cdot \sElm[i]) =
			\asgFun(\xSym)(\sSym[0] \cdot \sElm[1 - i])$, for all $i \in \{ 0, 1 \}$.

			Now, let $\GName$ be a generic \CGS.
			If $\GName \not\models \varphi$, by Theorem~\ref{thm:ngsl(elmchr)} of
			\NGSL\ elementary coherence, it holds that $\GName \not\emodels \varphi$.
			Otherwise, we have that $\GName \models \varphi$ and, in particular,
			$\GName \models \varphi_{1}$, which means that $\GName \models \qpSym
			(\psi_{1} \wedge \psi_{2})$.
			At this point, to prove that $\GName \not\emodels \varphi$, we show that,
			for all elementary dependence maps $\spcFun \in \ESpcSet[
			{\StrSet(\sElm[0])} ](\qpSym)$, there exists an assignment $\asgFun \in
			\AsgSet(\{ \xSym \}, \sElm[0])$ such that $\GName, \spcFun(\asgFun),
			\sElm[0] \not\emodels \psi_{1} \wedge \psi_{2}$.
			To do this, let us fix an elementary dependence map $\spcFun$ and an
			assignment $\asgFun$.
			Also, assume $\sElm[1] \defeq \trnFun(\sElm[0], \allowbreak
			\emptyset[\alpha \mapsto \asgFun(\xSym)(\sElm[0])])$ and \mbox{$\sElm[2]
			\defeq \trnFun(\sElm[0], \emptyset[\alpha \mapsto
			\spcFun(\asgFun)(\ySym)(\sElm[0])])$.
			Now, we distinguish between two cases.}
			\begin{itemize}
				\item
					$\pSym \in \labFun(\sElm[1])$ iff $\pSym \in \labFun(\sElm[2])$.
					In this case, we can easily observe that $\GName, \spcFun(\asgFun),
					\sElm[0] \not\models \psi_{1}$ and consequently, by
					Theorem~\ref{thm:ngsl(elmchr)}, it holds that $\GName,
					\spcFun(\asgFun), \sElm[0] \not\emodels \psi_{1} \wedge \psi_{2}$.
					So, we are done.
				\item
					$\pSym \in \labFun(\sElm[1])$ iff $\pSym \not\in \labFun(\sElm[2])$.
					If $\GName, \spcFun(\asgFun), \sElm[0] \not\models \psi_{2}$ then, by
					Theorem~\ref{thm:ngsl(elmchr)}, it holds that $\GName,
					\spcFun(\asgFun), \sElm[0] \not\emodels \psi_{1} \wedge \psi_{2}$.
					So, we are done.
					Otherwise, let $\sElm[3] \defeq \trnFun(\sElm[1], \emptyset[\alpha
					\mapsto \asgFun(\xSym)(\sElm[0] \cdot \sElm[1])])$ and $\sElm[4]
					\defeq \trnFun(\sElm[2], \emptyset[\alpha \mapsto
					\spcFun(\asgFun)(\ySym)(\sElm[0] \cdot \sElm[2])])$.
					Then, it holds that $\pSym \in \labFun(\sElm[3])$ iff $\pSym \in
					\labFun(\sElm[4])$.
					Now, consider a new assignment $\asgFun' \in \AsgSet(\{ \xSym \},
					\sElm[0])$ such that $\asgFun'(\xElm)(\sElm[0] \cdot \sElm[2]) =
					\asgFun(\xElm)(\sElm[0] \cdot \sElm[2])$ and $\pSym \in
					\labFun(\sElm[3]')$ iff $\pSym \not\in \labFun(\sElm[4])$, where
					$\sElm[3]' \defeq \trnFun(\sElm[1], \emptyset[\alpha \mapsto
					\asgFun'(\xSym)(\sElm[0] \cdot \sElm[1])])$.
					Observe that the existence of such an assignment, with particular
					reference to the second condition, is ensured by the fact that $\GName
					\models \varphi_{2}$.
					At this point, due to the elementariness of the dependence map
					$\spcFun$, we have that $\spcFun(\asgFun')(\ySym)(\sElm[0] \cdot
					\sElm[2]) = \spcFun(\asgFun)(\ySym)(\sElm[0] \cdot \sElm[2])$.
					Consequently, it holds that $\sElm[4] = \trnFun(\sElm[2],
					\emptyset[\alpha \mapsto \spcFun(\asgFun')(\ySym)(\sElm[0] \cdot
					\sElm[2])])$.
					Thus, $\GName, \spcFun(\asgFun'), \sElm[0] \not\models \psi_{2}$,
					which implies, by Theorem~\ref{thm:ngsl(elmchr)}, that $\GName,
					\spcFun(\asgFun'), \sElm[0] \not\emodels \psi_{1} \wedge \psi_{2}$.
					So, we are done.
			\end{itemize}
			Thus, the thesis of the theorem holds.
		\end{proof}

		The following corollary is an immediate consequence of the previous theorem.
		It is interesting to note that, at the moment, we do not know if such a
		result can be extended to the simpler \GL\ fragment.
		\begin{corollary}[\BGSL\ Non-Elementariness]
			\label{cor:bgsl(nelm)}
			There exists a satisfiable \BGSL[$1$-ag, $2$-var, $1$-alt] sentence in
			\pnf\ that is not elementarily satisfiable.
		\end{corollary}

		\begin{remark}[Kinds of Non-Elementariness]
			\label{rmk:kndnelm}
			It is worth remarking that the kind of non-elemen\-tariness of the
			sentence $\varphi$ shown in the above theorem can be called
			\emph{essential}, i.e., it cannot be eliminated, due to the fact that
			$\varphi$ is satisfiable but not elementarily satisfiable.
			However, there are different sentences, such as the conjunct
			$\varphi_{1}$ in $\varphi$, having both models on which they are
			elementarily satisfiable and models, like the \CGS\ $\GName[Rdc]$, on
			which they are only non-elementarily satisfiable.
			Such a kind of non-elementariness can be called \emph{non-essential},
			since it can be eliminated by an opportune choice of the underlying model.
			Note that a similar reasoning can be done for the dual concept of
			elementariness, which we call \emph{essential} \mbox{if all models
			satisfying a given sentence elementarily satisfy it as well.}
		\end{remark}

		Before continuing, we want to show the reason why we have redefined the
		concepts of implication and equivalence in the context of elementary
		semantics.
		Consider the \BGSL[$1$-ag, $2$-var, $1$-alt] sentence $\varphi_{1}$ used in
		Theorem~\ref{thm:tbbgsl(nelm)} of \TBBGSL\ non-elementariness.
		It is not hard to see that it is equivalent to the \OGSL[$1$-ag, $1$-var,
		$0$-alt] $\varphi' \defeq (\EExs{\xSym} (\alpha, \xSym) \psi_{1}
		\leftrightarrow \EExs{\xSym} (\alpha, \xSym) \psi_{2}) \wedge (\EExs{\xSym}
		(\alpha, \xSym) \psi_{3} \leftrightarrow \EExs{\xSym} (\alpha, \xSym)
		\psi_{4})$, where $\psi_{1} \defeq \X (\pSym \wedge \X \pSym)$, $\psi_{2}
		\defeq \X (\neg \pSym \wedge \X \pSym)$, $\psi_{3} \defeq \X (\pSym \wedge
		\X \neg \pSym)$, and $\psi_{4} \defeq \X (\neg \pSym \wedge \X \neg \pSym)$.
		Note that $\varphi'$ is in turn equivalent to the \CTLS\ formula $(\E
		\psi_{1} \leftrightarrow \E \psi_{2}) \wedge (\E \psi_{3} \leftrightarrow \E
		\psi_{4})$.
		However, $\varphi_{1}$ and $\varphi'$ are not elementarily equivalent, since
		we have that $\GName[Rdc] \not\emodels \varphi_{1}$ but $\GName[Rdc]
		\emodels \varphi'$, where $\GName[Rdc]$ is the \CGS\ of
		Figure~\vref{fig:lmm:qptl(rdc)}.

		At this point, we can proceed with the proof of the elementariness of
		satisfiability for \OGSL.
		It is important to note that there is no gap, in our knowledge, between the
		logics that are elementarily satisfiable and those that are not, since the
		fragment \BGSL[$1$-ag, $2$-var, $1$-alt] used in the previous theorem
		cannot be further reduced, due to the fact that otherwise it collapses into
		\OGSL.
		Before starting, we have to describe some notation regarding classic
		two-player games on infinite words~\cite{PP04}, which are used here as a
		technical tool.
		Note that we introduce the names of scheme and match in place of the more
		usual strategy and play, in order to avoid confusion between the concepts
		related to a \CGS\ and those related to the tool.

		A \emph{two-player arena} (\TPA, for short) is a tuple $\AName \defeq
		\TPAStruct$, where $\NdESet$ and $\NdOSet$ are non-empty non-intersecting
		sets of \emph{nodes} for player \emph{even} and \emph{odd}, respectively,
		$\EdgRel \defeq \EdgERel \cup \EdgORel$, with $\EdgERel \subseteq \NdESet
		\times \NdOSet$ and $\EdgORel \subseteq \NdOSet \times \NdESet$, is the
		\emph{edge relation} between nodes, and $\nElm[0] \in \NdOSet$ is a
		designated \emph{initial node}.

		An \emph{even position} in $\AName$ is a finite non-empty sequence of nodes
		$\posElm \in \NdESet^{+}$ such that $(\posElm)_{0} = \nElm[0]$ and, for all
		$i \in \numco{0}{\card{\posElm} - 1}\!$, there exists a node $\nElm \in
		\NdOSet$ for which $((\posElm)_{i}, \nElm) \in \EdgERel$ and $(\nElm,
		(\posElm)_{i + 1}) \in \EdgORel$ hold.
		In addition, an \emph{odd position} in $\AName$ is a finite non-empty
		sequence of nodes $\posElm = \posElm' \cdot \nElm \in \NdESet^{+} \cdot
		\NdOSet$, with $\nElm \in \NdOSet$, such that $\posElm'$ is an even position
		and $(\lst{\posElm'}, \nElm) \in \EdgERel$.
		\mbox{By $\PosESet$ and $\PosOSet$ we denote, respectively, the sets of even
		and odd positions.}

		An \emph{even} (resp., \emph{odd}) \emph{scheme} in $\AName$ is a function
		$\scheFun : \PosESet \to \NdOSet$ (resp., $\schoFun : \PosOSet \to
		\NdESet$) that maps each even (resp., odd) position to an odd (resp., even)
		node in a way that is compatible with the edge relation $\EdgERel$ (resp.,
		$\EdgORel$), i.e., for all $\posElm \in \PosESet$ (resp., $\posElm \in
		\PosOSet$), it holds that $(\lst{\posElm}, \scheFun(\posElm)) \in \EdgERel$
		(resp., $(\lst{\posElm}, \schoFun(\posElm)) \in \EdgORel$).
		By $\SchESet$ (resp., $\SchOSet$) we indicate the sets of even (resp., odd)
		schemes.

		A \emph{match} in $\AName$ is an infinite sequence of nodes $\mtcElm \in
		\NdESet^{\omega}$ such that $(\mtcElm)_{0} = \nElm[0]$ and, for all $i \in
		\SetN$, there exists a node $\nElm \in \NdOSet$ such that $((\mtcElm)_{i},
		\nElm) \in \EdgERel$ and $(\nElm, (\mtcElm)_{i + 1}) \in \EdgORel$.
		By $\MtcSet$ we denote the set of all matches.
		A \emph{match map} $\mtcFun : \SchESet \times \SchOSet \to \MtcSet$ is a
		function that, given two schemes $\scheFun \in \SchESet$ and $\schoFun \in
		\SchOSet$, returns the unique match $\mtcElm = \mtcFun(\scheFun, \schoFun)$
		such that, for all $i \in \SetN$, it holds that $(\mtcElm)_{i + 1} =
		\schoFun((\mtcElm)_{\leq i} \cdot \scheFun((\mtcElm)_{\leq i}))$.

		A \emph{two-player game} (\TPG, for short) is a tuple $\HName \defeq
		\TPGStruct$, where $\AName$ is a \TPA\ and $\WinSet \subseteq \MtcSet$.
		On one hand, we say that player even wins $\HName$ if there exists an even
		scheme $\scheFun \in \SchESet$ such that, for all odd schemes $\schoFun \in
		\SchOSet$, it holds that $\mtcFun(\scheFun, \schoFun) \in \WinSet$.
		On the other hand, we say that player odd wins $\HName$ if there exists an
		odd scheme $\schoFun \in \SchOSet$ such that, for all even schemes $\scheFun
		\in \SchESet$, it holds that $\mtcFun(\scheFun, \schoFun) \not\in \WinSet$.

		In the following, for a given binding prefix $\bpElm \in \BndSet(\VSet)$
		with $\VSet \subseteq \VarSet$, we denote by $\bndFun[\bpElm] : \AgSet \to
		\VSet$ the function associating with each agent the related variable in
		$\bpElm$, i.e., for all $\aElm \in \AgSet$, there is $i \in
		\numco{0}{\card{\bpElm}}$ such that $(\bpElm)_{i} = (\aElm,
		\bndFun[\bpElm](\aElm))$.

		As first step towards the proof of the elementariness of \OGSL, we have to
		give a construction of a two-player game, based on an a priori chosen \CGS,
		in which the players are explicitly viewed one as a dependence map and the
		other as a valuation, both over actions.
		This construction results to be a deep technical evolution of the proof
		method used for the dualization of alternating automata on infinite
		objects~\cite{MS87}.
		\begin{definition}[Dependence-vs-Valuation Game]
			\label{def:spcvaltpg}
			Let $\GName$ be a \CGS, $\sElm \in \StSet$ one of its states, $\PSet
			\subseteq \PthSet(\sElm)$ a set of paths, $\qpElm \in \QPSet(\VSet)$ a
			quantification prefix over a set $\VSet \subseteq \VarSet$ of variables,
			and $\bpElm \in \BndSet(\VSet)$ a binding.
			Then, the dependence-vs-valuation game for $\GName$ in $\sElm$ over
			$\PSet$ w.r.t.\ $\qpElm$ and $\bpElm$ is the \TPG\ $\HName(\GName, \sElm,
			\PSet, \qpElm, \bpElm) \defeq \TPGTuple {\AName(\GName, \sElm, \qpElm,
			\bpElm)} {\PSet}$, where the \TPA\ $\AName(\GName, \sElm, \qpElm, \bpElm)
			\defeq \TPATuple {\StSet} {\StSet \times \SpcSet[\AcSet](\qpElm)}
			{\EdgRel} {\sElm}$ has the edge relations defined as follows:
			\begin{itemize}
				\item
					$\EdgERel \defeq \set{ (\tElm, (\tElm, \spcFun)) }{ \tElm \in \StSet
					\land \spcFun \in \SpcSet[\AcSet](\qpElm) }$;
				\item
					$\EdgORel \defeq \set{ ((\tElm, \spcFun), \trnFun(\tElm,
					\spcFun(\valFun) \cmp \bndFun[\bpElm])) }{ \tElm \in \StSet \land
					\spcFun \in \SpcSet[\AcSet](\qpElm) \land \valFun \in
					\ValSet[\AcSet](\QPAVSet{\qpElm}) }$~\footnote{By $\gFun[2] \cmp
					\gFun[1] : \XSet \to \ZSet$ we denote the operation of
					\emph{composition} of two functions $\gFun[1] : \XSet \to \YSet[1]$
					and $\gFun[2] : \YSet[2] \to \ZSet$ with $\YSet[1] \subseteq
					\YSet[2]$.}.
			\end{itemize}
		\end{definition}

		In the next lemma we state a fundamental relationship between
		dependence-vs-valuation games and their duals.
		Basically, we prove that if a player wins the game then the opposite player
		can win the dual game, and vice versa.
		\mbox{This represents one of the two crucial steps in our elementariness
		proof.}
		\begin{lemma}[Dependence-vs-Valuation Duality]
			\label{lmm:spcvaldlt}
			Let $\GName$ be a \CGS, $\sElm \in \StSet$ one of its states, $\PSet
			\subseteq \PthSet(\sElm)$ a set of paths, $\qpElm \in \QPSet(\VSet)$ a
			quantification prefix over a set $\VSet \subseteq \VarSet$ of variables,
			and $\bpElm \in \BndSet(\VSet)$ a binding.
			Then, player even wins the \TPG\ $\HName(\GName, \sElm, \PSet, \qpElm,
			\bpElm)$ iff player odd wins the dual \TPG\ $\HName(\GName, \sElm,
			\PthSet(\sElm) \setminus \PSet, \dual{\qpElm}, \bpElm)$.
		\end{lemma}

		Now, we are going to give the definition of the important concept of
		\emph{encasement}.
		Informally, an encasement is a particular subset of paths in a given \CGS\
		that ``works to encase'' an elementary dependence map on strategies, in the
		sense that it contains all plays obtainable by complete assignments derived
		from the evaluation of the above mentioned dependence map.
		In our context, this concept is used to summarize all \mbox{relevant
		information needed to verify the elementary satisfiability of a sentence.}
		\begin{definition}[Encasement]
			\label{def:encasement}
			Let $\GName$ be a \CGS, $\sElm \in \StSet$ one of its states, $\PSet
			\subseteq \PthSet(\sElm)$ a set of paths, $\qpElm \in \QPSet(\VSet)$ a
			quantification prefix over a set $\VSet \subseteq \VarSet$ of variables,
			and $\bpElm \in \BndSet(\VSet)$ a binding.
			Then, $\PSet$ is an \emph{encasement} w.r.t.\ $\qpElm$ and $\bpElm$ if
			there exists an elementary dependence map $\spcFun \!\in\! \ESpcSet[
			{\StrSet(\sElm)} ](\qpElm)$ such that, for all assignments $\asgFun
			\!\in\! \AsgSet(\QPAVSet{\qpElm}, \sElm)$, it holds that
			$\playFun(\spcFun(\asgFun) \cmp \bndFun[\bpElm], \sElm) \!\in\! \PSet$.
		\end{definition}

		In the next lemma, we give the second of the two crucial steps in our
		elementariness proof.
		In particular, we are able to show a one-to-one relationship between the
		wining in the dependence-vs-valuation game of player even and the
		verification of the encasement property of the associated winning set.
		Moreover, in the case that the latter is a Borelian set, by using Martin's
		Determinacy Theorem~\cite{Mar75}, we obtain a complete characterization of
		the winning concept by means of that of encasements.
		\begin{lemma}[Encasement Characterization]
			\label{lmm:encasement}
			Let $\GName$ be a \CGS, $\sElm \in \StSet$ one of its states, $\PSet
			\subseteq \PthSet(\sElm)$ a set of paths, $\qpElm \in \QPSet(\VSet)$ a
			quantification prefix over a set $\VSet \subseteq \VarSet$ of variables,
			and $\bpElm \in \BndSet(\VSet)$ a binding.
			Then, the following hold:
			\begin{enumerate}[(i)]
				\item\label{lmm:encasement(ewin)}
					player even wins $\HName(\GName, \sElm, \PSet, \qpElm, \bpElm)$ iff
					$\PSet$ is an encasement w.r.t.\ $\qpElm$ and $\bpElm$;
				\item\label{lmm:encasement(owin-dir)}
					if player odd wins $\HName(\GName, \sElm, \PSet, \qpElm, \bpElm)$ then
					$\PSet$ is not an encasement w.r.t.\ $\qpElm$ and $\bpElm$;
				\item\label{lmm:encasement(owin-inv)}
					if $\PSet$ is a Borelian set and it is not an encasement w.r.t.\
					$\qpElm$ and $\bpElm$ then player odd wins $\HName(\GName, \sElm,
					\PSet, \qpElm, \bpElm)$.
			\end{enumerate}
		\end{lemma}

		Finally, we have all technical tools useful to prove the elementariness of
		the satisfiability for \OGSL.
		Intuitively, we describe a bidirectional reduction of the problem of
		interest to the verification of the winning in the dependence-vs-valuation
		game.
		The idea behind this construction resides in the strong similarity between
		the statement of Corollary~\ref{cor:sl(strqnt)} of \SL\ strategy
		quantification and the definition of the winning condition in a two-player
		game.
		Indeed, on one hand, we say that a sentence is satisfiable iff ``there
		exists a dependence map such that, for all all assignments, it holds that
		...''.
		On the other hand, we say that player even wins a game iff ``there exists
		an even scheme such that, for all odd schemes, it holds that ...''.
		In particular, for the \OGSL\ fragment, we can resolve the gap between these
		two formulations, by using the concept of elementary quantification.
		\begin{theorem}[\OGSL\ Elementariness]
			\label{thm:ogsl(elm)}
			Let $\GName$ be a \CGS, $\varphi$ an \OGSL\ formula, $\sElm \in \StSet$ a
			state, and $\asgFun \in \AsgSet(\sElm)$ an $\sElm$-total assignment with
			$\free{\varphi} \subseteq \dom{\asgFun}$.
			Then, it holds that $\GName, \asgFun, \sElm \models \varphi$ iff $\GName,
			\asgFun, \sElm \emodels \varphi$.
		\end{theorem}
		\begin{proof}
			The proof proceeds by induction on the structure of the formula.
			For the sake of succinctness, we only show the most important inductive
			case of principal subsentences $\phi \in \psnt{\varphi}$, i.e., when
			$\phi$ is of the form $\qpSym \bpElm \psi$, where $\qpElm \in
			\QPSet(\VSet)$ and $\bpElm \in \BPSet(\VSet)$ are, respectively, a
			quantification and binding prefix over a set $\VSet \subseteq \VarSet$ of
			variables, and $\psi$ is a variable-closed formula.

			\emph{[If].}
			The proof of this direction is practically the same of the one used in
			Theorem~\ref{thm:ngsl(elmchr)} of \NGSL\ elementary coherence.
			So, we omit to report it here.

			\emph{[Only if].}
			Assume that $\GName, \emptyfun, \sElm \models \qpSym \bpElm \psi$.
			Then, it is easy to see that, for all elementary dependence maps $\spcFun
			\in \ESpcSet[ {\StrSet(\sElm)} ](\dual{\qpElm})$, there is an assignment
			$\asgFun \in \AsgSet(\QPAVSet{\dual{\qpElm}}, \sElm)$ such that $\GName,
			\spcFun(\asgFun) \cmp \bndFun[\bpElm], \sElm \models \psi$.
			Indeed, suppose by contradiction that there exists an elementary
			dependence map $\spcFun \in \ESpcSet[ {\StrSet(\sElm)} ](\dual{\qpElm})$
			such that, for all assignments $\asgFun \in
			\AsgSet(\QPAVSet{\dual{\qpElm}}, \sElm)$, it holds that $\GName,
			\spcFun(\asgFun) \cmp \bndFun[\bpElm], \sElm \not\models \psi$, i.e.,
			$\GName, \spcFun(\asgFun) \cmp \bndFun[\bpElm], \sElm \models \neg \psi$,
			and so $\GName, \spcFun(\asgFun), \sElm \models \bpElm \neg \psi$.
			Then, by Corollary~\ref{cor:sl(strqnt)} of \SL\ strategy quantification,
			we have that $\GName, \emptyfun, \sElm \models \dual{\qpElm} \bpElm \neg
			\psi$, i.e., $\GName, \emptyfun, \sElm \models \neg \qpElm \bpElm \psi$,
			and so $\GName, \emptyfun, \sElm \not\models \qpElm \bpElm \psi$, which is
			impossible.

			Now, let $\PSet \defeq \set{ \playFun(\asgFun, \sElm) \in \PthSet(\sElm)
			}{ \asgFun \in \AsgSet(\AgSet, \sElm) \land \GName, \asgFun, \sElm
			\not\models \psi}$.
			Then, it is evident that, for all elementary dependence maps $\spcFun \in
			\ESpcSet[ {\StrSet(\sElm)} ](\dual{\qpElm})$, there is an assignment
			$\asgFun \in \AsgSet(\QPAVSet{\dual{\qpElm}}, \sElm)$ such that
			$\playFun(\spcFun(\asgFun) \cmp \bndFun[\bpElm], \sElm) \not\in \PSet$.

			At this point, by Definition~\ref{def:encasement} of encasement, it is
			clear that $\PSet$ is not an encasement w.r.t.\ $\dual{\qpElm}$ and
			$\bpElm$.
			Moreover, since $\psi$ describes a regular language, the derived set
			$\PSet$ is Borelian~\cite{PP04}.
			Consequently, by Item~\ref{lmm:encasement(owin-inv)} of
			Lemma~\ref{lmm:encasement} of encasement characterization, we have that
			player odd wins the \TPG\ $\HName(\GName, \sElm, \PSet, \dual{\qpElm},
			\bpElm)$.
			Thus, by Lemma~\ref{lmm:spcvaldlt} of dependence-vs-valuation duality,
			player even wins the dual \TPG\ $\HName(\GName, \sElm, \PthSet(\sElm)
			\setminus \PSet, \qpElm, \bpElm)$.
			Hence, by Item~\ref{lmm:encasement(ewin)} of Lemma~\ref{lmm:encasement},
			we have that $\PthSet(\sElm) \setminus \PSet$ is an encasement w.r.t.\
			$\qpElm$ and $\bpElm$.
			Finally, again by Definition~\ref{def:encasement}, there exists an
			elementary dependence map $\spcFun \in \ESpcSet[ {\StrSet(\sElm)}
			](\qpElm)$ such that, for all assignments $\asgFun \in
			\AsgSet(\QPAVSet{\qpElm}, \sElm)$, it holds that
			$\playFun(\spcFun(\asgFun) \cmp \bndFun[\bpElm], \sElm) \in \PthSet(\sElm)
			\setminus \PSet$.

			Now, it is immediate to observe that $\PthSet(\sElm) \setminus \PSet =
			\set{ \playFun(\asgFun, \sElm) \in \PthSet(\sElm) }{ \asgFun \in
			\AsgSet(\AgSet, \sElm) \land \GName, \asgFun, \sElm \models \psi}$.
			So, by the inductive hypothesis, we have that $\PthSet(\sElm) \setminus
			\PSet = \set{ \playFun(\asgFun, \sElm) \in \PthSet(\sElm) }{ \asgFun \in
			\AsgSet(\AgSet, \sElm) \land \GName, \asgFun, \sElm \emodels \psi}$, from
			which we derive that there exists an elementary dependence map $\spcFun
			\in \ESpcSet[ {\StrSet(\sElm)} ](\qpElm)$ such that, for all assignments
			$\asgFun \in \AsgSet(\QPAVSet{\qpElm}, \sElm)$, it holds that $\GName,
			\spcFun(\asgFun) \cmp \bndFun[\bpElm], \sElm \emodels \psi$.
			Consequently, by Definition~\ref{def:ngsl(elementarysemantics)} of \NGSL\
			elementary semantics, we have that $\GName, \emptyfun, \sElm \emodels
			\qpSym \bpElm \psi$.
		\end{proof}

		As an immediate consequence of the previous theorem, we derive the following
		fundamental corollary.
		\begin{corollary}[\OGSL\ Elementariness]
			\label{cor:ogsl(elm)}
			Let $\GName$ be a \CGS\ and $\varphi$ an \OGSL\ sentence.
			Then, $\GName \models \varphi$ iff $\GName \emodels \varphi$.
		\end{corollary}

		It is worth to observe that the elementariness property for \OGSL\ is a
		crucial difference w.r.t.\ \BGSL, which allows us to obtain an elementary
		decision procedure for the related model-checking problem, as described in
		the last part of the next section.

	\end{subsection}

\end{section}




\begin{section}{Model-Checking Procedures}
	\label{sec:modchkprc}

	In this section, we study the model-checking problem for \SL\ and show that,
	in general, it is non-elementarily decidable, while, in the particular case of
	\OGSL\ sentences, it is just 2\ExpTimeC, as for \ATLS.
	For the algorithmic procedures, we follow an \emph{automata-theoretic
	approach}~\cite{KVW00}, reducing the decision problem for the logics to the
	emptiness problem of an automaton.
	In particular, we use a bottom-up technique through which we recursively
	label each state of the \CGS\ of interest by all principal subsentences of the
	specification that are satisfied on it, starting from the innermost
	subsentences and terminating with the sentence under exam.
	In this way, at a given step of the recursion, since the satisfaction of all
	subsentences of the given principal sentence has already been determined, we
	can assume that the matrix of the latter is only composed by Boolean
	combinations and nesting of goals whose temporal part is simply \LTL.
	The procedure we propose here extends that used for \ATLS\ in~\cite{AHK02} by
	means of a richer structure of the automata involved in.

	The rest of this section is organized as follows.
	In Subsection~\ref{subsec:altaut}, we recall the definition of alternating
	parity tree automata.
	Then, in Subsection~\ref{subsec:slmodchk}, we build an automaton accepting a
	tree encoding of a \CGS\ iff this satisfies the formula of interest, which is
	used to prove the main result about \SL\ and \NGSL\ model checking.
	Finally, in Subsection~\ref{subsec:ogslmodchk}, we refine the previous result
	to obtain an elementary decision procedure for \OGSL.

	\begin{subsection}{Alternating tree automata}
		\label{subsec:altaut}

		\emph{Nondeterministic tree automata} are a generalization to infinite trees
		of the classical \emph{nondeterministic word automata} on infinite words
		(see~\cite{Tho90}, for an introduction).
		\emph{Alternating tree automata} are a further generalization of
		nondeterministic tree automata~\cite{MS87}.
		Intuitively, on visiting a node of the input tree, while the latter sends
		exactly one copy of itself to each of the successors of the node, the former
		can send several own copies to the same successor.
		Here we use, in particular, \emph{alternating parity tree automata}, which
		are alternating tree automata along with a \emph{parity acceptance
		condition} (see~\cite{GTW02}, for a survey).

		We now give the formal definition of alternating tree automata.
		\begin{definition}[Alternating Tree Automata]
			\label{def:ata}
			An \emph{alternating tree automaton} (\emph{\ATA}, for short) is a tuple
			$\AName \defeq \ATAStruct$, where $\LabSet$, $\DirSet$, and $\QSet$ are,
			respectively, non-empty finite sets of \emph{input symbols},
			\emph{directions}, and \emph{states}, $\qElm[0] \in \QSet$ is an
			\emph{initial state}, $\aleph$ is an \emph{acceptance condition} to be
			defined later, and $\delta : \QSet \times \LabSet \to \PBoolSet(\DirSet
			\times \QSet)$ is an \emph{alternating transition function} that maps each
			pair of states and input symbols to a positive Boolean combination on the
			set of propositions of the form $(\dElm, \qElm) \in \DirSet \times \QSet$,
			a.k.a. \emph{moves}.
		\end{definition}

		On one side, a \emph{nondeterministic tree automaton} (\emph{\NTA}, for
		short) is a special case of \ATA\ in which each conjunction in the
		transition function $\delta$ has exactly one move $(\dElm, \qElm)$
		associated with each direction $\dElm$.
		This means that, for all states $\qElm \in \QSet$ and symbols $\sigma \in
		\LabSet$, we have that $\atFun(\qElm, \sigma)$ is equivalent to a Boolean
		formula of the form $\bigvee_{i} \bigwedge_{\dElm \in \DirSet} (\dElm,
		\qElm[i, \dElm])$.
		On the other side, a \emph{universal tree automaton} (\emph{\UTA}, for
		short) is a special case of \ATA\ in which all the Boolean combinations that
		appear in $\delta$ are conjunctions of moves.
		Thus, we have that $\atFun(\qElm, \sigma) = \bigwedge_{i} (\dElm[i],
		\qElm[i])$, for all states $\qElm \in \QSet$ and symbols $\sigma \in
		\LabSet$.

		The semantics of the \ATA s is given through the following concept of run.
		\begin{definition}[\ATA\ Run]
			\label{def:ata(run)}
			A \emph{run} of an \ATA\ $\AName = \ATAStruct$ on a $\LabSet$-labeled
			$\DirSet$-tree $\TName = \LTStruct$ is a $(\DirSet \times \QSet)$-tree
			$\RSet$ such that, for all nodes $\xElm \in \RSet$, where $\xElm =
			\prod_{i = 1}^{n} (\dElm[i], \qElm[i])$ and $\yElm \defeq \prod_{i =
			1}^{n} \dElm[i]$ with $n \in \numco{0}{\omega}$, it holds that \emph{(i)}
			$\yElm \in \TSet$ and \emph{(ii)}, there is a set of moves $\SSet
			\subseteq \DirSet \times \QSet$ with $\SSet \models \delta(\qElm[n],
			\vFun(\yElm))$ such that $\xElm \cdot (\dElm, \qElm) \in \RSet$, for all
			$(\dElm, \qElm) \in \SSet$.
		\end{definition}

		In the following, we consider \ATA s along with the \emph{parity acceptance
		condition} (\emph{\APT}, for short) $\aleph \defeq (\FSet_{1}, \ldots,
		\FSet_{k}) \in (\pow{\QSet})^{+}$ with $\FSet_{1} \subseteq \ldots \subseteq
		\FSet_{k} = \QSet$ (see~\cite{KVW00}, for more).
		The number $k$ of sets in the tuple $\aleph$ is called the \emph{index} of
		the automaton.
		We also consider \ATA s with the \emph{co-B\"uchi acceptance condition}
		(\emph{\ACT}, for short) that is the special parity condition with index
		$2$.

		Let $\RSet$ be a run of an \ATA\ $\AName$ on a tree $\TName$ and $\wElm$ one
		of its branches.
		Then, by $\infFun(\wElm) \defeq \set{ \qElm \in \QSet }{ \card{\set{ i \in
		\SetN }{ \exists \dElm \in \DirSet . (w)_{i} = (\dElm, \qElm) }} = \omega }$
		we denote the set of states that occur infinitely often as the second
		component of the letters along the branch $w$.
		Moreover, we say that $w$ satisfies the parity acceptance condition $\aleph
		\!=\! (\FSet_{1}, \ldots, \FSet_{k})$ if the least index $i \!\in\!
		\numcc{1}{k}$ for which $\infFun(w) \cap \FSet_{i} \neq \emptyset$ is even.

		At this point, we can define the concept of language accepted by an \ATA.
		\begin{definition}[\ATA\ Acceptance]
			\label{def:ata(acp)}
			An \ATA\ $\AName = \ATAStruct$ \emph{accepts} a $\LabSet$-labeled
			$\DirSet$-tree $\TName$ iff is there exists a run $\RSet$ of $\AName$ on
			$\TName$ such that all its infinite branches satisfy the acceptance
			condition $\aleph$.
		\end{definition}
		By $\LangSet(\AName)$ we denote the language accepted by the \ATA\ $\AName$,
		i.e., the set of trees $\TName$ accepted by $\AName$.
		Moreover, $\AName$ is said to be \emph{empty} if $\LangSet(\AName) =
		\emptyset$.
		The \emph{emptiness problem} for $\AName$ is to decide whether
		$\LangSet(\AName) = \emptyset$.

		We finally show a simple but useful result about the \APT\ direction
		projection.
		To do this, we first need to introduce an extra notation.
		Let $\fElm \in \BoolSet(\PSet)$ be a Boolean formula on a set of
		propositions $\PSet$.
		Then, by $\fElm[][\pElm / \qElm \:\!|\:\! \pElm \in \PSet']$ we denote the
		formula in which all occurrences of the propositions $\pElm \in \PSet'
		\subseteq \PSet$ in $\fElm$ are replaced by the proposition $\qElm$
		belonging to a possibly different set.
		\begin{theorem}[\APT\ Direction Projection]
			\label{thm:apt(dirprj)}
			Let $\AName \defeq \TATuple {\LabSet \times \DirSet} {\DirSet} {\QSet}
			{\atFun} {\qElm[0]} {\aleph}$ be an \APT\ over a set of $m$ directions
			with $n$ states and index $k$.
			Moreover, let $\dElm[0] \in \DirSet$ be a distinguished direction.
			Then, there exists an \NPT\ $\NName^{\dElm[0]} \defeq \TATuple {\LabSet}
			{\DirSet} {\QSet'} {\atFun'} {\qElm[0 | ']} {\aleph'}$ with $m \cdot
			2^{\AOmicron{k \cdot n \cdot \log n}}$ states and index $\AOmicron{k \cdot
			n \cdot \log n}$ such that, for all $\LabSet$-labeled $\DirSet$-tree
			$\TName \defeq \LTStruct$, it holds that $\TName \in
			\LangSet(\NName^{\dElm[0]})$ iff $\TName' \in \LangSet(\AName)$, where
			$\TName'$ is the $(\LabSet \times \DirSet)$-labeled $\DirSet$-tree
			$\LTTuple{}{}{\TSet}{\vFun'}$ such that $\vFun'(\tElm) \defeq (\vFun(t),
			\lst{\dElm[0] \cdot t})$, for all $\tElm \in \TSet$.
		\end{theorem}
		\begin{proof}
			As first step, we use the well-known nondeterminization procedure for
			\APT s~\cite{MS95} in order to transform the \APT\ $\AName$ into an
			equivalent \NPT\ $\NName = \TATuple {\LabSet \times \DirSet} {\DirSet}
			{\QSet''} {\atFun''} {\qElm[0 | '']} {\aleph''}$ with $2^{\AOmicron{k
			\cdot n \cdot \log n}}$ states and index $k' = \AOmicron{k \cdot n \cdot
			\log n}$.
			Then, we transform the latter into the new \NPT\ $\NName^{\dElm[0]} \defeq
			\TATuple {\LabSet} {\DirSet} {\QSet'} {\atFun'} {\qElm[0 | ']} {\aleph'}$
			with $m \cdot 2^{\AOmicron{k \cdot n \cdot \log n}}$ states and same index
			$k'$, where $\QSet' \defeq \QSet'' \times \DirSet$, $\qElm[0 | '] \defeq
			(\qElm[0 | ''], \dElm[0])$, $\aleph' \defeq (\FSet[1] \times \DirSet,
			\ldots, \FSet[k'] \times \DirSet)$ with $\aleph'' \defeq (\FSet[1],
			\ldots, \FSet[k'])$, and $\atFun'((\qElm, \dElm), \sigma) \defeq
			\atFun''(\qElm, (\sigma, \dElm)) [(\dElm', \qElm') / (\dElm', (\qElm',
			\dElm')) \:\!|\:\! (\dElm', \qElm') \in \DirSet \times \QSet'']$, for all
			$(\qElm, \dElm) \in \QSet'$ and $\sigma \in \LabSet$.
			Now, it easy to see that $\NName^{\dElm[0]}$ satisfies the declared
			statement.
		\end{proof}

	\end{subsection}

	\begin{subsection}{\SL\ Model Checking}
		\label{subsec:slmodchk}

		A first step towards our construction of an algorithmic procedure for the
		solution of the \SL\ model-checking problem is to define, for each possible
		formula $\varphi$, an alternating parity tree automaton $\AName[\varphi |
		^{\GName}]$ that recognizes a tree encoding $\TName$ of a \CGS\
		$\GName$, containing the information on an assignment $\asgFun$ on the free
		variables/agents of $\varphi$, iff $\GName$ is a model of $\varphi$ under
		$\asgFun$.
		The high-level idea at the base of this construction is an evolution and
		merging of those behind the translations of \QPTL\ and \LTL, respectively,
		\mbox{into nondeterministic~\cite{SVW87} and alternating~\cite{MSS88}
		B\"uchi automata.}

		To proceed with the formal description of the model-checking procedure, we
		have to introduce a concept of encoding for the assignments of a \CGS.
		\begin{definition}[Assignment-State Encoding]
			\label{def:asgstenc}
			Let $\GName$ be a \CGS, $\sElm \in \StSet[\GName]$ one of its states, and
			$\asgFun \in \AsgSet[\GName](\VSet, \sElm)$ an assignment defined on the
			set $\VSet \subseteq \VarSet \cup \AgSet$.
			Then, a $(\ValSet[ {\AcSet[\GName]} ](\VSet) \times
			\StSet[\GName])$-labeled $\StSet[\GName]$-tree $\TName \defeq
			\LTTuple{}{}{\TSet}{\uFun}$, where $\TSet \defeq \set{ \trkElm_{\geq 1} }{
			\trkElm \in \TrkSet[\GName](\sElm) }$, is an \emph{assignment-state
			encoding} for $\asgFun$ if it holds that $\uFun(\tElm) \defeq
			(\flip{\asgFun}(\sElm \cdot \tElm), \lst{\sElm \cdot \tElm})$, for all
			$\tElm \in \TSet$.
		\end{definition}
		Observe that there is a unique assignment-state encoding for each given
		assignment.

		In the next lemma, we prove the existence of an \APT\ for each \CGS\ and
		\SL\ formula that is able to recognize all the assignment-state encodings of
		an a priori given assignment, made the assumption that the formula is
		satisfied on the \CGS\ under this assignment.
		\begin{lemma}[\SL\ Formula Automaton]
			\label{lmm:sl(frmaut)}
			Let $\GName$ be a \CGS\ and $\varphi$ an \SL\ formula.
			Then, there exists an \APT\ $\AName[\varphi | ^{\GName}] \defeq \TATuple
			{\ValSet[ {\AcSet[\GName]} ](\free{\varphi}) \times \StSet[\GName]}
			{\StSet[\GName]} {\QSet[\varphi]} {\atFun[\varphi]} {\qElm[0\varphi]}
			{\aleph_{\varphi}}$ such that, for all states $\sElm \in \StSet[\GName]$
			and assignments $\asgFun \in \AsgSet[\GName](\free{\varphi}, \sElm)$, it
			holds that $\GName, \asgFun, \sElm \models \varphi$ iff $\TName \in
			\LangSet(\AName[\varphi | ^{\GName}])$, where $\TName$ is the
			assignment-state encoding for $\asgFun$.
		\end{lemma}
		\begin{proof}
			The construction of the \APT\ $\AName[\varphi | ^{\GName}]$ is done
			recursively on the structure of the formula $\varphi$, which w.l.o.g.\
			is supposed to be in \enf, by using a variation of the transformation,
			via alternating tree automata, of the S$1$S and S$k$S logics into
			nondeterministic B\"uchi word and tree automata recognizing all models of
			the formula of interest~\cite{Buc62,Rab69}.

			The detailed construction of $\AName[\varphi | ^{\GName}]$, by a case
			analysis on $\varphi$, follows.
			\begin{itemize}
				\item
					If $\varphi \in \APSet$, the automaton has to verify if the atomic
					proposition is locally satisfied or not.
					To do this, we set $\AName[\varphi | ^{\GName}] \defeq \TATuple
					{\ValSet[ {\AcSet[\GName]} ](\emptyset) \times \StSet[\GName]}
					{\StSet[\GName]} {\{ \varphi \}} {\atFun[\varphi]} {\varphi} {(\{
					\varphi \})}$, where $\atFun[\varphi](\varphi, (\valFun, \sElm))
					\defeq \Tt$, if $\varphi \in \labFun[\GName](\sElm)$, and
					$\atFun[\varphi](\varphi, (\valFun, \sElm)) \defeq \Ff$, otherwise.
					Intuitively, $\AName[\varphi | ^{\GName}]$ only verifies that the
					state $\sElm$ in the labeling of the root of the assignment-state
					encoding of the empty assignment $\emptyfun$ satisfies $\varphi$.
				\item
					The boolean case $\varphi = \neg \varphi'$ is treated in the classical
					way, by simply dualizing the automaton $\AName[\varphi' | ^{\GName}] =
					\TATuple {\ValSet[ {\AcSet[\GName]} ](\free{\varphi'}) \times
					\StSet[\GName]} {\StSet[\GName]} {\QSet[\varphi']} {\atFun[\varphi']}
					{\qElm[0\varphi']} {\aleph_{\varphi'}}$~\cite{MS87}.
				\item
					The boolean cases $\varphi = \varphi_{1} \Opr \varphi_{2}$, with $\Opr
					\in \{ \wedge, \vee \}$, are treated in a way that is similar to the
					classical one, by simply merging the two automata $\AName[\varphi_{1}
					| ^{\GName}] = \TATuple {\ValSet[ {\AcSet[\GName]}
					](\free{\varphi_{1}}) \times \StSet[\GName]} {\StSet[\GName]}
					{\QSet[\varphi_{1}]} {\atFun[\varphi_{1}]} {\qElm[0\varphi_{1}]}
					{\aleph_{\varphi_{1}}}$ and $\AName[\varphi_{2} | ^{\GName}] =
					\TATuple {\ValSet[ {\AcSet[\GName]} ](\free{\varphi_{2}}) \times
					\StSet[\GName]} {\StSet[\GName]} {\QSet[\varphi_{2}]}
					{\atFun[\varphi_{2}]} {\qElm[0\varphi_{2}]} {\aleph_{\varphi_{2}}}$
					into the automaton $\AName[\varphi | ^{\GName}] \defeq \TATuple
					{\ValSet[ {\AcSet[\GName]} ](\free{\varphi}) \times \StSet[\GName]}
					{\StSet[\GName]} {\QSet[\varphi]} {\atFun[\varphi]} {\qElm[0\varphi]}
					{\aleph_{\varphi}}$, where the following hold:
					\begin{itemize}
						\item
							$\QSet[\varphi] \defeq \{ \qElm[0\varphi] \} \cup
							\QSet[\varphi_{1}] \cup \QSet[\varphi_{2}]$, with $\qElm[0\varphi]
							\not\in \QSet[\varphi_{1}] \cup \QSet[\varphi_{2}]$;
						\item
							$\atFun[\varphi](\qElm[0\varphi], (\valFun, \sElm)) \defeq
							\atFun[\varphi_{1}](\qElm[0\varphi_{1}], (\valFun_{\rst
							\free{\varphi_{1}}}, \sElm)) \:\Opr
							\atFun[\varphi_{2}](\qElm[0\varphi_{2}], (\valFun_{\rst
							\free{\varphi_{2}}}, \sElm))$, for all $(\valFun, \sElm) \in
							\ValSet[ {\AcSet[\GName]} ](\free{\varphi}) \times
							\StSet[\GName]$;
						\item
							$\atFun[\varphi](\qElm, (\valFun, \sElm)) \defeq
							\atFun[\varphi_{1}](\qElm, (\valFun_{\rst \free{\varphi_{1}}},
							\sElm))$, if $\qElm \in \QSet[\varphi_{1}]$, and
							$\atFun[\varphi](\qElm, (\valFun, \sElm)) \defeq
							\atFun[\varphi_{2}](\qElm, (\valFun_{\rst \free{\varphi_{2}}},
							\sElm))$, otherwise, for all $\qElm \in \QSet[\varphi_{1}] \cup
							\QSet[\varphi_{2}]$ and $(\valFun, \sElm) \in \ValSet[
							{\AcSet[\GName]} ](\free{\varphi}) \times \StSet[\GName]$;
						\item
							$\aleph_{\varphi} \defeq (\FSet[1\varphi], \ldots,
							\FSet[k\varphi])$, where \emph{(i)} $\aleph_{\varphi_{1}} \defeq
							(\FSet[1\varphi_{1}], \ldots, \FSet[k_{1}\varphi_{1}])$ and
							$\aleph_{\varphi_{2}} \defeq (\FSet[1\varphi_{2}], \ldots,
							\FSet[k_{2}\varphi_{2}])$, \emph{(ii)} $h = \min \{ k_{1}, k_{2}
							\}$ and $k = \max \{ k_{1}, k_{2} \}$, \emph{(iii)}
							$\FSet[i\varphi] \defeq \FSet[i\varphi_{1}] \cup
							\FSet[i\varphi_{2}]$, for $i \in \numcc{1}{h}$, \emph{(iv)}
							$\FSet[i\varphi] \defeq \FSet[i\varphi_{j}]$, for $i \in
							\numcc{h + 1}{k - 1}$ with $k_{j} = k$, and \emph{(v)}
							$\FSet[k\varphi] \defeq \QSet[\varphi]$.
					\end{itemize}
				\item
					The case $\varphi = \X \varphi'$ is solved by running the automaton
					$\AName[\varphi' | ^{\GName}] = \TATuple {\ValSet[ {\AcSet[\GName]}
					](\free{\varphi'}) \times \StSet[\GName]} {\StSet[\GName]}
					{\QSet[\varphi']} {\atFun[\varphi']} {\qElm[0\varphi']}
					{\aleph_{\varphi'}}$ on the successor node of the root of the
					assignment-state encoding in the direction individuated by the
					assignment itself.
					To do this, we use the automaton $\AName[\varphi | ^{\GName}] \defeq
					\TATuple {\ValSet[ {\AcSet[\GName]} ](\free{\varphi}) \times
					\StSet[\GName]} {\StSet[\GName]} {\QSet[\varphi]} {\atFun[\varphi]}
					{\qElm[0\varphi]} {\aleph_{\varphi}}$, where the following hold:
					\begin{itemize}
						\item
							$\QSet[\varphi] \defeq \{ \qElm[0\varphi] \} \cup
							\QSet[\varphi']$, with $\qElm[0\varphi] \not\in \QSet[\varphi']$;
						\item
							$\atFun[\varphi](\qElm[0\varphi], (\valFun, \sElm)) \defeq
							(\trnFun[\GName](\sElm, \valFun_{\rst \AgSet}),
							\qElm[0\varphi'])$, for all $(\valFun, \sElm) \in \ValSet[
							{\AcSet[\GName]} ](\free{\varphi}) \times \StSet[\GName]$;
						\item
							$\atFun[\varphi](\qElm, (\valFun, \sElm)) \defeq
							\atFun[\varphi'](\qElm, (\valFun_{\rst \free{\varphi'}}, \sElm))$,
							for all $\qElm \in \QSet[\varphi']$ and $(\valFun, \sElm) \in
							\ValSet[ {\AcSet[\GName]} ](\free{\varphi}) \times
							\StSet[\GName]$;
						\item
							$\aleph_{\varphi} \defeq (\FSet[1\varphi'], \ldots,
							\FSet[k\varphi'] \cup \{ \qElm[0\varphi] \})$, where
							$\aleph_{\varphi'} \defeq (\FSet[1\varphi'], \ldots,
							\FSet[k\varphi'])$.
					\end{itemize}
				\item
					To handle the case $\varphi = \varphi_{1} \U \varphi_{2}$, we use
					the automaton $\AName[\varphi | ^{\GName}] \defeq \TATuple {\ValSet[
					{\AcSet[\GName]} ](\free{\varphi}) \times \StSet[\GName]}
					{\StSet[\GName]} {\QSet[\varphi]} {\atFun[\varphi]} {\qElm[0\varphi]}
					{\aleph_{\varphi}}$ that verifies the truth of the until operator
					using its one-step unfolding equivalence $\varphi_{1} \U \varphi_{2}
					\equiv \varphi_{2} \vee \varphi_{1} \wedge \X \varphi_{1} \U
					\varphi_{2}$, by appropriately running the two automata
					$\AName[\varphi_{1} | ^{\GName}] = \TATuple {\ValSet[ {\AcSet[\GName]}
					](\free{\varphi_{1}}) \times \StSet[\GName]} {\StSet[\GName]}
					{\QSet[\varphi_{1}]} {\atFun[\varphi_{1}]} {\qElm[0\varphi_{1}]}
					{\aleph_{\varphi_{1}}}$ and $\AName[\varphi_{2} | ^{\GName}] =
					\TATuple {\ValSet[ {\AcSet[\GName]} ](\free{\varphi_{2}}) \times
					\StSet[\GName]} {\StSet[\GName]} {\QSet[\varphi_{2}]}
					{\atFun[\varphi_{2}]} {\qElm[0\varphi_{2}]} {\aleph_{\varphi_{2}}}$
					for the inner formulas $\varphi_{1}$ and $\varphi_{2}$.
					\mbox{The definitions of $\AName[\varphi | ^{\GName}]$ components
					follows:}
					\begin{itemize}
						\item
							$\QSet[\varphi] \defeq \{ \qElm[0\varphi] \} \cup
							\QSet[\varphi_{1}] \cup \QSet[\varphi_{2}]$, with $\qElm[0\varphi]
							\not\in \QSet[\varphi_{1}] \cup \QSet[\varphi_{2}]$;
						\item
							$\atFun[\varphi](\qElm[0\varphi], (\valFun, \sElm)) \defeq
							\atFun[\varphi_{2}](\qElm[0\varphi_{2}], (\valFun_{\rst
							\free{\varphi_{2}}}, \sElm)) \vee
							\atFun[\varphi_{1}](\qElm[0\varphi_{1}], (\valFun_{\rst
							\free{\varphi_{1}}}, \sElm)) \wedge (\trnFun[\GName](\sElm,
							\valFun_{\rst \AgSet}), \qElm[0\varphi])$, for all $(\valFun,
							\sElm) \in \ValSet[ {\AcSet[\GName]} ](\free{\varphi}) \times
							\StSet[\GName]$;
						\item
							$\atFun[\varphi](\qElm, (\valFun, \sElm)) \defeq
							\atFun[\varphi_{1}](\qElm, (\valFun_{\rst \free{\varphi_{1}}},
							\sElm))$, if $\qElm \in \QSet[\varphi_{1}]$, and
							$\atFun[\varphi](\qElm, (\valFun, \sElm)) \defeq
							\atFun[\varphi_{2}](\qElm, (\valFun_{\rst \free{\varphi_{2}}},
							\sElm))$, otherwise, for all $\qElm \in \QSet[\varphi_{1}] \cup
							\QSet[\varphi_{2}]$ and $(\valFun, \sElm) \in \ValSet[
							{\AcSet[\GName]} ](\free{\varphi}) \times \StSet[\GName]$;
						\item
							$\aleph_{\varphi} \defeq (\FSet[1\varphi], \ldots,
							\FSet[k\varphi])$, where \emph{(i)} $\aleph_{\varphi_{1}} \defeq
							(\FSet[1\varphi_{1}], \ldots, \FSet[k_{1}\varphi_{1}])$ and
							$\aleph_{\varphi_{2}} \defeq (\FSet[1\varphi_{2}], \ldots,
							\FSet[k_{2}\varphi_{2}])$, \emph{(ii)} $h = \min \{ k_{1}, k_{2}
							\}$ and $k = \max \{ k_{1}, k_{2} \}$, \emph{(iii)}
							$\FSet[i\varphi] \defeq \{ \qElm[0\varphi] \} \cup
							\FSet[i\varphi_{1}] \cup \FSet[i\varphi_{2}]$, for $i \in
							\numcc{1}{h}$, \emph{(iv)} $\FSet[i\varphi] \defeq \{
							\qElm[0\varphi] \} \cup \FSet[i\varphi_{j}]$, for $i \in \numcc{h
							+ 1}{k - 1}$ with $k_{j} = k$, and \emph{(v)} $\FSet[k\varphi]
							\defeq \QSet[\varphi]$.
					\end{itemize}
					It is important to observe that the initial state $\qElm[0\varphi]$ is
					included in all sets of the parity acceptance condition, in particular
					in $\FSet[1\varphi]$, in order to avoid its regeneration for an
					infinite number of times.
				\item
					To handle the case $\varphi = \varphi_{1} \R \varphi_{2}$, we use
					the automaton $\AName[\varphi | ^{\GName}] \defeq \TATuple {\ValSet[
					{\AcSet[\GName]} ](\free{\varphi}) \times \StSet[\GName]}
					{\StSet[\GName]} {\QSet[\varphi]} {\atFun[\varphi]} {\qElm[0\varphi]}
					{\aleph_{\varphi}}$ that verifies the truth of the release operator
					using its one-step unfolding equivalence $\varphi_{1} \R \varphi_{2}
					\equiv \varphi_{2} \wedge (\varphi_{1} \vee \X \varphi_{1} \R
					\varphi_{2})$, by appropriately running the two automata
					$\AName[\varphi_{1} | ^{\GName}] = \TATuple {\ValSet[ {\AcSet[\GName]}
					](\free{\varphi_{1}}) \times \StSet[\GName]} {\StSet[\GName]}
					{\QSet[\varphi_{1}]} {\atFun[\varphi_{1}]} {\qElm[0\varphi_{1}]}
					{\aleph_{\varphi_{1}}}$ and $\AName[\varphi_{2} | ^{\GName}] =
					\TATuple {\ValSet[ {\AcSet[\GName]} ](\free{\varphi_{2}}) \times
					\StSet[\GName]} {\StSet[\GName]} {\QSet[\varphi_{2}]}
					{\atFun[\varphi_{2}]} {\qElm[0\varphi_{2}]} {\aleph_{\varphi_{2}}}$
					for the inner formulas $\varphi_{1}$ and $\varphi_{2}$.
					\mbox{The definitions of $\AName[\varphi | ^{\GName}]$ components
					follows:}
					\begin{itemize}
						\item
							$\QSet[\varphi] \defeq \{ \qElm[0\varphi] \} \cup
							\QSet[\varphi_{1}] \cup \QSet[\varphi_{2}]$, with $\qElm[0\varphi]
							\not\in \QSet[\varphi_{1}] \cup \QSet[\varphi_{2}]$;
						\item
							$\atFun[\varphi](\qElm[0\varphi], (\valFun, \sElm)) \defeq
							\atFun[\varphi_{2}](\qElm[0\varphi_{2}], (\valFun_{\rst
							\free{\varphi_{2}}}, \sElm)) \wedge
							(\atFun[\varphi_{1}](\qElm[0\varphi_{1}], (\valFun_{\rst
							\free{\varphi_{1}}}, \sElm)) \vee (\trnFun[\GName](\sElm,
							\valFun_{\rst \AgSet}), \qElm[0\varphi]))$, for all $(\valFun,
							\sElm) \in \ValSet[ {\AcSet[\GName]} ](\free{\varphi}) \times
							\StSet[\GName]$;
						\item
							$\atFun[\varphi](\qElm, (\valFun, \sElm)) \defeq
							\atFun[\varphi_{1}](\qElm, (\valFun_{\rst \free{\varphi_{1}}},
							\sElm))$, if $\qElm \in \QSet[\varphi_{1}]$, and
							$\atFun[\varphi](\qElm, (\valFun, \sElm)) \defeq
							\atFun[\varphi_{2}](\qElm, (\valFun_{\rst \free{\varphi_{2}}},
							\sElm))$, otherwise, for all $\qElm \in \QSet[\varphi_{1}] \cup
							\QSet[\varphi_{2}]$ and $(\valFun, \sElm) \in \ValSet[
							{\AcSet[\GName]} ](\free{\varphi}) \times \StSet[\GName]$;
						\item
							$\aleph_{\varphi} \defeq (\FSet[1\varphi], \ldots,
							\FSet[k\varphi])$, where \emph{(i)} $\aleph_{\varphi_{1}} \defeq
							(\FSet[1\varphi_{1}], \ldots, \FSet[k_{1}\varphi_{1}])$ and
							$\aleph_{\varphi_{2}} \defeq (\FSet[1\varphi_{2}], \ldots,
							\FSet[k_{2}\varphi_{2}])$, \emph{(ii)} $h = \min \{ k_{1}, k_{2}
							\}$ and $k = \max \{ k_{1}, k_{2} \}$, \emph{(iii)}
							$\FSet[1\varphi] \defeq \FSet[1\varphi_{1}] \cup
							\FSet[1\varphi_{2}]$, \emph{(iv)} $\FSet[i\varphi] \defeq \{
							\qElm[0\varphi] \} \cup \FSet[i\varphi_{1}] \cup
							\FSet[i\varphi_{2}]$, for $i \!\in\! \numcc{2}{h}$, \emph{(iv)}
							$\FSet[i\varphi] \!\defeq\! \{ \qElm[0\varphi] \} \cup
							\FSet[i\varphi_{j}]$, for $i \!\in\! \numcc{h + 1}{k - 1}$ with
							$k_{j} \!=\! k$, and \emph{(v)} $\FSet[k\varphi] \!\defeq\!
							\QSet[\varphi]$.
					\end{itemize}
					It is important to observe that, differently from the case of the
					until operator, the initial state $\qElm[0\varphi]$ is included in all
					sets of the parity acceptance condition but $\FSet[1\varphi]$, in
					order to allow its regeneration for an infinite number of time.
				\item
					The case $\varphi = (\aElm, \xElm) \varphi'$ is solved by simply
					transforming the transition function of the automaton $\AName[\varphi'
					| ^{\GName}] = \TATuple {\ValSet[ {\AcSet[\GName]} ](\free{\varphi'})
					\times \StSet[\GName]} {\StSet[\GName]} {\QSet[\varphi']}
					{\atFun[\varphi']} {\qElm[0\varphi']} {\aleph_{\varphi'}}$, by setting
					the value of the valuations in input w.r.t.\ the agent $\aElm$ to the
					value of the same valuation w.r.t.\ the variable $\xElm$.
					The definitions of the transition function for $\AName[\varphi |
					^{\GName}] \defeq \TATuple {\ValSet[ {\AcSet[\GName]}
					](\free{\varphi}) \times \StSet[\GName]} {\StSet[\GName]}
					{\QSet[\varphi']} {\atFun[\varphi]} {\qElm[0\varphi']}
					{\aleph_{\varphi'}}$ follows: $\atFun[\varphi](\qElm, (\valFun,
					\sElm)) \defeq \atFun[\varphi'](\qElm, (\valFun', \sElm))$, where
					$\valFun' = \valFun[][\aElm \mapsto \valFun(\xElm)]_{\rst
					\free{\varphi'}}$, if $\aElm \in \free{\varphi'}$, and $\valFun' =
					\valFun$, otherwise, for all $\qElm \in \QSet[\varphi']$ and
					$(\valFun, \sElm) \in \ValSet[ {\AcSet[\GName]} ](\free{\varphi})
					\times \StSet[\GName]$.
				\item
					To handle the case $\varphi = \EExs{\xElm} \varphi'$, assuming that
					$\xElm \in \free{\varphi'}$, we use the operation of existential
					projection for nondeterministic tree automata.
					To do this, we have first to nondeterminize the \APT\ $\AName[\varphi'
					| ^{\GName}]$, by applying the classic transformation~\cite{MS95}.
					In this way, we obtain an equivalent \NPT\ $\NName[\varphi' |
					^{\GName}] = \TATuple {\ValSet[ {\AcSet[\GName]} ](\free{\varphi'})
					\times \StSet[\GName]} {\StSet[\GName]} {\QSet[\varphi']}
					{\atFun[\varphi']} {\qElm[0\varphi']} {\aleph_{\varphi'}}$.
					Now, we make the projection, by defining the new \NPT\ $\AName[\varphi
					| ^{\GName}] \defeq \TATuple {\ValSet[ {\AcSet[\GName]}
					](\free{\varphi}) \times \StSet[\GName]} {\StSet[\GName]}
					{\QSet[\varphi']} {\atFun[\varphi]} {\qElm[0\varphi']}
					{\aleph_{\varphi'}}$ where $\atFun[\varphi](\qElm, (\valFun, \sElm))
					\defeq \bigvee_{\cElm \in \AcSet[\GName]} \atFun[\varphi'](\qElm,
					(\valFun[][\xElm \mapsto \cElm], \sElm))$, for all $\qElm \in
					\QSet[\varphi']$ and $(\valFun, \sElm) \in \ValSet[ {\AcSet[\GName]}
					](\free{\varphi}) \times \StSet[\GName]$.
			\end{itemize}

			At this point, it only remains to prove that, for all states $\sElm \in
			\StSet[\GName]$ and assignments $\asgFun \in
			\AsgSet[\GName](\free{\varphi}, \sElm)$, it holds that $\GName, \asgFun,
			\sElm \models \varphi$ iff $\TName \in \LangSet(\AName[\varphi |
			^{\GName}])$, where $\TName$ is the assignment-state encoding for
			$\asgFun$.
			The proof can be developed by a simple induction on the structure of the
			formula $\varphi$ and is left to the reader as a simple exercise.
		\end{proof}

		We now have the tools to describe the recursive model-checking procedure on
		nested subsentences for \SL\ and its fragments under the general semantics.

		To proceed, we have first to prove the following theorem that represents the
		core of our automata-theoretic approach.
		\begin{theorem}[\SL\ Sentence Automaton]
			\label{thm:sl(sntaut)}
			Let $\GName$ be a \CGS, $\sElm \in \StSet[\GName]$ one of its states, and
			$\varphi$ an \SL\ sentence.
			Then, there exists an \NPT\ $\NName[\varphi | ^{\GName, \sElm}]$ such that
			$\GName, \emptyfun, \sElm \models \varphi$ iff $\LangSet(\NName[\varphi |
			^{\GName, \sElm}]) \neq \emptyset$.
		\end{theorem}
		\begin{proof}
			To construct the \NPT\ $\NName[\varphi | ^{\GName, \sElm}]$ we apply
			Theorem~\ref{thm:apt(dirprj)} of \APT\ direction projection with
			distinguished direction $\sElm$ to the \APT\ $\AName[\varphi | ^{\GName}]$
			derived by Lemma~\ref{lmm:sl(frmaut)} of \SL\ formula automaton.
			In this way, we can ensure that the state labeling of nodes of the
			assignment-state encoding is coherent with the node itself.
			Observe that, since $\varphi$ is a sentence, we have that $\free{\varphi}
			= \emptyset$, and so, the unique assignment-state encoding of interest is
			that related to the empty assignment $\emptyfun$.

			\emph{[Only if].}
			Suppose that $\GName, \emptyfun, \sElm \models \varphi$.
			Then, by Lemma~\ref{lmm:sl(frmaut)}, we have that $\TName \in
			\LangSet(\AName[\varphi | ^{\GName}])$, where $\TName$ is the elementary
			dependence-state encoding for $\emptyfun$.
			Hence, by Theorem~\ref{thm:apt(dirprj)}, it holds that
			$\LangSet(\NName[\varphi | ^{\GName, \sElm}]) \neq \emptyset$.

			\emph{[If].}
			Suppose that $\LangSet(\NName[\varphi | ^{\GName, \sElm}]) \neq
			\emptyset$.
			Then, by Theorem~\ref{thm:apt(dirprj)}, there exists an $( \{ \emptyfun \}
			\times \StSet[\GName])$-labeled $\StSet[\GName]$-tree $\TName$ such that
			$\TName \in \LangSet(\AName[\varphi | ^{\GName}])$.
			Now, it is immediate to see that $\TName$ is the assignment-state encoding
			for $\emptyfun$.
			Hence, by Lemma~\ref{lmm:sl(frmaut)}, we have that $\GName, \emptyfun,
			\sElm \models \varphi$.
		\end{proof}

		Before continuing, we define the length $\lng{\varphi}$ of an \SL\ formula
		$\varphi$ as the number $\card{\sub{\varphi}}$ of its subformulas.
		We also introduce a generalization of the Knuth's double arrow notation in
		order to represents a tower of exponentials: $a \uparrow\uparrow_{b} 0
		\defeq b$ and $a \uparrow\uparrow_{b} (c + 1) \defeq a^{a
		\uparrow\uparrow_{b} c}$, for all $a, b, c \in \SetN$.

		At this point, we prove the main theorem about the non-elementary complexity
		of \SL\ model-checking problem.
		\begin{theorem}[\SL\ Model Checking]
			\label{thm:sl(modchk)}
			The model-checking problem for \SL\ is \PTimeC\ w.r.t.\ the size of the
			model and \NElmTime\ w.r.t.\ the size of the specification.
		\end{theorem}
		\begin{proof}
			By Theorem~\ref{thm:sl(sntaut)} of \SL\ sentence automaton, to verify that
			$\GName, \emptyfun, \sElm \models \varphi$, we simply calculate the
			emptiness of the \NPT\ $\NName[\varphi | ^{\GName, \sElm}]$ having
			$\card{\StSet[\GName]} \cdot (2 \uparrow\uparrow_{m} m)$ states and index
			$2 \uparrow\uparrow_{m} m$, where $m = \AOmicron{\lng{\varphi} \cdot \log
			\lng{\varphi}}$.
			It is well-known that the emptiness problem for such a kind of automaton
			with $n$ states and index $h$ is solvable in time
			$\AOmicron{n^{h}}$~\cite{KV98}.
			Thus, we get that the time complexity of checking whether $\GName,
			\emptyfun, \sElm \models \varphi$ is $\card{\StSet[\GName]}^{2
			\uparrow\uparrow_{m} m}$.
			Hence, the membership of the model-checking problem for \SL\ in \PTime\
			w.r.t.\ the size of the model and \NElmTime\ w.r.t.\ the size of the
			specification directly follows.
			Finally, by getting the relative lower bound on the model from the same
			problem for \ATLS~\cite{AHK02}, the thesis is proved.
		\end{proof}

		Finally, we show a refinement of the previous result, when we consider
		sentences of the \NGSL\ fragment.
		\begin{theorem}[\NGSL\ Model Checking]
			\label{thm:ngsl(modchk)}
			The model-checking problem for \NGSL\ is \PTimeC\ w.r.t.\ the size of the
			model and $(k + 1)$-\ExpTime\ w.r.t.\ the maximum alternation $k$ of the
			specification.
		\end{theorem}
		\begin{proof}
			By Theorem~\ref{thm:sl(sntaut)} of \SL\ sentence automaton, to verify that
			$\GName, \emptyfun, \sElm \models \qpElm \psi$, where $\qpElm \psi$ is an
			\NGSL\ principal sentence without proper subsentences, we can simply
			calculate the emptiness of the \NPT\ $\NName[\qpElm \psi | ^{\GName,
			\sElm}]$ having $\card{\StSet[\GName]} \cdot (2 \uparrow\uparrow_{m} k)$
			states and index $2 \uparrow\uparrow_{m} k$, where $m =
			\AOmicron{\lng{\psi} \cdot \log \lng{\psi}}$ and $k = \alt{\qpElm \psi}$.
			Thus, we get that the time complexity of checking whether $\GName,
			\emptyfun, \sElm \models \qpElm \psi$ is $\card{\StSet[\GName]}^{2
			\uparrow\uparrow_{m} k}$.
			At this point, since we have to do this verification for each possible
			state $\sElm \in \StSet[\GName]$ and principal subsentence $\qpElm \psi
			\in \psnt{\varphi}$ of the whole \NGSL\ specification $\varphi$, we derive
			that the bottom-up model-checking procedure requires time
			$\card{\StSet[\GName]}^{2 \uparrow\uparrow_{\lng{\varphi}} k}$, where $k =
			\max \set{ \alt{\qpElm \psi} }{ \qpElm \psi \in \psnt{\varphi} }$.
			Hence, the membership of the model-checking problem for \SL\ in \PTime\
			w.r.t.\ the size of the model and $(k + 1)$-\ExpTime\ w.r.t.\ the maximum
			alternation $k$ of the specification directly follows.
			Finally, by getting the relative lower bound on the model from the same
			\mbox{problem for \ATLS~\cite{AHK02}, the thesis is proved.}
		\end{proof}

	\end{subsection}

	\begin{subsection}{\OGSL\ Model Checking}
		\label{subsec:ogslmodchk}

		We now show how the concept of elementariness of dependence maps over
		strategies can be used to enormously reduce the complexity of the
		model-checking procedure for the \OGSL\ fragment.
		The idea behind our approach is to avoid the use of projections used to
		handle the strategy quantifications, by reducing them to action
		quantifications, which can be managed locally on each state of the model
		without a tower of exponential blow-ups.

		To start with the description of the ad-hoc procedure for \OGSL, we first
		prove the existence of a \UCT\ for each \CGS\ and \OGSL\ goal $\bpElm \psi$
		that recognizes all the assignment-state encodings of an a
		priori given assignment, \mbox{made the assumption that the goal is
		satisfied on the \CGS\ under this assignment.}
		\begin{lemma}[\OGSL\ Goal Automaton]
			\label{lmm:ogsl(golaut)}
			Let $\GName$ be a \CGS\ and $\bpElm \psi$ an \OGSL\ goal without principal
			subsentences.
			Then, there exists an \UCT\ $\UName[\bpElm \psi | ^{\GName}] \defeq
			\TATuple {\ValSet[ {\AcSet[\GName]} ](\free{\bpElm \psi}) \times
			\StSet[\GName]} {\StSet[\GName]} {\QSet[\bpElm \psi]} {\atFun[\bpElm
			\psi]} {\qElm[\bpElm \psi]} {\aleph_{\bpElm \psi}}$ such that, for all
			states $\sElm \in \StSet[\GName]$ and assignments $\asgFun \in
			\AsgSet[\GName](\free{\bpElm \psi}, \sElm)$, it holds that $\GName,
			\asgFun, \sElm \models \bpElm \psi$ iff $\TName \in \LangSet(\UName[\bpElm
			\psi | ^{\GName}])$, where $\TName$ is the assignment-state encoding for
			$\asgFun$.
		\end{lemma}
		\begin{proof}
			A first step in the construction of the \UCT\ $\UName[\bpElm \psi |
			^{\GName}]$ is to consider the \UCW\ $\UName[\psi] \defeq \WATuple
			{\pow{\APSet}} {\QSet[\psi]} {\atFun[\psi]} {\QSet[0\psi]}
			{\aleph_{\psi}}$ obtained by dualizing the \NBW\ resulting from the
			application of the classic Vardi-Wolper construction to the \LTL\ formula
			$\neg \psi$~\cite{VW86b}.
			Observe that $\LangSet(\UName[\psi]) = \LangSet(\psi)$, i.e.,
			$\UName[\psi]$ recognizes all infinite words on the alphabet
			$\pow{\APSet}$ that satisfy the \LTL\ formula $\psi$.
			Then, define the components of $\UName[\bpElm \psi | ^{\GName}] \defeq
			\TATuple {\ValSet[ {\AcSet[\GName]} ](\free{\bpElm \psi}) \times
			\StSet[\GName]} {\StSet[\GName]} {\QSet[\bpElm \psi]} {\atFun[\bpElm
			\psi]} {\qElm[0\bpElm\psi]} {\aleph_{\bpElm \psi}}$ as follows:
			\begin{itemize}
				\item
					$\QSet[\bpElm \psi] \defeq \{ \qElm[0\bpElm\psi] \} \cup \QSet[\psi]$,
					with $\qElm[0\bpElm\psi] \not\in \QSet[\psi]$;
				\item
					$\atFun[\bpElm \psi](\qElm[0\bpElm\psi], (\valFun, \sElm)) \defeq
					\bigwedge_{\qElm \in \QSet[0\psi]} \atFun[\bpElm \psi](\qElm,
					(\valFun, \sElm))$, for all $(\valFun, \sElm) \in \ValSet[
					{\AcSet[\GName]} ](\free{\bpElm \psi}) \times \StSet[\GName]$;
				\item
					$\atFun[\bpElm \psi](\qElm, (\valFun, \sElm)) \!\defeq\!
					\bigwedge_{\qElm' \!\in\! \atFun[\psi](\qElm, \labFun[\GName](\sElm))}
					(\trnFun[\GName](\sElm, \valFun \cmp \bndFun[\bpElm]), \qElm')$, for
					all $\qElm \!\in\! \QSet[\psi]$ and $(\valFun, \sElm) \!\in\! \ValSet[
					{\AcSet[\GName]} ](\free{\bpElm \psi}) \times \StSet[\GName]$;
				\item
					$\aleph_{\bpElm \psi} \defeq \aleph_{\psi}$.
			\end{itemize}
			Intuitively, the \UCT\ $\UName[\bpElm \psi | ^{\GName}]$ simply runs the
			\UCW\ $\UName[\psi]$ on the branch of the encoding individuated by the
			assignment in input.
			Thus, it is easy to see that, for all states $\sElm \in \StSet[\GName]$
			and assignments $\asgFun \in \AsgSet[\GName](\free{\bpElm \psi}, \sElm)$,
			it holds that $\GName, \asgFun, \sElm \models \bpElm \psi$ iff $\TName \in
			\LangSet(\UName[\bpElm \psi | ^{\GName}])$, where $\TName$ is the
			assignment-state encoding for $\asgFun$.
		\end{proof}

		Now, to describe our modified technique, we introduce a new concept of
		encoding regarding the elementary dependence maps over strategies.
		\begin{definition}[Elementary Dependence-State Encoding]
			\label{def:elmspcstenc}
			Let $\GName$ be a \CGS, $\sElm \in \StSet[\GName]$ one of its states, and
			$\spcFun \in \ESpcSet[ {\StrSet[\GName](\sElm)} ](\qpElm)$ an elementary
			dependence map over strategies for a quantification prefix $\qpElm \in
			\QPSet(\VSet)$ over the set $\VSet \subseteq \VarSet$.
			Then, a $(\SpcSet[ {\AcSet[\GName]} ](\qpElm) \times
			\StSet[\GName])$-labeled $\StSet[\GName]$-tree $\TName \defeq
			\LTTuple{}{}{\TSet}{\uFun}$, where $\TSet \defeq \set{ \trkElm_{\geq 1} }{
			\trkElm \in \TrkSet[\GName](\sElm) }$, is an \emph{elementary
			dependence-state encoding} for $\spcFun$ if it holds that $\uFun(\tElm)
			\defeq (\adj{\spcFun}(\sElm \cdot \tElm), \lst{\sElm \cdot \tElm})$, for
			all $\tElm \in \TSet$.
		\end{definition}
		Observe that there exists a unique elementary dependence-state encoding for
		each elementary dependence map over strategies.

		In the next lemma, we show how to handle locally the strategy
		quantifications on each state of the model, by simply using a quantification
		over actions, which is modeled by the choice of an action dependence map.
		Intuitively, we guess in the labeling what is the right part of the
		dependence map over strategies for each node of the tree and then verify
		that, for all assignments of universal variables, the corresponding complete
		assignment satisfies the inner formula.
		\begin{lemma}[\OGSL\ Sentence Automaton]
			\label{lmm:ogsl(sntaut)}
			Let $\GName$ be a \CGS\ and $\qpElm \bpElm \psi$ an \OGSL\ principal
			sentence without principal subsentences.
			Then, there exists a \UCT\ $\UName[\qpElm \bpElm \psi | ^{\GName}] \defeq
			\TATuple {\SpcSet[ {\AcSet[\GName]} ](\qpElm) \times \StSet[\GName]}
			{\StSet[\GName]} {\QSet[\qpElm \bpElm \psi]} {\atFun[\qpElm \bpElm \psi]}
			{\qElm[0\qpElm\bpElm\psi]} {\aleph_{\qpElm \bpElm \psi}}$ such that, for
			all states $\sElm \in \StSet[\GName]$ and elementary dependence maps over
			strategies $\spcFun \in \ESpcSet[ {\StrSet[\GName](\sElm)} ](\qpElm)$, it
			holds that $\GName, \spcFun(\asgFun), \sElm \emodels \bpElm \psi$, for all
			$\asgFun \in \AsgSet[\GName](\QPAVSet{\qpElm}, \sElm)$, iff $\TName \in
			\LangSet(\UName[\qpElm \bpElm \psi | ^{\GName}])$, where $\TName$ is the
			elementary dependence-state encoding for $\spcFun$.
		\end{lemma}
		\begin{proof}
			By Lemma~\ref{lmm:ogsl(golaut)} of \OGSL\ goal automaton, there is an
			\UCT\ $\UName[\bpElm \psi | ^{\GName}] = \TATuple {\ValSet[
			{\AcSet[\GName]} ](\free{\bpElm \psi}) \times \StSet[\GName]}
			{\StSet[\GName]} {\QSet[\bpElm \psi]} {\atFun[\bpElm \psi]}
			{\qElm[0\bpElm\psi]} {\aleph_{\bpElm \psi}}$ such that, for all states
			$\sElm \!\in\! \StSet[\GName]$ and assignments $\asgFun \!\in\!
			\AsgSet[\GName](\free{\bpElm \psi}, \sElm)$, it holds that $\GName,
			\asgFun, \sElm \models \bpElm \psi$ iff $\TName \in \LangSet(\UName[\bpElm
			\psi | ^{\GName}])$, where $\TName$ is the assignment-state encoding for
			$\asgFun$.

			Now, transform $\UName[\bpElm \psi | ^{\GName}]$ into the new \UCT\
			$\UName[\qpElm \bpElm \psi | ^{\GName}] \defeq \TATuple {\SpcSet[
			{\AcSet[\GName]} ](\qpElm) \times \StSet[\GName]} {\StSet[\GName]}
			{\QSet[\qpElm \bpElm \psi]} {\atFun[\qpElm \bpElm \psi]}
			{\qElm[0\qpElm\bpElm\psi]} {\aleph_{\qpElm \bpElm \psi}}$, with
			$\QSet[\qpElm \bpElm \psi] \defeq \QSet[\bpElm \psi]$,
			$\qElm[0\qpElm\bpElm\psi] \defeq \qElm[0\bpElm\psi]$, and
			$\aleph_{\qpElm \bpElm \psi} \defeq \aleph_{\bpElm \psi}$, which is used
			to handle the quantification prefix $\qpElm$ atomically, where the
			transition function is defined as follows: $\atFun[\qpElm \bpElm
			\psi](\qElm, (\spcFun, \sElm)) \defeq \bigwedge_{\valFun \in \ValSet[
			{\AcSet[\GName]} ](\QPAVSet{\qpElm})} \atFun[\bpElm \psi](\qElm,
			(\spcFun(\valFun), \sElm))$, for all $\qElm \in \QSet[\qpElm \bpElm \psi]$
			and $(\spcFun, \sElm) \in \SpcSet[ {\AcSet[\GName]} ](\qpElm) \times
			\StSet[\GName]$.
			Intuitively, $\UName[\qpElm \bpElm \psi | ^{\GName}]$ reads an action
			dependence map $\spcFun$ on each node of the input tree $\TName$ labeled
			with a state $s$ of $\GName$ and simulates the execution of the transition
			function $\atFun[\bpElm \psi](\qElm, (\valFun, \sElm))$ of $\UName[\bpElm
			\psi | ^{\GName}]$, for each possible valuation $\valFun =
			\spcFun(\valFun')$ on $\free{\bpElm \psi}$ obtained from $\spcFun$ by a
			universal valuation $\valFun' \in \ValSet[ {\AcSet[\GName]}
			](\QPAVSet{\qpElm})$.
			It is important to observe that we cannot move the component set $\SpcSet[
			{\AcSet[\GName]} ](\qpElm)$ from the input alphabet to the states of
			$\UName[\qpElm \bpElm \psi | ^{\GName}]$, by making a related guessing of
			the dependence map $\spcFun$ in the transition function, since we have to
			ensure that all states in a given node of the tree $\TName$, i.e., in each
			track of the original model $\GName$, make the same choice for $\spcFun$.

			Finally, it remains to prove that, for all states $\sElm \in
			\StSet[\GName]$ and elementary dependence map over strategies $\spcFun \in
			\ESpcSet[ {\StrSet[\GName](\sElm)} ](\qpElm)$, it holds that $\GName,
			\spcFun(\asgFun), \sElm \emodels \bpElm \psi$, for all $\asgFun \in
			\AsgSet[\GName](\QPAVSet{\qpElm}, \sElm)$, iff $\TName \in
			\LangSet(\UName[\qpElm \bpElm \psi | ^{\GName}])$, where $\TName$ is the
			elementary dependence-state encoding for $\spcFun$.

			\emph{[Only if].}
			Suppose that $\GName, \spcFun(\asgFun), \sElm \emodels \bpElm \psi$, for
			all $\asgFun \in \AsgSet[\GName](\QPAVSet{\qpElm}, \sElm)$.
			Since $\psi$ does not contain principal subsentences, we have that
			$\GName, \spcFun(\asgFun), \sElm \models \bpElm \psi$.
			So, due to the property of $\UName[\bpElm \psi | ^{\GName}]$, it follows
			that there exists an assignment-state encoding $\TName[\asgFun] \in
			\LangSet(\UName[\bpElm \psi | ^{\GName}])$, which implies the existence of
			an $(\StSet[\GName] \times \QSet[\bpElm \psi])$-tree $\RSet[\asgFun]$ that
			is an accepting run for $\UName[\bpElm \psi | ^{\GName}]$ on
			$\TName[\asgFun]$.
			At this point, let $\RSet \defeq \bigcup_{\asgFun \in
			\AsgSet[\GName](\QPAVSet{\qpElm}, \sElm)} \RSet[\asgFun]$ be the union of
			all runs.
			Then, due to the particular definition of the transition function of
			$\UName[\qpElm \bpElm \psi | ^{\GName}]$, it is not hard to see that
			$\RSet$ is an accepting run for $\UName[\qpElm \bpElm \psi | ^{\GName}]$
			on $\TName$.
			Hence, $\TName \in \LangSet(\UName[\qpElm \bpElm \psi | ^{\GName}])$.

			\emph{[If].}
			Suppose that $\TName \in \LangSet(\UName[\qpElm \bpElm \psi |
			^{\GName}])$.
			Then, there exists an $(\StSet[\GName] \times \QSet[\qpElm \bpElm
			\psi])$-tree $\RSet$ that is an accepting run for $\UName[\qpElm \bpElm
			\psi | ^{\GName}]$ on $\TName$.
			Now, for each $\asgFun \in \AsgSet[\GName](\QPAVSet{\qpElm}, \sElm)$, let
			$\RSet[\asgFun]$ be the run for $\UName[\bpElm \psi | ^{\GName}]$ on the
			assignment-state encoding $\TName[\asgFun]$ for $\spcFun(\asgFun)$.
			Due to the particular definition of the transition function of
			$\UName[\qpElm \bpElm \psi | ^{\GName}]$, it is easy to see that
			$\RSet[\asgFun] \subseteq \RSet$.
			Thus, since $\RSet$ is accepting, we have that $\RSet[\asgFun]$ is
			accepting as well.
			So, $\TName[\asgFun] \in \LangSet(\UName[\bpElm \psi | ^{\GName}])$.
			At this point, due to the property of $\UName[\bpElm \psi | ^{\GName}]$,
			it follows that $\GName, \spcFun(\asgFun), \sElm \models \bpElm \psi$.
			Now, since $\psi$ does not contain principal subsentences, we have that
			$\GName, \spcFun(\asgFun), \sElm \emodels \bpElm \psi$, for all $\asgFun
			\in \AsgSet[\GName](\QPAVSet{\qpElm}, \sElm)$.
		\end{proof}

		At this point, we can prove the following theorem that is at the base of the
		elementary model-checking procedure for \OGSL.
		\begin{theorem}[\OGSL\ Sentence Automaton]
			\label{thm:ogsl(sntaut)}
			Let $\GName$ be a \CGS, $\sElm \in \StSet[\GName]$ one of its states, and
			$\qpElm \bpElm \psi$ an \OGSL\ principal sentence without principal
			subsentences.
			Then, there exists an \NPT\ $\NName[\qpElm \bpElm \psi | ^{\GName,
			\sElm}]$ such that $\GName, \emptyfun, \sElm \models \qpElm \bpElm \psi$
			iff $\LangSet(\NName[\qpElm \bpElm \psi | ^{\GName, \sElm}]) \neq
			\emptyset$.
		\end{theorem}
		\begin{proof}
			As in the general case of \SL\ sentence automaton, we have to ensure that
			the state labeling of nodes of the elementary dependence-state encoding is
			coherent with the node itself.
			To do this, we apply Theorem~\ref{thm:apt(dirprj)} of \APT\ direction
			projection with distinguished direction $\sElm$ to the \UPT\
			$\UName[\qpElm \bpElm \psi | ^{\GName}]$ derived by
			Lemma~\ref{lmm:ogsl(sntaut)} of the \OGSL\ sentence automaton, thus
			obtaining the required \NPT\ $\NName[\qpElm \bpElm \psi | ^{\GName,
			\sElm}]$.

			\emph{[Only if].}
			Suppose that $\GName, \emptyfun, \sElm \models \qpElm \bpElm \psi$.
			By Corollary~\ref{cor:ogsl(elm)} of \OGSL\ elementariness, it means that
			$\GName, \emptyfun, \sElm \emodels \qpElm \bpElm \psi$.
			Then, by Definition~\ref{def:ngsl(elementarysemantics)} of \NGSL\
			elementary semantics, there exists an elementary dependence map $\spcFun
			\in \ESpcSet[ {\StrSet[\GName](\sElm)} ](\qpElm)$ such that $\GName,
			\spcFun(\asgFun), \sElm \emodels \bpElm \psi$, for all $\asgFun \in
			\AsgSet[\GName](\QPAVSet{\qpElm}, \sElm)$.
			Thus, by Lemma~\ref{lmm:ogsl(sntaut)}, we have that $\TName \in
			\LangSet(\UName[\qpElm \bpElm \psi | ^{\GName}])$, where $\TName$ is the
			elementary dependence-state encoding for $\spcFun$.
			Hence, by Theorem~\ref{thm:apt(dirprj)}, it holds that
			$\LangSet(\NName[\qpElm \bpElm \psi | ^{\GName, \sElm}]) \neq \emptyset$.

			\emph{[If].}
			Suppose that $\LangSet(\NName[\qpElm \bpElm \psi | ^{\GName, \sElm}]) \neq
			\emptyset$.
			Then, by Theorem~\ref{thm:apt(dirprj)}, there exists an $(\SpcSet[
			{\AcSet[\GName]} ](\qpElm) \times \StSet[\GName])$-labeled
			$\StSet[\GName]$-tree $\TName$ such that $\TName \in
			\LangSet(\UName[\qpElm \bpElm \psi | ^{\GName}])$.
			Now, it is immediate to see that there is an elementary dependence map
			$\spcFun \in \ESpcSet[ {\StrSet[\GName](\sElm)} ](\qpElm)$ for which
			$\TName$ is the elementary dependence-state encoding.
			Thus, by Lemma~\ref{lmm:ogsl(sntaut)}, we have that $\GName,
			\spcFun(\asgFun), \sElm \emodels \bpElm \psi$, for all $\asgFun \in
			\AsgSet[\GName](\QPAVSet{\qpElm}, \sElm)$.
			By Definition~\ref{def:ngsl(elementarysemantics)} of \NGSL\ elementary
			semantics, it holds that $\GName, \emptyfun, \sElm \emodels \qpElm \bpElm
			\psi$.
			Hence, by Corollary~\ref{cor:ogsl(elm)} of \OGSL\ elementariness, it means
			that $\GName, \emptyfun, \sElm \models \qpElm \bpElm \psi$.
		\end{proof}

		Finally, we show in the next fundamental theorem the precise complexity of
		the model-checking for \OGSL.
		\begin{theorem}[\OGSL\ Model Checking]
			\label{thm:ogsl(modchk)}
			The model-checking problem for \OGSL\ is \PTimeC\ w.r.t.\ the size of the
			model and 2\ExpTimeC\ w.r.t.\ the size of the specification.
		\end{theorem}
		\begin{proof}
			By Theorem~\ref{thm:ogsl(sntaut)} of \OGSL\ sentence automaton, to verify
			that $\GName, \emptyfun, \sElm \models \qpElm \bpElm \psi$, we simply
			calculate the emptiness of the \NPT\ $\NName[\qpElm \bpElm \psi |
			^{\GName, \sElm}]$.
			This automaton is obtained by the operation of direction projection on the
			\UCT\ $\UName[\qpElm \bpElm \psi | ^{\GName}]$, which is in turn derived
			by the \UCT\ $\UName[\bpElm \psi | ^{\GName}]$.
			Now, it is easy to see that the number of states of $\UName[\bpElm \psi |
			^{\GName}]$, and consequently of $\UName[\qpElm \bpElm \psi | ^{\GName}]$,
			is $2^{\AOmicron{\lng{\psi}}}$.
			So, $\NName[\qpElm \bpElm \psi | ^{\GName, \sElm}]$ has
			$\card{\StSet[\GName]} \cdot 2^{2^{\AOmicron{\lng{\psi}}}}$ states and
			index $2^{\AOmicron{\lng{\psi}}}$.

			The emptiness problem for such a kind of automaton with $n$ states and
			index $h$ is solvable in time $\AOmicron{n^{h}}$~\cite{KV98}.
			Thus, we get that the time complexity of checking whether $\GName,
			\emptyfun, \sElm \models \qpElm \bpElm \psi$ is
			$\card{\StSet[\GName]}^{2^{\AOmicron{\lng{\psi}}}}$.
			At this point, since we have to do this verification for each possible
			state $\sElm \in \StSet[\GName]$ and principal subsentence $\qpElm \bpElm
			\psi \in \psnt{\varphi}$ of the whole \OGSL\ specification $\varphi$, we
			derive that the whole bottom-up model-checking procedure requires time
			$\card{\StSet[\GName]}^{2^{\AOmicron{\lng{\varphi}}}}$.
			Hence, the membership of the model-checking problem for \OGSL\ in \PTime\
			w.r.t.\ the size of the model and 2\ExpTime\ w.r.t.\ the size of the
			specification directly follows.
			Finally the thesis is proved, by getting the relative lower bounds from
			the same problem for \ATLS~\cite{AHK02}.
		\end{proof}

	\end{subsection}

\end{section}





\begin{section}{Conclusion}
	\label{sec:conclusion}

	In this paper, we introduced and studied \SL\ as a very powerful logic
	formalism to reasoning about strategic behaviors of multi-agent concurrent
	games.
	In particular, we proved that it subsumes the classical temporal and game
	logics not using explicit fix-points.
	As one of the main results about \SL, we shown that the relative
	model-checking problem is decidable but non-elementary hard.
	As further and interesting practical results, we investigated several of its
	syntactic fragments.
	The most appealing one is \OGSL, which is obtained by restricting \SL\ to deal
	with one temporal goal at a time.
	Interestingly, \OGSL\ strictly extends \ATLS, while maintaining all its
	positive properties.
	In fact, the model-checking problem is 2\ExpTimeC, hence not harder than the
	one for \ATLS.
	Moreover, although for the sake of space it is not reported in this paper, we
	shown that it is invariant under bisimulation and decision-unwinding, and
	consequently, it has the decision-tree model property.
	The main reason why \OGSL\ has all these positive properties is that it
	satisfies a special model property, which we name ``\emph{elementariness}''.
	Informally, this property asserts that all strategy quantifications in a
	sentence can be reduced to a set of quantifications over actions, which turn
	out to be easier to handle.
	We remark that among all \SL\ fragments we investigated, \OGSL\ is the only
	one that satisfies this property.
	As far as we know, \OGSL\ is the first significant proper extension of \ATLS\
	having an elementary model-checking problem, and even more, with the same
	computational complexity.
	All these positive aspects make us strongly believe that \OGSL\ is a valid
	alternative to \ATLS\ to be used in the field of formal verification for
	multi-agent concurrent systems.

	As another interesting fragment we investigated in this paper, we recall
	\BGSL.
	This logic allows us to express important game-theoretic properties, such as
	Nash equilibrium, which cannot be defined in \OGSL.
	Unfortunately, we do not have an elementary model-checking procedure for it,
	neither we can exclude it.
	We leave to investigate this as future work.


	Last but not least, from a theoretical point of view, we are convinced that
	our framework can be used as a unifying basis for logic reasonings about
	strategic behaviors in multi-agent scenarios and their relationships.
	In particular, it can be used to study variations and extensions of \OGSL\ in
	a way similar as it has been done in the literature for \ATLS.
	For example, it could be interesting to investigate memoryful \OGSL, by
	inheriting and extending the ``memoryful'' concept used for \ATLS\ and \CHPSL\
	and investigated in~\cite{MMV10a} and~\cite{FKL10}, respectively.
	Also, we recall that this concept implicitly allows to deal with backwards
	temporal modalities.
	As another example, it would be interesting to investigate the graded
	extension of \OGSL, in a way similar as it has been done
	in~\cite{BMM09,BMM10,BMM12} and~\cite{KSV02,BLMV08} for \CTL\ and \MuCalculus,
	respectively.
	We recall that graded quantifiers in branching-time temporal logics allow to
	count how many equivalent classes of paths satisfy a given property.
	This concept in \OGSL\ would further allow the counting of strategies and so
	to succinctly check the existence of more than one nonequivalent winning
	strategy for a given agent, in one shot.
	We hope to lift to graded \OGSL\ questions left open about graded
	branching-time temporal logic, such as the precise satisfiability complexity
	of graded full computation tree logic~\cite{BMM12}.

\end{section}


	\appendix



\begin{section}{Mathematical Notation}
	\label{app:mthnot}

	In this short reference appendix, we report the classical mathematical
	notation and some common definitions that are used along the whole work.

	\begin{paragraph}*{Classic objects}

		We consider $\SetN$ as the set of \emph{natural numbers} and $\numcc{m}{n}
		\defeq \set{ k \in \SetN }{ m \leq k \leq n }$, $\numco{m}{n} \defeq \set{ k
		\in \SetN }{ m \leq k < n }$, $\numoc{m}{n} \defeq \set{ k \in \SetN }{ m <
		k \leq n }$, and $\numoo{m}{n} \defeq \set{ k \in \SetN }{ m < k < n }$ as
		its \emph{interval} subsets, with $m \in \SetN$ and $n \in \SetNI \defeq
		\SetN \cup \{ \omega \}$, where $\omega$ is the \emph{numerable infinity},
		i.e., the \emph{least infinite ordinal}.
		Given a \emph{set} $\XSet$ of \emph{objects}, we denote by $\card{\XSet} \in
		\SetNI \cup \{ \infty \}$ the \emph{cardinality} of $\XSet$, i.e., the
		number of its elements, where $\infty$ represents a \emph{more than
		countable} cardinality, and by $\pow{\XSet} \defeq \set{ \YSet }{ \YSet
		\subseteq \XSet }$ the \emph{powerset} of $\XSet$, i.e., the set of all its
		subsets.

	\end{paragraph}

	\begin{paragraph}*{Relations}

		By $\RRel \subseteq \XSet \times \YSet$ we denote a \emph{relation} between
		the \emph{domain} $\dom{\RRel} \defeq \XSet$ and \emph{codomain}
		$\cod{\RRel} \defeq \YSet$, whose \emph{range} is indicated by $\rng{\RRel}
		\defeq \set{ \yElm \in \YSet }{ \exists \xElm \in \XSet .\: (\xElm, \yElm)
		\in \RRel }$.
		We use $\RRel^{-1} \defeq \set{ (\yElm, \xElm) \in \YSet \times \XSet }{
		(\xElm, \yElm) \in \RRel }$ to represent the \emph{inverse} of $\RRel$
		itself.
		Moreover, by $\SRel \cmp \RRel$, with $\RRel \subseteq \XSet \times \YSet$
		and $\SRel \subseteq \YSet \times \ZSet$, we denote the \emph{composition}
		of $\RRel$ with $\SRel$, i.e., the relation $\SRel \cmp \RRel \defeq \set{
		(\xElm, \zElm) \in \XSet \times \ZSet }{ \exists \yElm \in \YSet .\: (\xElm,
		\yElm) \in \RRel \land (\yElm, \zElm) \in \SRel }$.
		We also use $\RRel^{n} \defeq \RRel^{n - 1} \cmp \RRel$, with $n \in
		\numco{1}{\omega}\!$, to indicate the \emph{$n$-iteration} of $\RRel
		\subseteq \XSet \times \YSet$, where $\YSet \subseteq \XSet$ and $\RRel^{0}
		\defeq \set{ (\yElm, \yElm) }{ \yElm \in \YSet }$ is the \emph{identity} on
		$\YSet$.
		With $\RRel^{+} \defeq \bigcup_{n = 1}^{< \omega} \RRel^{n}$ and $\RRel^{*}
		\defeq \RRel^{+} \cup \RRel^{0}$ we denote, respectively, the
		\emph{transitive} and \emph{reflexive-transitive closure} of $\RRel$.
		Finally, for an \emph{equivalence} relation $\RRel \subseteq \XSet \times
		\XSet$ on $\XSet$, we represent with $\class{ \XSet }{ \:\RRel } \defeq
		\set{ [\xElm]_{\RRel} }{ \xElm \in \XSet }$, where $[\xElm]_{\RRel} \defeq
		\set{ \xElm' \in \XSet }{ (\xElm, \xElm') \in \RRel }$, the \emph{quotient}
		set of $\XSet$ w.r.t.\ $\RRel$, i.e., the set of all related equivalence
		\emph{classes} $[\cdot]_{\RRel}$.

	\end{paragraph}

	\begin{paragraph}*{Functions}

		We use the symbol $\YSet^{\XSet} \subseteq \pow{\XSet \times \YSet}$ to
		denote the set of \emph{total functions} $\fFun$ from $\XSet$ to $\YSet$,
		i.e., the relations $\fFun \subseteq \XSet \times \YSet$ such that for all
		$\xElm \in \dom{\fFun}$ there is exactly one element $\yElm \in \cod{\fFun}$
		such that $(\xElm, \yElm) \in \fFun$.
		Often, we write $\fFun : \XSet \to \YSet$ and $\fFun : \XSet \pto \YSet$ to
		indicate, respectively, $\fFun \in \YSet^{\XSet}$ and $\fFun \in
		\bigcup_{\XSet' \subseteq \XSet} \YSet^{\XSet'}$.
		Regarding the latter, note that we consider $\fFun$ as a \emph{partial
		function} from $\XSet$ to $\YSet$, where $\dom{\fFun} \subseteq \XSet$
		contains all and only the elements for which $\fFun$ is defined.
		Given a set $\ZSet$, by $\fFun[\rst \ZSet] \defeq \fFun \cap (\ZSet \times
		\YSet)$ we denote the \emph{restriction} of $\fFun$ to the set $\XSet \cap
		\ZSet$, i.e., the function $\fFun[\rst \ZSet] : \XSet \cap \ZSet \pto \YSet$
		such that, for all $\xElm \in \dom{\fFun} \cap \ZSet$, it holds that
		$\fFun[\rst \ZSet](\xElm) = \fFun(\xElm)$.
		Moreover, with $\emptyfun$ we indicate a generic \emph{empty function},
		i.e., a function with empty domain.
		Note that $\XSet \cap \ZSet = \emptyset$ implies $\fFun[\rst \ZSet] =
		\emptyfun$.
		Finally, for two partial functions $\fFun, \gFun : \XSet \pto \YSet$, we use
		$\fFun \Cup \gFun$ and $\fFun \Cap \gFun$ to represent, respectively, the
		\emph{union} and \emph{intersection} of these functions defined as follows:
		$\dom{\fFun \Cup \gFun} \defeq \dom{\fFun} \cup \dom{\gFun} \setminus \set{
		\xElm \in \dom{\fFun} \cap \dom{\gFun} }{ \fFun(\xElm) \neq \gFun(\xElm) }$,
		$\dom{\fFun \Cap \gFun} \defeq \set{ \xElm \in \dom{\fFun} \cap \dom{\gFun}
		}{ \fFun(\xElm) = \gFun(\xElm) }$, $(\fFun \Cup \gFun)(\xElm) =
		\fFun(\xElm)$ for $\xElm \in \dom{\fFun \Cup \gFun} \cap \dom{\fFun}$,
		$(\fFun \Cup \gFun)(\xElm) = \gFun(\xElm)$ for $\xElm \in \dom{\fFun \Cup
		\gFun} \cap \dom{\gFun}$, and $(\fFun \Cap \gFun)(\xElm) = \fFun(\xElm)$ for
		$\xElm \in \dom{\fFun \Cap \gFun}$.

	\end{paragraph}

	\begin{paragraph}*{Words}

		By $\XSet^{n}$, with $n \in \SetN$, we denote the set of all
		\emph{$n$-tuples} of elements from $\XSet$, by $\XSet^{*} \defeq \bigcup_{n
		= 0}^{< \omega} \XSet^{n}$ the set of \emph{finite words} on the
		\emph{alphabet} $\XSet$, by $\XSet^{+} \defeq \XSet^{*} \setminus \{
		\epsilon \}$ the set of \emph{non-empty words}, and by $\XSet^{\omega}$ the
		set of \emph{infinite words}, where, as usual, $\epsilon \in \XSet^{*}$ is
		the \emph{empty word}.
		The \emph{length} of a word $\wElm \in \XSet^{\infty} \defeq \XSet^{*} \cup
		\XSet^{\omega}$ is represented with $\card{\wElm} \in \SetNI$.
		By $(\wElm)_{i}$ we indicate the \emph{$i$-th letter} of the finite word
		$\wElm \in \XSet^{+}$, with $i \in \numco{0}{\card{\wElm}}$.
		Furthermore, by $\fst{\wElm} \defeq (\wElm)_{0}$ (resp., $\lst{\wElm} \defeq
		(\wElm)_{\card{\wElm} - 1}$), we denote the \emph{first} (resp.,
		\emph{last}) letter of $\wElm$.
		In addition, by $(\wElm)_{\leq i}$ (resp., $(\wElm)_{> i}$), we indicate the
		\emph{prefix} up to (resp., \emph{suffix} after) the letter of index $i$ of
		$\wElm$, i.e., the finite word built by the first $i + 1$ (resp., last
		$\card{\wElm} - i - 1$) letters $(\wElm)_{0}, \ldots, (\wElm)_{i}$ (resp.,
		$(\wElm)_{i + 1}, \ldots, (\wElm)_{\card{\wElm} - 1}$).
		We also set, $(\wElm)_{< 0} \defeq \epsilon$, $(\wElm)_{< i} \defeq
		(\wElm)_{\leq i - 1}$, $(\wElm)_{\geq 0} \defeq \wElm$, and $(\wElm)_{\geq
		i} \defeq (\wElm)_{> i - 1}$, for $i \in \numco{1}{\card{\wElm}}$.
		Mutatis mutandis, the notations of $i$-th letter, first, prefix, and suffix
		apply to infinite words too.
		Finally, by $\pfx{\wElm[1], \wElm[2]} \in \XSet^{\infty}$ we denote the
		\emph{maximal common prefix} of two different words $\wElm[1], \wElm[2] \in
		\XSet^{\infty}$, i.e., the finite word $\wElm \in \XSet^{*}$ for which
		there are two words $\wElm[1|'], \wElm[2|'] \in \XSet^{\infty}$ such that
		$\wElm[1] = \wElm \cdot \wElm[1|']$, $\wElm[2] = \wElm \cdot \wElm[2|']$,
		and $\fst{\wElm[1|']} \neq \fst{\wElm[2|']}$.
		By convention, we set $\pfx{\wElm, \wElm} \defeq \wElm$.

	\end{paragraph}

	\begin{paragraph}*{Trees}

		For a set $\DirSet$ of objects, called \emph{directions}, a
		\emph{$\DirSet$-tree} is a set $\TSet \subseteq \DirSet^{*}$ closed under
		prefix, i.e., if $\tElm \cdot \dElm \in \TSet$, with $\dElm \in \DirSet$,
		then also $\tElm \in \TSet$.
		We say that it is \emph{complete} if it holds that $\tElm \cdot \dElm' \in
		\TSet$ whenever $\tElm \cdot \dElm \in \TSet$, for all $\dElm' < \dElm$,
		where $< \: \subseteq \DirSet \times \DirSet$ is an a priori fixed strict
		total order on the set of directions that is clear from the context.
		Moreover, it is \emph{full} if $\TSet = \DirSet^{*}$.
		The elements of $\TSet$ are called \emph{nodes} and the empty word
		$\epsilon$ is the \emph{root} of $\TSet$.
		For every $\tElm \in \TSet$ and $\dElm \in \DirSet$, the node $\tElm \cdot
		\dElm \in \TSet$ is a \emph{successor} of $\tElm$ in $\TSet$.
		The tree is $b$-\emph{bounded} if the maximal number $b$ of its successor
		nodes is finite, i.e., $b = \max_{\tElm \in \TSet} \card{\set{ \tElm \cdot
		\dElm \in \TSet }{ \dElm \in \DirSet }} < \omega$.
		A \emph{branch} of the tree is an infinite word $\wElm \in \DirSet^{\omega}$
		such that $(\wElm)_{\leq i} \in \TSet$, for all $i \in \SetN$.
		For a finite set $\LabSet$ of objects, called \emph{symbols}, a
		\emph{$\LabSet$-labeled $\DirSet$-tree} is a quadruple
		$\LTDef{\LabSet}{\DirSet}$, where $\TSet$ is a $\DirSet$-tree and $\vFun :
		\TSet \to \LabSet$ is a \emph{labeling function}.
		When $\DirSet$ and $\LabSet$ are clear from the context, we call $\LTStruct$
		simply a (labeled) tree.

	\end{paragraph}

\end{section}




\begin{section}{Proofs of Section~\ref{sec:strqnt}}
	\label{app:strqnt}

	In this appendix, we report the proofs of lemmas needed to prove the
	elementariness of \OGSL.
	Before this, we describe two relevant properties that link together dependence
	maps of a given quantification prefix with those of the dual one.
	These properties report, in the dependence maps framework, what is known to
	hold, in an equivalent way, for first and second order logic.
	In particular, they result to be two key points towards a complete
	understanding of the strategy quantifications of our logic.

	The first of these properties enlighten the fact that two arbitrary dual
	dependence maps $\spcFun$ and $\dual{\spcFun}$ always share a common valuation
	$\valFun$.
	To better understand this concept, consider for instance the functions
	$\spcFun[1]$ and $\dual[6]{\spcFun}$ of the examples illustrated just after
	Definition~\ref{def:qntspc} of dependence maps.
	Then, it is easy to see that the valuation $\valFun \in
	\ValSet[\DSet](\VSet)$ with $\valFun(\xSym) = \valFun(\ySym) = 1$ and
	$\valFun(\zSym) = 0$ resides in both the ranges of $\spcFun[1]$ and
	$\dual[6]{\spcFun}$, i.e., $\valFun \in \rng{\spcFun[1]} \cap
	\rng{\dual[6]{\spcFun}}$.
	\begin{lemma}[Dependence Incidence]
		\label{lmm:scpinc}
		Let $\qpElm \in \QPSet(\VSet)$ be a quantification prefix over a set of
		variables $\VSet \subseteq \VarSet$ and $\DSet$ a generic set.
		Moreover, let $\spcFun \in \SpcSet[\DSet](\qpElm)$ and $\dual{\spcFun} \in
		\SpcSet[\DSet](\dual{\qpElm})$ be two dependence maps.
		Then, there exists a valuation $\valFun \in \ValSet[\DSet](\VSet)$ such
		that $\valFun = \spcFun(\valFun_{\rst \QPAVSet{\qpElm}}) =
		\dual{\spcFun}(\valFun_{\rst \QPAVSet{\dual{\qpElm}}})$.
	\end{lemma}
	\begin{proof}
		W.l.o.g., suppose that $\qpElm$ starts with an existential quantifier.
		If this is not the case, the dual prefix $\dual{\qpElm}$ necessarily
		satisfies the above requirement, so, we can simply shift our reasoning on
		it.

		The whole proof proceeds by induction on the alternation number
		$\alt{\qpElm}$ of $\qpElm$.
		As base case, if $\alt{\qpElm} = 0$, we define $\valFun \defeq
		\spcFun(\emptyfun)$, since $\QPAVSet{\qpElm} = \emptyset$.
		Obviously, it holds that $\valFun = \spcFun(\valFun_{\rst
		\QPAVSet{\qpElm}}) = \dual{\spcFun}(\valFun_{\rst
		\QPAVSet{\dual{\qpElm}}})$, due to the fact that $\valFun_{\rst
		\QPAVSet{\qpElm}} = \emptyfun$ and $\valFun_{\rst
		\QPAVSet{\dual{\qpElm}}} = \valFun$.
		Now, as inductive case, suppose that the statement is true for all
		prefixes $\qpElm' \in \QPSet(\VSet')$ with $\alt{\qpElm'} = n$, where
		$\VSet' \subset \VSet$.
		Then, we prove that it is true for all prefixes $\qpElm \in \QPSet(\VSet)$
		with $\alt{\qpElm} = n + 1$ too.
		To do this, we have to uniquely split $\qpElm = \qpElm' \cdot \qpElm''$
		into the two prefixes $\qpElm' \in \QPSet(\VSet')$ and $\qpElm'' \in
		\QPSet(\VSet \setminus \VSet')$ such that $\alt{\qpElm'} = n$ and
		$\alt{\qpElm''} = 0$.
		At this point, the following two cases can arise.
		\begin{itemize}
			\item
				If $n$ is even, it is immediate to see that $\QPEVSet{\qpElm''} =
				\emptyset$.
				So, consider the dependence maps $\spcFun' \in \SpcSet[\DSet](\qpElm')$
				and $\dual{\spcFun'} \in \SpcSet[\DSet](\dual{\qpElm'})$ such that
				$\spcFun'(\valFun_{\rst \QPAVSet{\qpElm'}}) = \spcFun(\valFun)_{\rst
				\VSet'}$ and $\dual{\spcFun'}(\dual{\valFun}) =
				\dual{\spcFun}(\dual{\valFun})_{\rst \VSet'}$, for all valuations
				$\valFun \in \ValSet[\DSet](\QPAVSet{\qpElm})$ and $\dual{\valFun} \in
				\ValSet[\DSet](\QPAVSet{\dual{\qpElm}}) =
				\ValSet[\DSet](\QPAVSet{\dual{\qpElm'}})$.
				By the inductive hypothesis, there exists a valuation $\valFun' \in
				\ValSet[\DSet](\VSet')$ such that $\valFun' = \spcFun'(\valFun'_{\rst
				\QPAVSet{\qpElm'}}) = \dual{\spcFun'}(\valFun'_{\rst
				\QPAVSet{\dual{\qpElm'}}})$.
				So, set $\valFun \defeq \dual{\spcFun}(\valFun'_{\rst
				\QPAVSet{\dual{\qpElm}}})$.
			\item
				If $n$ is odd, it is immediate to see that $\QPAVSet{\qpElm''} =
				\emptyset$.
				So, consider the dependence maps $\spcFun' \in \SpcSet[\DSet](\qpElm')$
				and $\dual{\spcFun'} \in \SpcSet[\DSet](\dual{\qpElm'})$ such that
				$\spcFun'(\valFun) = \spcFun(\valFun)_{\rst \VSet'}$ and
				$\dual{\spcFun'}(\dual{\valFun}_{\rst \QPAVSet{\dual{\qpElm'}}}) =
				\dual{\spcFun}(\dual{\valFun})_{\rst \VSet'}$, for all valuations
				$\valFun \in \ValSet[\DSet](\QPAVSet{\qpElm}) =
				\ValSet[\DSet](\QPAVSet{\qpElm'})$ and $\dual{\valFun} \in
				\ValSet[\DSet](\QPAVSet{\dual{\qpElm}})$.
				By the inductive hypothesis, there exists a valuation $\valFun' \in
				\ValSet[\DSet](\VSet')$ such that $\valFun' = \spcFun'(\valFun'_{\rst
				\QPAVSet{\qpElm'}}) = \dual{\spcFun'}(\valFun'_{\rst
				\QPAVSet{\dual{\qpElm'}}})$.
				So, set $\valFun \defeq \spcFun(\valFun'_{\rst
				\QPAVSet{\qpElm}})$.
		\end{itemize}
		Now, it is easy to see that in both cases the valuation $\valFun$
		satisfies the thesis, i.e., $\valFun = \spcFun(\valFun_{\rst
		\QPAVSet{\qpElm}}) = \dual{\spcFun}(\valFun_{\rst
		\QPAVSet{\dual{\qpElm}}})$.
	\end{proof}

	The second property we are going to prove describes the fact that, if all
	dependence maps $\spcFun$ of a given prefix $\qpElm$, for a dependent specific
	universal valuation $\valFun$, share a given property then there is a dual
	dependence maps $\dual{\spcFun}$ that has the same property, for all universal
	valuations $\dual{\valFun}$.
	To have a better understanding of this idea, consider again the examples
	reported just after Definition~\ref{def:qntspc} and let $\PSet \defeq \{ (0,
	0, 1), (0, 1, 0) \} \subset \ValSet[\DSet](\VSet)$, where the triple $(\lElm,
	\mElm, \nElm)$ stands for the valuation that assigns $\lElm$ to $\xSym$,
	$\mElm$ to $\ySym$, and $\nElm$ to $\zSym$.
	Then, it is easy to see that all ranges of the dependence maps $\spcFun[i]$
	for $\qpSym$ intersect $\PSet$, i.e., for all $i \in \numcc{0}{3}$, there is
	$\valFun \in \ValSet[\DSet](\QPAVSet{\qpSym})$ such that $\spcFun[i](\valFun)
	\in \PSet$.
	Moreover, consider the dual dependence maps $\dual[2]{\spcFun}$ for
	$\dual{\qpSym}$.
	Then, it is not hard to see that $\dual[2]{\spcFun}(\dual{\valFun}) \in
	\PSet$, for all $\dual{\valFun} \in
	\ValSet[\DSet](\QPAVSet{\dual{\qpSym}})$.
	\begin{lemma}[Dependence Dualization]
		\label{lmm:scpdlz}
		Let $\qpElm \in \QPSet(\VSet)$ be a quantification prefix over a set of
		variables $\VSet \subseteq \VarSet$, $\DSet$ a generic set, and $\PSet
		\subseteq \ValSet[\DSet](\VSet)$ a set of valuations of $\VSet$ over
		$\DSet$.
		Moreover, suppose that, for all dependence maps $\spcFun \in
		\SpcSet[\DSet](\qpElm)$, there is a valuation $\valFun \in
		\ValSet[\DSet](\QPAVSet{\qpElm})$ such that $\spcFun(\valFun) \in \PSet$.
		Then, there exists a dependence map $\dual{\spcFun} \in
		\SpcSet[\DSet](\dual{\qpElm})$ such that, for all valuations
		$\dual{\valFun} \in \ValSet[\DSet](\QPAVSet{\dual{\qpElm}})$, it holds
		that $\dual{\spcFun}(\dual{\valFun}) \in \PSet$.
	\end{lemma}
	\begin{proof}
		The proof easily proceeds by induction on the length of the prefix
		$\qpElm$.
		As base case, when $\card{\qpElm} = 0$, we have that
		$\SpcSet[\DSet](\qpElm) = \SpcSet[\DSet](\dual{\qpElm}) = \{ \emptyfun
		\}$, i.e., the only possible dependence maps is the empty function, which
		means that the statement is vacuously verified.
		As inductive case, we have to distinguish between two cases, as follows.
		\begin{itemize}
			\item
				$\qpElm = \EExs{\xElm} \cdot \qpElm'$.

				As first thing, note that $\QPAVSet{\qpElm} = \QPAVSet{\qpElm'}$ and,
				for all elements $\eElm \in \DSet$, consider the projection
				$\PSet[\eElm] \defeq \set{ \valFun' \in
				\ValSet[\DSet](\QPVSet(\qpElm'))}{ \valFun'[\xElm \mapsto \eElm] \in
				\PSet }$ of $\PSet$ on	the variable $\xElm$ with value $\eElm$.

				\hspace{0.75em}
				Then, by hypothesis, we can derive that, for all $\eElm \in \DSet$ and
				$\spcFun' \in \SpcSet[\DSet](\qpElm')$, there exists $\valFun' \in
				\ValSet[\DSet](\QPAVSet{\qpElm'})$ such that $\spcFun'(\valFun') \in
				\PSet[\eElm]$.
				Indeed, let $\eElm \in \DSet$ and $\spcFun' \in
				\SpcSet[\DSet](\qpElm')$, and build the function $\spcFun :
				\ValSet[\DSet](\QPAVSet{\qpElm}) \to \ValSet[\DSet](\VSet)$ given by
				$\spcFun(\valFun') \defeq \spcFun'(\valFun')[\xElm \mapsto \eElm]$,
				for all $\valFun' \in \ValSet[\DSet](\QPAVSet{\qpElm}) =
				\ValSet[\DSet](\QPAVSet{\qpElm'})$.
				It is immediate to see that $\spcFun \in \SpcSet[\DSet](\qpElm)$.
				So, by the hypothesis, there is $\valFun' \in
				\ValSet[\DSet](\QPAVSet{\qpElm})$ such that $\spcFun(\valFun') \in
				\PSet$, which implies $\spcFun'(\valFun')[\xElm \mapsto \eElm] \in
				\PSet$, and so, $\spcFun'(\valFun') \in	\PSet[\eElm]$.

				\hspace{0.75em}
				Now, by the inductive hypothesis, for all elements $\eElm \in \DSet$,
				there exists $\dual[\eElm]{\spcFun'} \in
				\SpcSet[\DSet](\dual{\qpElm'})$ such that, for all	$\dual{\valFun'}
				\in \ValSet[\DSet](\QPAVSet{\dual{\qpElm'}})$, it holds that
				$\dual[\eElm]{\spcFun'}(\dual{\valFun'}) \in \PSet[\eElm]$, i.e.,
				$\dual[\eElm]{\spcFun'}(\dual{\valFun'})[\xElm \mapsto \eElm] \in
				\PSet$.

				\hspace{0.75em}
				At this point, consider the function $\dual{\spcFun} :
				\ValSet[\DSet](\QPAVSet{\dual{\qpElm}}) \to \ValSet[\DSet](\VSet)$
				given by $\dual{\spcFun}(\dual{\valFun}) \defeq
				\dual[{\dual{\valFun}(\xElm)}]{\spcFun'}(\dual{\valFun}_{\rst
				\QPAVSet{\dual{\qpElm'}}})[\xElm \mapsto \dual{\valFun}(\xElm)]$,
				for all $\dual{\valFun} \in
				\ValSet[\DSet](\QPAVSet{\dual{\qpElm}})$.
				Then, it is possible to verify that $\dual{\spcFun} \in
				\SpcSet[\DSet](\dual{\qpElm})$.
				Indeed, for each $\yElm \in \QPAVSet{\dual{\qpElm}}$ and
				$\dual{\valFun} \in \ValSet[\DSet](\QPAVSet{\dual{\qpElm}})$, we
				have that $\dual{\spcFun}(\dual{\valFun})(\yElm) =
				\dual[{\dual{\valFun}(\xElm)}]{\spcFun'}(\dual{\valFun}_{\rst
				\QPAVSet{\dual{\qpElm'}}})[\xElm \mapsto
				\dual{\valFun}(\xElm)](\yElm)$.
				Now, if $\yElm = \xElm$ then $\dual{\spcFun}(\dual{\valFun})(\yElm)
				= \dual{\valFun}(\yElm)$.
				Otherwise, since $\dual[\dual{\valFun}(\xElm)]{\spcFun'}$ is a
				dependence map, it holds that $\dual{\spcFun}(\dual{\valFun})(\yElm) =
				\dual[{\dual{\valFun}(\xElm)}]{\spcFun'}(\dual{\valFun}_{\rst
				\QPAVSet{\dual{\qpElm'}}})(\yElm) = \dual{\valFun}_{\rst
				\QPAVSet{\dual{\qpElm'}}}(\yElm) = \dual{\valFun}(\yElm)$.
				So, Item~\ref{def:qntspc(aqnt)} of Definition~\ref{def:qntspc} of
				dependence maps is verified.
				It only remains to prove Item~\ref{def:qntspc(eqnt)}.
				Let $\yElm \in \QPEVSet{\dual{\qpElm}}$ and $\dual{\valFun[1]},
				\dual{\valFun[2]} \in \ValSet[\DSet](\QPAVSet{\dual{\qpElm}})$, with
				$\dual{\valFun[1]}_{\rst \QPDepSet(\dual{\qpElm}, \yElm)} =
				\dual{\valFun[2]}_{\rst \QPDepSet(\dual{\qpElm}, \yElm)}$.
				It is immediate to see that $\xElm \in \QPDepSet(\dual{\qpElm},
				\yElm)$, so, $\dual{\valFun[1]}(\xElm) = \dual{\valFun[2]}(\xElm)$,
				which implies	that $\dual[\dual{\valFun[1]}(\xElm)]{\spcFun'} =
				\dual[\dual{\valFun[2]}(\xElm)]{\spcFun'}$.
				At this point, again for the fact that
				$\dual[\dual{\valFun}(\xElm)]{\spcFun'}$ is a dependence map, for each
				$\dual{\valFun} \in \ValSet[\DSet](\QPAVSet{\dual{\qpElm}})$, we have
				that $\dual[\dual{\valFun[1]}(\xElm)]{\spcFun'}(\dual{\valFun[1]}_{\rst
				\QPAVSet{\dual{\qpElm'}}})(\yElm) =
				\dual[\dual{\valFun[2]}(\xElm)]{\spcFun'}(\dual{\valFun[2]}_{\rst
				\QPAVSet{\dual{\qpElm'}}})(\yElm)$.
				Thus, it holds that $\dual{\spcFun}(\dual{\valFun[1]})(\yElm) \!=\!
				\dual[\dual{\valFun[1]}(\xElm)]{\spcFun'}(\dual{\valFun[1]}_{\rst
				\QPAVSet{\dual{\qpElm'}}})[\xElm \mapsto
				\dual{\valFun[1]}(\xElm)](\yElm) =
				\dual[\dual{\valFun[2]}(\xElm)]{\spcFun'}(\dual{\valFun[2]}_{\rst
				\QPAVSet{\dual{\qpElm'}}})[\xElm \mapsto
				\dual{\valFun[2]}(\xElm)](\yElm) =
				\dual{\spcFun}(\dual{\valFun[2]})(\yElm)$.

				\hspace{0.75em}
				Finally, it is enough to observe that, by construction,
				$\dual{\spcFun}(\dual{\valFun}) \in \PSet$, for all $\dual{\valFun} \in
				\ValSet[\DSet](\QPAVSet{\dual{\qpElm}})$, since
				$\dual[{\dual{\valFun}(\xElm)}]{\spcFun'}(\dual{\valFun}_{\rst
				\QPAVSet{\dual{\qpElm'}}}) \in \PSet[{\dual{\valFun}(\xElm)}]$.
				Thus, the thesis holds for this case.
			\item
				$\qpElm = \AAll{\xElm} \cdot \qpElm'$.

				We first show that there exists $\eElm \in \DSet$ such that, for all
				$\spcFun' \in \SpcSet[\DSet](\qpElm')$, there is $\valFun' \in
				\ValSet[\DSet](\QPAVSet{\qpElm'})$ for which $\spcFun'(\valFun')
				\in \PSet[\eElm]$ holds, where the set $\PSet[\eElm]$ is defined as
				above.

				\hspace{0.75em}
				To do this, suppose by contradiction that, for all $\eElm \in \DSet$,
				there is a $\spcFun[\eElm|'] \in \SpcSet[\DSet](\qpElm')$ such that,
				for all $\valFun' \in \ValSet[\DSet](\QPAVSet{\qpElm'})$, it holds
				that $\spcFun[\eElm|'](\valFun') \not\in \PSet[\eElm]$.
				Also, consider the function $\spcFun :
				\ValSet[\DSet](\QPAVSet{\qpElm}) \to \ValSet[\DSet](\VSet)$ given by
				$\spcFun(\valFun) \defeq \spcFun[\valFun(\xElm)|'](\valFun_{\rst
				\QPAVSet{\qpElm'}})[\xElm \mapsto \valFun(\xElm)]$, for all $\valFun
				\in \ValSet[\DSet](\QPAVSet{\qpElm})$.
				Then, is possible to verify that $\spcFun \in \SpcSet[\DSet](\qpElm)$.
				Indeed, for each $\yElm \in \QPAVSet{\qpElm}$ and $\valFun \in
				\ValSet[\DSet](\QPAVSet{\qpElm})$, we have that
				$\spcFun(\valFun)(\yElm) = \spcFun[\valFun(\xElm)|'](\valFun_{\rst
				\QPAVSet{\qpElm'}})[\xElm \mapsto \valFun(\xElm)](\yElm)$.
				Now, if $\yElm = \xElm$ then $\spcFun(\valFun)(\yElm) =
				\valFun(\yElm)$.
				Otherwise, since $\spcFun[\valFun(\xElm)|']$ is a dependence map, it
				holds that $\spcFun(\valFun)(\yElm) =
				\spcFun[\valFun(\xElm)|'](\valFun_{\rst \QPAVSet{\qpElm'}})(\yElm) =
				\valFun_{\rst \QPAVSet{\qpElm'}}(\yElm) = \valFun(\yElm)$.
				So, Item~\ref{def:qntspc(aqnt)} of Definition~\ref{def:qntspc} of
				dependence maps is verified.
				It only remains to prove Item~\ref{def:qntspc(eqnt)}.
				Let $\yElm \in \QPEVSet{\qpElm}$ and $\valFun[1], \valFun[2] \in
				\ValSet[\DSet](\QPAVSet{\qpElm})$, with $\valFun[1]_{\rst
				\QPDepSet(\qpElm, \yElm)} = \valFun[2]_{\rst \QPDepSet(\qpElm,
				\yElm)}$.
				It is immediate to see that $\xElm \in \QPDepSet(\qpElm, \yElm)$, so,
				$\valFun[1](\xElm) = \valFun[2](\xElm)$, which implies	that
				$\spcFun[{\valFun[1](\xElm)}|'] = \spcFun[{\valFun[2](\xElm)}|']$.
				At this point, again for the fact that $\spcFun[\valFun(\xElm)|']$ is a
				dependence map, for each $\valFun \in \ValSet[\DSet](\QPAVSet{\qpElm})$,
				we have that $\spcFun[{\valFun[1](\xElm)}|'](\valFun[1]_{\rst
				\QPAVSet{\qpElm'}})(\yElm) =
				\spcFun[{\valFun[2](\xElm)}|'](\valFun[2]_{\rst
				\QPAVSet{\qpElm'}})(\yElm)$.
				Thus, it holds that $\spcFun(\valFun[1])(\yElm) =
				\spcFun[{\valFun[1](\xElm)}|'](\valFun[1]_{\rst
				\QPAVSet{\qpElm'}})[\xElm \mapsto \valFun[1](\xElm)](\yElm) =
				\spcFun[{\valFun[2](\xElm)}|'](\valFun[2]_{\rst
				\QPAVSet{\qpElm'}})[\xElm \mapsto \valFun[2](\xElm)](\yElm) =
				\spcFun(\valFun[2])(\yElm)$.
				Now, by the contradiction hypothesis, we have that $\spcFun(\valFun)
				\not\in \PSet$, for all $\valFun \in \ValSet(\QPAVSet{\qpElm})$, since
				$\spcFun[\valFun(\xElm)|'](\valFun_{\rst \QPAVSet{\qpElm'}}) \not\in
				\PSet[\valFun(\xElm)]$, which is in evident contradiction with the
				hypothesis.

				\hspace{0.75em}
				At this point, by the inductive hypothesis, there exists
				$\dual{\spcFun'} \in \SpcSet[\DSet](\dual{\qpElm'})$ such that, for
				all $\dual{\valFun'} \in \ValSet[\DSet](\QPAVSet{\dual{\qpElm'}})$, it
				holds that $\dual{\spcFun'}(\dual{\valFun'}) \in \PSet[\eElm]$, i.e.,
				$\dual{\spcFun'}(\dual{\valFun'})[\xElm \mapsto \eElm] \in \PSet$.

				\hspace{0.75em}
				Finally, build the function $\dual{\spcFun} :
				\ValSet[\DSet](\QPAVSet{\dual{\qpElm}}) \to \ValSet[\DSet](\VSet)$
				given by $\dual{\spcFun}(\dual{\valFun}) \defeq
				\dual{\spcFun'}(\dual{\valFun})[\xElm \mapsto \eElm]$, for all
				$\dual{\valFun} \in \ValSet[\DSet](\QPAVSet{\dual{\qpElm}}) =
				\ValSet[\DSet](\QPAVSet{\dual{\qpElm'}})$.
				It is immediate to see that $\dual{\spcFun} \in
				\SpcSet[\DSet](\dual{\qpElm})$.
				Moreover, for all valuations $\dual{\valFun} \in
				\ValSet[\DSet](\QPAVSet{\dual{\qpElm}})$, it holds that
				$\dual{\spcFun}(\dual{\valFun}) \in \PSet$.
				Thus, the thesis holds for this case too.
		\end{itemize}
		Hence, we have done with the proof of the lemma.
	\end{proof}

	At this point, we are able to give the proofs of Lemma~\ref{lmm:adjspc} of
	adjoint dependence maps, Lemma~\ref{lmm:spcvaldlt} of dependence-vs-valuation
	duality, and Lemma~\ref{lmm:encasement} of encasement characterization.

	\def\thexlemma{\ref{lmm:adjspc}}
	\begin{lemma}[Adjoint Dependence Maps]
		Let $\qpElm \in \QPSet(\VSet)$ be a quantification prefix over a set of
		variables $\VSet \subseteq \VarSet$, $\DSet$ and $\TSet$ two generic sets,
		and $\spcFun : \ValSet[\TSet \to \DSet](\QPAVSet{\qpElm}) \to
		\ValSet[\TSet \to \DSet](\VSet)$ and $\adj{\spcFun} : \TSet \to
		(\ValSet[\DSet](\QPAVSet{\qpElm}) \to \ValSet[\DSet](\VSet))$ two
		functions such that $\adj{\spcFun}$ is the adjoint of $\spcFun$.
		Then, $\spcFun \in \SpcSet[\TSet \to \DSet](\qpElm)$ iff, for all $t \in
		\TSet$, it holds that $\adj{\spcFun}(t) \in \SpcSet[\DSet](\qpElm)$.
	\end{lemma}
	\begin{proof}
		To prove the statement, it is enough to show, separately, that
		Items~\ref{def:qntspc(aqnt)} and~\ref{def:qntspc(eqnt)} of
		Definition~\ref{def:qntspc} of dependence maps hold for $\spcFun$ if the
		$\adj{\spcFun}(\tElm)$ satisfies the same items, for all $\tElm \in \TSet$,
		and vice versa.

		\emph{[Item~\ref{def:qntspc(aqnt)}, if].}
		Assume that $\adj{\spcFun}(\tElm)$ satisfies Item~\ref{def:qntspc(aqnt)},
		for each $\tElm \in \TSet$, i.e., $\adj{\spcFun}(\tElm)(\valFun)_{\rst
		\QPAVSet{\qpElm}} = \valFun$, for all $\valFun \in
		\ValSet[\DSet](\QPAVSet{\qpElm})$.
		Then, we have that $\adj{\spcFun}(\tElm)(\flip{\gFun}(\tElm)) =
		\flip{\gFun}(\tElm)$, so,
		$\adj{\spcFun}(\tElm)(\flip{\gFun}(\tElm))(\xElm) =
		\flip{\gFun}(\tElm)(\xElm)$, for all $\gFun \in \ValSet[\TSet \to
		\DSet](\QPAVSet{\qpElm})$ and $\xElm \in \QPAVSet{\qpElm}$.
		By hypothesis, we have that $\spcFun(\gFun)(\xElm)(\tElm) =
		\adj{\spcFun}(\tElm)(\flip{\gFun}(\tElm))(\xElm)$, thus
		$\spcFun(\gFun)(\xElm)(\tElm) = \flip{\gFun}(\tElm)(\xElm) =
		\gFun(\xElm)(\tElm)$, which means that $\spcFun(\gFun)_{\rst
		\QPAVSet{\qpElm}} = \gFun$, for all $\gFun \in \ValSet[\TSet \to
		\DSet](\QPAVSet{\qpElm})$.

		\emph{[Item~\ref{def:qntspc(aqnt)}, only if].}
		Assume now that $\spcFun$ satisfies Item~\ref{def:qntspc(aqnt)},
		i.e., $\spcFun(\gFun)_{\rst \QPAVSet{\qpElm}} = \gFun$, for all $\gFun \in
		\ValSet[\TSet \to \DSet](\QPAVSet{\qpElm})$.
		Then, we have that $\spcFun(\gFun)(\xElm)(\tElm) = \gFun(\xElm)(\tElm)$,
		for all $\xElm \in \QPAVSet{\qpElm}$ and $\tElm \in \TSet$.
		By hypothesis, we have that
		$\adj{\spcFun}(\tElm)(\flip{\gFun}(\tElm))(\xElm) =
		\spcFun(\gFun)(\xElm)(\tElm)$, so,
		$\adj{\spcFun}(\tElm)(\flip{\gFun}(\tElm))(\xElm) = \gFun(\xElm)(\tElm) =
		\flip{\gFun}(\tElm)(\xElm)$, which means that
		$\adj{\spcFun}(\tElm)(\flip{\gFun}(\tElm))_{\rst \QPAVSet{\qpElm}} =
		\flip{\gFun}(\tElm)$.
		Now, since for each $\valFun \in \ValSet[\DSet](\QPAVSet{\qpElm})$, there
		is an $\gFun \in \ValSet[\TSet \to \DSet](\QPAVSet{\qpElm})$ such that
		$\flip{\gFun}(\tElm) = \valFun$, we obtain that
		$\adj{\spcFun}(\tElm)(\valFun)_{\rst \QPAVSet{\qpElm}} = \valFun$, for all
		$\valFun \in \ValSet[\DSet](\QPAVSet{\qpElm})$ and $\tElm \in \TSet$.

		\emph{[Item~\ref{def:qntspc(eqnt)}, if].}
		Assume that $\adj{\spcFun}(\tElm)$ satisfies Item~\ref{def:qntspc(eqnt)},
		for each $\tElm \in \TSet$, i.e., $\adj{\spcFun}(\tElm)(\valFun[1])(\xElm)
		= \adj{\spcFun}(\tElm)(\valFun[2])(\xElm)$, for all $\valFun[1],
		\valFun[2] \in \ValSet[\DSet](\QPAVSet{\qpElm})$ and $\xElm \in
		\QPEVSet{\qpElm}$ such that $\valFun[1]_{\rst \QPDepSet(\qpElm, \xElm)} =
		\valFun[2]_{\rst \QPDepSet(\qpElm, \xElm)}$.
		Then, we have that $\adj{\spcFun}(\tElm)(\flip{\gFun[1]}(\tElm))(\xElm) =
		\adj{\spcFun}(\tElm)(\flip{\gFun[2]}(\tElm))(\xElm)$, for all $\gFun[1],
		\gFun[2] \in \ValSet[\TSet \to \DSet](\QPAVSet{\qpElm})$ such that
		$\gFun[1]_{\rst \QPDepSet(\qpElm, \xElm)} = \gFun[2]_{\rst
		\QPDepSet(\qpElm, \xElm)}$.
		By hypothesis, we have that $\spcFun(\gFun[1])(\xElm)(\tElm) =
		\adj{\spcFun}(\tElm)(\flip{\gFun[1]}(\tElm))(\xElm)$ and
		$\adj{\spcFun}(\tElm)(\flip{\gFun[2]}(\tElm))(\xElm) =
		\spcFun(\gFun[2])(\xElm)(\tElm)$, thus $\spcFun(\gFun[1])(\xElm)(\tElm) =
		\spcFun(\gFun[2])(\xElm)(\tElm)$.
		Hence, $\spcFun(\gFun[1])(\xElm) \!=\! \spcFun(\gFun[2])(\xElm)$, for all
		$\gFun[1], \gFun[2] \in \ValSet[\TSet \to \DSet](\QPAVSet{\qpElm})$ and
		$\xElm \in \QPEVSet{\qpElm}$ such that $\gFun[1]_{\rst \QPDepSet(\qpElm,
		\xElm)} = \gFun[2]_{\rst \QPDepSet(\qpElm, \xElm)}$.

		\emph{[Item~\ref{def:qntspc(eqnt)}, only if].}
		Assume that $\spcFun$ satisfies Item~\ref{def:qntspc(eqnt)}, i.e.,
		$\spcFun(\gFun[1])(\xElm) = \spcFun(\gFun[2])(\xElm)$, for all $\gFun[1],
		\gFun[2] \in \ValSet[\TSet \to \DSet](\QPAVSet{\qpElm})$ and $\xElm \in
		\QPEVSet{\qpElm}$ such that $\gFun[1]_{\rst \QPDepSet(\qpElm, \xElm)} =
		\gFun[2]_{\rst \QPDepSet(\qpElm, \xElm)}$.
		Then, we have that $\spcFun(\gFun[1])(\xElm)(\tElm) =
		\spcFun(\gFun[2])(\xElm)(\tElm)$, for all $\tElm \in \TSet$.
		By hypothesis, we have that
		$\adj{\spcFun}(\tElm)(\flip{\gFun[1]}(\tElm))(\xElm) \allowbreak =
		\spcFun(\gFun[1])(\xElm)(\tElm)$ and $\spcFun(\gFun[2])(\xElm)(\tElm) =
		\adj{\spcFun}(\tElm)(\flip{\gFun[2]}(\tElm))(\xElm)$, hence
		$\adj{\spcFun}(\tElm)(\flip{\gFun[1]}(\tElm))(\xElm) =
		\adj{\spcFun}(\tElm)(\flip{\gFun[2]}(\tElm))(\xElm)$.
		Now, since for each $\valFun[1], \valFun[2] \in
		\ValSet[\DSet](\QPAVSet{\qpElm})$, with $\valFun[1]_{\rst
		\QPDepSet(\qpElm, \xElm)} = \valFun[2]_{\rst \QPDepSet(\qpElm, \xElm)}$,
		there are $\gFun[1], \gFun[2] \in \ValSet[\TSet \to
		\DSet](\QPAVSet{\qpElm})$ such that $\flip{\gFun[1]}(\tElm) = \valFun[1]$
		and $\flip{\gFun[2]}(\tElm) = \valFun[2]$, with $\gFun[1]_{\rst
		\QPDepSet(\qpElm, \xElm)} = \gFun[2]_{\rst \QPDepSet(\qpElm, \xElm)}$, we
		obtain that $\adj{\spcFun}(\tElm)(\valFun[1])(\xElm) =
		\adj{\spcFun}(\tElm)(\valFun[2])(\xElm)$, for all $\valFun[1], \valFun[2]
		\in \ValSet[\DSet](\QPAVSet{\qpElm})$ and $\xElm \in \QPEVSet{\qpElm}$
		such that $\valFun[1]_{\rst \QPDepSet(\qpElm, \xElm)} = \valFun[2]_{\rst
		\QPDepSet(\qpElm, \xElm)}$.
	\end{proof}

	\def\thexlemma{\ref{lmm:spcvaldlt}}
	\begin{lemma}[Dependence-vs-Valuation Duality]
		Let $\GName$ be a \CGS, $\sElm \in \StSet$ one of its states, $\PSet
		\subseteq \PthSet(\sElm)$ a set of paths, $\qpElm \in \QPSet(\VSet)$ a
		quantification prefix over a set of variables $\VSet \subseteq \VarSet$,
		and $\bpElm \in \BndSet(\VSet)$ a binding.
		Then, player even wins the \TPG\ $\HName(\GName, \sElm, \PSet, \qpElm,
		\bpElm)$ iff player odd wins the dual \TPG\ $\HName(\GName, \sElm,
		\PthSet(\sElm) \setminus \PSet, \dual{\qpElm}, \bpElm)$.
	\end{lemma}
	\begin{proof}
		Let $\AName$ and $\dual{\AName}$ be, respectively, the two \TPA s
		$\AName(\GName, \sElm, \qpElm, \bpElm)$ and $\AName(\GName, \sElm,
		\dual{\qpElm}, \bpElm)$.
		It is easy to observe that $\PosESet[\AName] = \PosESet[\dual{\AName}] =
		\TrkSet(\sElm)$.
		Moreover, it holds that $\PosOSet[\AName] = \set{ \trkElm \cdot
		(\lst{\trkElm}, \spcFun) }{ \trkElm \in \TrkSet(\sElm) \land \spcFun \in
		\SpcSet[\AcSet](\qpElm) }$ and $\PosOSet[\dual{\AName}] = \set{ \trkElm
		\cdot (\lst{\trkElm}, \dual{\spcFun}) }{ \trkElm \in \TrkSet(\sElm) \land
		\dual{\spcFun} \in \SpcSet[\AcSet](\dual{\qpElm}) }$.
		We now prove, separately, the two directions of the statement.

		\emph{[Only if].}
		Suppose that player even wins the \TPG\ $\HName(\GName, \sElm, \PSet,
		\qpElm, \bpElm)$.
		Then, there exists an even scheme $\scheFun \in \SchESet[\AName]$ such
		that, for all odd schemes $\schoFun \in \SchOSet[\AName]$, it holds that
		$\mtcFun[\AName](\scheFun, \schoFun) \in \PSet$.
		Now, to prove that odd wins the dual \TPG\ $\HName(\GName, \sElm,
		\PthSet(\sElm) \setminus \PSet, \dual{\qpElm}, \bpElm)$, we have to show
		that there exists an odd scheme $\dual{\schoFun} \in
		\SchOSet[\dual{\AName}]$ such that, for all even schemes $\dual{\scheFun}
		\in \SchESet[\dual{\AName}]$, it holds that
		$\mtcFun[\dual{\AName}](\dual{\scheFun}, \dual{\schoFun}) \in \PSet$.

		To do this, let us first consider a function $\zFun :
		\SpcSet[\AcSet](\qpElm) \times \SpcSet[\AcSet](\dual{\qpElm}) \to
		\ValSet[\AcSet](\VSet)$ such that $\zFun(\spcFun, \dual{\spcFun}) =
		\spcFun(\zFun(\spcFun, \dual{\spcFun})_{\rst \QPAVSet{\qpElm}}) =
		\dual{\spcFun}(\zFun(\spcFun, \dual{\spcFun})_{\rst
		\QPAVSet{\dual{\qpElm}}})$, for all $\spcFun \in \SpcSet[\AcSet](\qpElm)$
		and $\dual{\spcFun} \in \SpcSet[\AcSet](\dual{\qpElm})$.
		The existence of such a function is ensured by Lemma~\ref{lmm:scpinc} on
		the dependence incidence.

		Now, define the odd scheme $\dual{\schoFun} \in \SchOSet[\dual{\AName}]$
		in $\dual{\AName}$ as follows: $\dual{\schoFun}(\trkElm \cdot
		(\lst{\trkElm}, \dual{\spcFun})) \defeq \trnFun(\lst{\trkElm}, \allowbreak
		\zFun(\spcFun, \dual{\spcFun}) \cmp \bndFun[\bpElm])$, for all $\trkElm
		\in \TrkSet(\sElm)$ and $\dual{\spcFun} \in
		\SpcSet[\AcSet](\dual{\qpElm})$, where $\spcFun \in
		\SpcSet[\AcSet](\qpElm)$ is such that $\scheFun(\trkElm) = (\lst{\trkElm},
		\spcFun)$.
		Moreover, let $\dual{\scheFun} \in \SchESet[\dual{\AName}]$ be a generic
		even scheme in $\dual{\AName}$ and consider the derived odd scheme
		$\schoFun \in \SchOSet[\AName]$ in $\AName$ defined as follows:
		$\schoFun(\trkElm \cdot (\lst{\trkElm}, \spcFun)) \defeq
		\trnFun(\lst{\trkElm}, \zFun(\spcFun, \dual{\spcFun}) \cmp
		\bndFun[\bpElm])$, for all $\trkElm \in \TrkSet(\sElm)$ and $\spcFun \in
		\SpcSet[\AcSet](\qpElm)$, where $\dual{\spcFun} \in
		\SpcSet[\AcSet](\dual{\qpElm})$ is such that $\dual{\scheFun}(\trkElm) =
		(\lst{\trkElm}, \dual{\spcFun})$.

		At this point, it remains only to prove that $\mtcElm = \dual{\mtcElm}$,
		where $\mtcElm \defeq \mtcFun[\AName](\scheFun, \schoFun)$ and
		$\dual{\mtcElm} \defeq \mtcFun[\dual{\AName}](\dual{\scheFun},
		\dual{\schoFun})$.
		To do this, we proceed by induction on the prefixes of the matches, i.e.,
		we show that $(\mtcElm)_{\leq i} = (\dual{\mtcElm})_{\leq i}$, for all $i
		\in \SetN$.
		The base case is immediate by definition of match, since we have that
		$(\mtcElm)_{\leq 0} = \sElm = (\dual{\mtcElm})_{\leq 0}$.
		Now, as inductive case, suppose that $(\mtcElm)_{\leq i} =
		(\dual{\mtcElm})_{\leq i}$, for $i \in \SetN$.
		By the definition of match, we have that $(\mtcElm)_{i + 1} =
		\schoFun((\mtcElm)_{\leq i} \cdot \scheFun((\mtcElm)_{\leq i}))$ and
		$(\dual{\mtcElm})_{i + 1} = \dual{\schoFun}((\dual{\mtcElm})_{\leq i}
		\cdot \dual{\scheFun}((\dual{\mtcElm})_{\leq i}))$.
		Moreover, by the inductive hypothesis, it follows that
		$\schoFun((\mtcElm)_{\leq i} \cdot \scheFun((\mtcElm)_{\leq i})) =
		\schoFun((\dual{\mtcElm})_{\leq i} \cdot \scheFun((\dual{\mtcElm})_{\leq
		i}))$.
		At this point, let $\spcFun \in \SpcSet[\AcSet](\qpElm)$ and
		$\dual{\spcFun} \in \SpcSet[\AcSet](\dual{\qpElm})$ be two quantification
		dependence maps such that $\scheFun((\dual{\mtcElm})_{\leq i}) =
		((\dual{\mtcElm})_{i}, \spcFun)$ and
		$\dual{\scheFun}((\dual{\mtcElm})_{\leq i}) = ((\dual{\mtcElm})_{i},
		\dual{\spcFun})$.
		Consequently, by substituting the values of the even schemes $\scheFun$
		and $\dual{\scheFun}$, it holds that $(\mtcElm)_{i + 1} =
		\schoFun((\dual{\mtcElm})_{\leq i} \cdot ((\dual{\mtcElm})_{i}, \spcFun))$
		and $(\dual{\mtcElm})_{i + 1} = \dual{\schoFun}((\dual{\mtcElm})_{\leq i}
		\cdot ((\dual{\mtcElm})_{i}, \dual{\spcFun}))$.
		Furthermore, by the definition of the odd schemes $\schoFun$ and
		$\dual{\schoFun}$, it follows that $\schoFun((\dual{\mtcElm})_{\leq i}
		\cdot ((\dual{\mtcElm})_{i}, \spcFun)) = \trnFun((\dual{\mtcElm})_{i},
		\zFun(\spcFun, \dual{\spcFun}) \cmp \bpFun[\bpElm]) =
		\dual{\schoFun}((\dual{\mtcElm})_{\leq i} \cdot ((\dual{\mtcElm})_{i},
		\dual{\spcFun}))$.
		Thus, we have that $(\mtcElm)_{i + 1} = (\dual{\mtcElm})_{i + 1}$, which
		implies $(\mtcElm)_{\leq i + 1} = (\dual{\mtcElm})_{\leq i + 1}$.

		\emph{[If].}
		Suppose that player odd wins the dual \TPG\ $\HName(\GName, \sElm,
		\PthSet(\sElm) \setminus \PSet, \dual{\qpElm}, \bpElm)$.
		Then, there exists an odd scheme $\dual{\schoFun} \in
		\SchOSet[\dual{\AName}]$ such that, for all even schemes $\dual{\scheFun}
		\in \SchESet[\dual{\AName}]$, it holds that
		$\mtcFun[\dual{\AName}](\dual{\scheFun}, \dual{\schoFun}) \in \PSet$.
		Now, to prove that even wins the \TPG\ $\HName(\GName, \sElm, \PSet,
		\qpElm, \bpElm)$, we have to show that there exists an even scheme
		$\scheFun \in \SchESet[\AName]$ such that, for all odd schemes $\schoFun
		\in \SchOSet[\AName]$, it holds that $\mtcFun[\AName](\scheFun, \schoFun)
		\in \PSet$.

		To do this, let us first consider the two functions $\gFun :
		\TrkSet(\sElm) \to \pow{\ValSet[\AcSet](\VSet)}$ and $\hFun :
		\TrkSet(\sElm) \to \pow{\StSet}$ such that $\gFun(\trkElm) \defeq \set{
		\dual{\spcFun}(\dual{\valFun}) }{ \dual{\spcFun} \in
		\SpcSet[\AcSet](\dual{\qpElm}) \land \dual{\valFun} \in
		\ValSet[\AcSet](\QPAVSet{\dual{\qpElm}}) \land \dual{\schoFun}(\trkElm
		\cdot (\lst{\trkElm}, \dual{\spcFun})) = \trnFun(\lst{\trkElm},
		\dual{\spcFun}(\dual{\valFun}) \cmp \bndFun[\bpElm]) }$ and
		$\hFun(\trkElm) \defeq \set{ \dual{\schoFun}(\trkElm \cdot (\lst{\trkElm},
		\dual{\spcFun})) }{ \dual{\spcFun} \in \SpcSet[\AcSet](\dual{\qpElm}) }$,
		for all $\trkElm \in \TrkSet(\sElm)$.
		Now, it is easy to see that, for each $\trkElm \in \TrkSet(\sElm)$ and
		$\dual{\spcFun} \in \SpcSet[\AcSet](\dual{\qpElm})$, there is
		$\dual{\valFun} \in \ValSet[\AcSet](\QPAVSet{\dual{\qpElm}})$ such that
		$\dual{\spcFun}(\dual{\valFun}) \in \gFun(\trkElm)$.
		Consequently, by Lemma~\ref{lmm:scpdlz} on dependence dualization, for all
		$\trkElm \in \TrkSet(\sElm)$, there is $\spcFun[\trkElm] \in
		\SpcSet[\AcSet](\qpElm)$ such that, for each $\valFun \in
		\ValSet[\AcSet](\QPAVSet{\qpElm})$, it holds that
		$\spcFun[\trkElm](\valFun) \in \gFun(\trkElm)$, and so,
		$\trnFun(\lst{\trkElm}, \spcFun[\trkElm](\valFun) \cmp \bndFun[\bpElm])
		\in \hFun(\trkElm)$.

		Now, define the even scheme $\scheFun \in \SchESet[\AName]$ in $\AName$ as
		follows: $\scheFun(\trkElm) \defeq (\lst{\trkElm}, \spcFun[\trkElm])$, for
		all $\trkElm \in \TrkSet(\sElm)$.
		Moreover, let $\schoFun \in \SchESet[\AName]$ be a generic odd scheme in
		$\AName$ and consider the derived even scheme $\dual{\scheFun} \in
		\SchESet[\dual{\AName}]$ in $\dual{\AName}$ defined as follows:
		$\dual{\scheFun}(\trkElm) \defeq (\lst{\trkElm},
		\dual[\trkElm]{\spcFun})$, for all $\trkElm \in \TrkSet(\sElm)$, where
		$\dual[\trkElm]{\spcFun} \in \SpcSet[\AcSet](\dual{\qpElm})$ is such that
		$\schoFun(\trkElm \cdot (\lst{\trkElm}, \spcFun[\trkElm])) =
		\dual{\schoFun}(\trkElm \cdot (\lst{\trkElm}, \dual[\trkElm]{\spcFun}))$.
		The existence of such a dependence map is ensure by the previous membership
		of the successor of $\lst{\trkElm}$ in $\hFun(\trkElm)$.

		At this point, it remains only to prove that $\mtcElm = \dual{\mtcElm}$,
		where $\mtcElm \defeq \mtcFun[\AName](\scheFun, \schoFun)$ and
		$\dual{\mtcElm} \defeq \mtcFun[\dual{\AName}](\dual{\scheFun},
		\dual{\schoFun})$.
		To do this, we proceed by induction on the prefixes of the matches, i.e.,
		we show that $(\mtcElm)_{\leq i} = (\dual{\mtcElm})_{\leq i}$, for all $i
		\in \SetN$.
		The base case is immediate by definition of match, since we have that
		$(\mtcElm)_{\leq 0} = \sElm = (\dual{\mtcElm})_{\leq 0}$.
		Now, as inductive case, suppose that $(\mtcElm)_{\leq i} =
		(\dual{\mtcElm})_{\leq i}$, for $i \in \SetN$.
		By the definition of match, we have that $(\mtcElm)_{i + 1} =
		\schoFun((\mtcElm)_{\leq i} \cdot \scheFun((\mtcElm)_{\leq i}))$ and
		$(\dual{\mtcElm})_{i + 1} = \dual{\schoFun}((\dual{\mtcElm})_{\leq i}
		\cdot \dual{\scheFun}((\dual{\mtcElm})_{\leq i}))$.
		Moreover, by the inductive hypothesis, it follows that
		$\schoFun((\mtcElm)_{\leq i} \cdot \scheFun((\mtcElm)_{\leq i})) =
		\schoFun((\dual{\mtcElm})_{\leq i} \cdot \scheFun((\dual{\mtcElm})_{\leq
		i}))$.
		Now, by substituting the values of the even schemes $\scheFun$ and
		$\dual{\scheFun}$, we have that $(\mtcElm)_{i + 1} =
		\schoFun((\dual{\mtcElm})_{\leq i} \cdot ((\dual{\mtcElm})_{i},
		\spcFun_{(\dual{\mtcElm})_{\leq i}}))$ and $(\dual{\mtcElm})_{i + 1} =
		\dual{\schoFun}((\dual{\mtcElm})_{\leq i} \cdot ((\dual{\mtcElm})_{i},
		\dual{\spcFun}_{\dual{\mtcElm}_{\leq i}}))$.
		At this point, due to the choice of the dependence map
		$\dual{\spcFun}_{(\dual{\mtcElm})_{\leq i}}$, it holds that
		$\schoFun((\dual{\mtcElm})_{\leq i} \cdot ((\dual{\mtcElm})_{i},
		\spcFun_{(\dual{\mtcElm})_{\leq i}})) =
		\dual{\schoFun}((\dual{\mtcElm})_{\leq i} \cdot ((\dual{\mtcElm})_{i},
		\dual{\spcFun}_{(\dual{\mtcElm})_{\leq i}}))$.
		Thus, we have that $(\mtcElm)_{i + 1} = (\dual{\mtcElm})_{i + 1}$, which
		implies $(\mtcElm)_{\leq i + 1} = (\dual{\mtcElm})_{\leq i + 1}$.
	\end{proof}

	\def\thexlemma{\ref{lmm:encasement}}
	\begin{lemma}[Encasement Characterization]
		Let $\GName$ be a \CGS, $\sElm \in \StSet$ one of its states, $\PSet
		\subseteq \PthSet(\sElm)$ a set of paths, $\qpElm \in \QPSet(\VSet)$ a
		quantification prefix over a set of variables $\VSet \subseteq \VarSet$,
		and $\bpElm \in \BndSet(\VSet)$ a binding.
		Then, the following hold:
		\begin{enumerate}[(i)]
			\item
				player even wins $\HName(\GName, \sElm, \PSet, \qpElm, \bpElm)$ iff
				$\PSet$ is an encasement w.r.t.\ $\qpElm$ and $\bpElm$;
			\item
				if player odd wins $\HName(\GName, \sElm, \PSet, \qpElm, \bpElm)$ then
				$\PSet$ is not an encasement w.r.t.\ $\qpElm$ and $\bpElm$;
			\item
				if $\PSet$ is a Borelian set and it is not an encasement w.r.t.\
				$\qpElm$ and $\bpElm$ then player odd wins $\HName(\GName, \sElm,
				\PSet, \qpElm, \bpElm)$.
		\end{enumerate}
	\end{lemma}
	\begin{proof}
		\emph{[Item~\ref{lmm:encasement(ewin)}, only if].}
		Suppose that player even wins the \TPG\ $\HName(\GName, \sElm, \PSet,
		\qpElm, \bpElm)$.
		Then, there exists an even scheme $\scheFun \in \SchESet$ such that, for
		all odd schemes $\schoFun \in \SchOSet$, it holds that $\mtcFun(\scheFun,
		\schoFun) \in \PSet$.
		Now, to prove the statement, we have to show that there exists an elementary
		dependence map $\spcFun \in \ESpcSet[ {\StrSet(\sElm)} ](\qpElm)$ such that,
		for all assignments $\asgFun \in \AsgSet(\QPAVSet{\qpElm}, \sElm)$, it holds
		that $\playFun(\spcFun(\asgFun) \cmp \bndFun[\bpElm], \sElm) \in \PSet$.

		To do this, consider the function $\wFun : \TrkSet(\sElm) \to
		\SpcSet[\AcSet](\qpElm)$ constituting the projection of $\scheFun$ on the
		second component of its codomain, i.e., for all $\trkElm \in
		\TrkSet(\sElm)$, it holds that $\scheFun(\trkElm) = (\lst{\trkElm},
		\wFun(\trkElm))$.
		By Lemma~\ref{lmm:adjspc} on adjoint dependence maps, there exists an
		elementary dependence map $\spcFun \in \ESpcSet[ {\StrSet(\sElm)} ](\qpElm)$
		for which $\wFun$ is the adjoint, i.e., $\wFun = \adj{\spcFun}$.
		Moreover, let $\asgFun \in \AsgSet(\QPAVSet{\qpElm}, \sElm)$ be a generic
		assignment and consider the derived odd scheme $\schoFun \in \SchOSet$
		defined ad follows: $\schoFun(\trkElm \cdot (\lst{\trkElm},
		\spcFun')) = \trnFun(\lst{\trkElm}, \spcFun'(\flip{\asgFun}(\trkElm)) \cmp
		\bndFun[\bpElm])$, for all $\trkElm \in \TrkSet(\sElm)$ and $\spcFun' \in
		\SpcSet[\AcSet](\qpElm)$.

		At this point, it remains only to prove that $\playElm = \mtcElm$, where
		$\playElm \defeq \playFun(\spcFun(\asgFun) \cmp \bndFun[\bpElm], \sElm)$
		and $\mtcElm \defeq \mtcFun(\scheFun, \schoFun)$.
		To do this, we proceed by induction on the prefixes of both the play and
		the match, i.e., we show that $(\playElm)_{\leq i} = (\mtcElm)_{\leq i}$,
		for all $i \in \SetN$.
		The base case is immediate by definition, since we have that
		$(\playElm)_{\leq 0} = \sElm = (\mtcElm)_{\leq 0}$.
		Now, as inductive case, suppose that $(\playElm)_{\leq i} =
		(\mtcElm)_{\leq i}$, for $i \in \SetN$.
		On one hand, by the definition of match, we have that $(\mtcElm)_{i + 1} =
		\schoFun((\mtcElm)_{\leq i} \cdot \scheFun((\mtcElm)_{\leq i}))$, from
		which, by substituting the value of the even scheme $\scheFun$, we derive
		$(\mtcElm)_{i + 1} = \schoFun((\mtcElm)_{\leq i} \cdot ((\mtcElm)_{i},
		\adj{\spcFun}((\mtcElm)_{\leq i})))$.
		On the other hand, by the definition of play, we have that $(\pthElm)_{i +
		1} = \trnFun((\pthElm)_{i}, \adj{\spcFun}((\pthElm)_{\leq
		i})(\flip{\asgFun}((\pthElm)_{\leq i}))\cmp \bpFun[\bpElm])$, from which,
		by using the definition of the odd scheme $\schoFun$, we derive
		$(\pthElm)_{i + 1} = \schoFun((\pthElm)_{\leq i} \cdot ((\pthElm)_{i},
		\adj{\spcFun}((\pthElm)_{\leq i})))$.
		Then, by the inductive hypothesis, we have that $(\mtcElm)_{i + 1} =
		\schoFun((\mtcElm)_{\leq i} \cdot ((\mtcElm)_{i},
		\adj{\spcFun}((\mtcElm)_{\leq i}))) = \schoFun((\pthElm)_{\leq i} \cdot
		((\pthElm)_{i}, \adj{\spcFun}((\pthElm)_{\leq i}))) = (\pthElm)_{i + 1}$,
		which implies $(\mtcElm)_{\leq i + 1} = (\pthElm)_{\leq i + 1}$.

		\emph{[Item~\ref{lmm:encasement(ewin)}, if].}
		Suppose that $\PSet$ is an encasement w.r.t.\ $\qpElm$ and $\bpElm$.
		Then, there exists an elementary dependence map $\spcFun \in \ESpcSet[
		{\StrSet(\sElm)} ](\qpElm)$ such that, for all assignments $\asgFun \in
		\AsgSet(\QPAVSet{\qpElm}, \sElm)$, it holds that $\playFun(\spcFun(\asgFun)
		\cmp \bndFun[\bpElm], \sElm) \in \PSet$.
		Now, to prove the statement, we have to show that there exists an even
		scheme $\scheFun \in \SchESet$ such that, for all odd schemes $\schoFun
		\in \SchOSet$, it holds that $\mtcFun(\scheFun, \schoFun) \in \PSet$.

		To do this, consider the even scheme $\scheFun \in \SchESet$ defined as
		follows: $\scheFun(\trkElm) \!\defeq\! (\lst{\trkElm},
		\adj{\spcFun}(\trkElm))$, for all $\trkElm \in \TrkSet(\sElm)$.
		Observe that, by Lemma~\ref{lmm:adjspc} on adjoint dependence maps, the
		definition is well-formed.
		Moreover, let $\schoFun \in \SchOSet$ be a generic odd scheme and consider
		a derived assignment $\asgFun \in \AsgSet(\QPAVSet{\qpElm}, \sElm)$
		satisfying the following property: $\flip{\asgFun}(\trkElm) \in \set{
		\valFun \in \ValSet[\AcSet](\QPAVSet{\qpElm}) }{ \schoFun(\trkElm \cdot
		(\lst{\trkElm}, \adj{\spcFun}(\trkElm))) = \trnFun(\lst{\trkElm},
		\adj{\spcFun}(\valFun) \cmp \bndFun[\bpElm]) }$, for all $\trkElm \in
		\TrkSet(\sElm)$.

		At this point, it remains only to prove that $\playElm = \mtcElm$, where
		$\playElm \defeq \playFun(\spcFun(\asgFun) \cmp \bndFun[\bpElm], \sElm)$
		and $\mtcElm \defeq \mtcFun(\scheFun, \schoFun)$.
		To do this, we proceed by induction on the prefixes of both the play and
		the match, i.e., we show that $(\playElm)_{\leq i} = (\mtcElm)_{\leq i}$,
		for all $i \in \SetN$.
		The base case is immediate by definition, since we have that
		$(\playElm)_{\leq 0} = \sElm = (\mtcElm)_{\leq 0}$.
		Now, as inductive case, suppose that $(\playElm)_{\leq i} =
		(\mtcElm)_{\leq i}$, for $i \in \SetN$.
		On one hand, by the definition of match, we have that $(\mtcElm)_{i + 1} =
		\schoFun((\mtcElm)_{\leq i} \cdot \scheFun((\mtcElm)_{\leq i}))$, from
		which, by the definition of the even scheme $\scheFun$, we derive
		$(\mtcElm)_{i + 1} = \schoFun((\mtcElm)_{\leq i} \cdot ((\mtcElm)_{i},
		\adj{\spcFun}((\mtcElm)_{\leq i})))$.
		On the other hand, by the definition of play, we have that $(\pthElm)_{i +
		1} = \trnFun((\pthElm)_{i}, \adj{\spcFun}((\pthElm)_{\leq
		i})(\flip{\asgFun}((\pthElm)_{\leq i}))\cmp \bpFun[\bpElm])$, from which,
		by the choice of the assignment $\asgFun$, we derive $(\pthElm)_{i + 1} =
		\schoFun((\pthElm)_{\leq i} \cdot ((\pthElm)_{i},
		\adj{\spcFun}((\pthElm)_{\leq i})))$.
		Then, by the inductive hypothesis, we have that $(\mtcElm)_{i + 1} =
		\schoFun((\mtcElm)_{\leq i} \cdot ((\mtcElm)_{i},
		\adj{\spcFun}((\mtcElm)_{\leq i}))) = \schoFun((\pthElm)_{\leq i} \cdot
		((\pthElm)_{i}, \adj{\spcFun}((\pthElm)_{\leq i}))) = (\pthElm)_{i + 1}$,
		which implies $(\mtcElm)_{\leq i + 1} = (\pthElm)_{\leq i + 1}$.

		\emph{[Item~\ref{lmm:encasement(owin-dir)}].}
		If player odd wins the \TPG\ $\HName(\GName, \sElm, \PSet, \qpElm,
		\bpElm)$, we have that player even does not win the same game.
		Consequently, by Item~\ref{lmm:encasement(ewin)}, it holds that $\PSet$ is
		not an encasement w.r.t.\ $\qpElm$ and $\bpElm$.

		\emph{[Item~\ref{lmm:encasement(owin-inv)}].}
		If $\PSet$ is not an encasement w.r.t.\ $\qpElm$ and $\bpElm$, by
		Item~\ref{lmm:encasement(ewin)}, we have that player even does not win the
		\TPG\ $\HName(\GName, \sElm, \PSet, \qpElm, \bpElm)$.
		Now, since $\PSet$ is Borelian, by the determinacy
		theorem~\cite{Mar75,Mar85}, it holds that player odd wins the same game.
	\end{proof}

\end{section}






\begin{ack}

	We wish to thank the authors of \cite{CLM10} for their helpful comments and
	discussions on a preliminary version of the paper.

\end{ack}


	\bibliographystyle{acmsmall}
	\bibliography{References.bib}

	\received{? 20?}{? 20?}{? 20?}

\end{document}